\documentclass[10pt]{article}
\pdfoutput=1

\usepackage{amsfonts,amsmath,amssymb}
\usepackage{mathtools}
\usepackage{stackrel}
\usepackage{leftidx}
\usepackage[amsmath,amsthm,thmmarks]{ntheorem}

\usepackage{euscript}
\usepackage{pxfonts}
\usepackage{dsfont}
\usepackage{bbold}

\usepackage{enumerate}

\usepackage{hyperref}
\usepackage{cleveref}

\usepackage{xcolor}
\usepackage{tcolorbox}
\usepackage{graphicx}
\usepackage{subfigure}

\usepackage[all,2cell,arrow,matrix]{xy} \UseAllTwocells \SilentMatrices
\usepackage{tikz-cd}
\usepackage{tikz}
\usetikzlibrary{arrows,arrows.meta,decorations.markings,decorations.pathreplacing,calc}

\usepackage{verbatim}
\usepackage{comment}

\usepackage{appendix}

\usepackage{epstopdf} 

\usepackage{sectsty}
\sectionfont{\Large}

\usepackage[nottoc]{tocbibind} 
\tikzset{->-/.style={decoration={markings,mark=at position #1 with {\arrow{Stealth}}},postaction={decorate}},->-/.default=0.55}
\colorlet{e_ext}{red}
\colorlet{m_ext}{blue!50}
\tikzset{e_str/.style={very thick,red!80}}
\tikzset{m_str/.style={very thick,blue!80}}
\tikzset{m_dual_str/.style={thick,dashed,blue}}
\tikzset{link_label/.style={scale=0.8,black}}

\theoremstyle{definition}

\newtheorem{thm}{Theorem}[section]

{
\theoremsymbol{\mbox{$\spadesuit$}}

}
{
\theoremsymbol{\mbox{$\blacklozenge$}}
\newtheorem{expl}[thm]{Example}
}
{
\theoremsymbol{\mbox{$\diamondsuit$}}
\newtheorem{rem}[thm]{Remark}
}
\qedsymbol{\mbox{$\square$}}

\numberwithin{equation}{section}
\numberwithin{thm}{section}

\newcommand\be            {\begin{equation}}
\newcommand\ee            {\end{equation}}
\newcommand\bea           {\begin{eqnarray}}
\newcommand\eea         {\end{eqnarray}}
\newcommand\bnu          {\begin{enumerate}}
\newcommand\enu          {\end{enumerate}}
\newcommand\bit          {\begin{itemize}}
\newcommand\eit          {\end{itemize}}

\newcommand{\pf}{\begin{proof}}
\newcommand{\epf}{\qed\end{proof}}

\DeclareMathOperator*{\medotimes}{\vcenter{\hbox{{\scalebox{0.8}{$\bigotimes$}}}}}

\DeclareMathAlphabet{\mathcal}{OMS}{cmsy}{m}{n}	
\DeclareMathAlphabet{\mathsf}{OT1}{cmss}{m}{n}	

\newcommand\Cb			{\mathbb{C}}

\newcommand\Zb			{\mathbb{Z}}
\newcommand\Z			{\mathcal{Z}}

\newcommand\CC			{\EuScript{C}}

\DeclareMathOperator{\Hom}{Hom}

\DeclareMathOperator{\id}{id}

\DeclareMathOperator{\ev}{ev}
\DeclareMathOperator{\coev}{coev}

\newcommand{\one}			{\mathbb{1}}

\newcommand\vect			{\mathrm{Vec}}

\newcommand\rep			{\mathrm{Rep}}

\newcommand\nao			{\mbox{$n$+1}}

\newcommand\toric {{\mathsf{TC}_3}}
\newcommand\core {\mathsf{TC}_3^0}

\newcommand\ot {\mathbb{1}_c}
\newcommand\mt {m_c}

\begin{document}

\begin{center} \LARGE
Defects in the 3-dimensional toric code model \\
form a braided fusion 2-category
\end{center}

\vspace{0.1cm}
\begin{center}
Liang Kong$^{a,b}$, \, Yin Tian$^{c}$, \, Zhi-Hao Zhang$^{d,a}$
~\footnote{Emails:{\tt \,\,\, kongl@sustech.edu.cn, yintian@mail.tsinghua.edu.cn, zzh31416@mail.ustc.edu.cn}}
\\[2em]
$^a$ Shenzhen Institute for Quantum Science and Engineering, \\
Southern University of Science and Technology, Shenzhen, 518055, China 
\\[0.8em]
$^b$ Guangdong Provincial Key Laboratory of Quantum Science and Engineering, \\
Southern University of Science and Technology, Shenzhen, 518055, China
\\[0.8em]
$^c$ Yau Mathematical Sciences Center, \\
Tsinghua University, Beijing 100084, China
\\[0.8em]
$^d$ Wu Wen-Tsun Key Laboratory of Mathematics of Chinese Academy of Sciences, \\ School of Mathematical Sciences, \\
University of Science and Technology of China, Hefei, 230026, China 

\end{center}

\vspace{0.5cm}
\begin{abstract}
It was well known that there are $e$-particles and $m$-strings in the 3-dimensional (spatial dimension) toric code model, which realizes the 3-dimensional $\Zb_2$ topological order. Recent mathematical result, however, shows that there are additional string-like topological defects in the 3-dimensional $\Zb_2$ topological order. In this work, we construct all  topological defects of codimension 2 and higher, and show that they form a braided fusion 2-category satisfying a braiding non-degeneracy condition. 
\end{abstract}

\vspace{0.2cm}
\tableofcontents


\section{Introduction}

In recent years, we have witnessed a surging interaction between the theory of topological orders in condensed matter physics and the category theory in mathematics. This interaction is mainly due to the fact that a topological order can be described by its topological defects (or excitations) up to invertible topological orders (see for examples \cite{Kit03,KW14}), and these topological defects form a higher category \cite{KW14,KWZ15,JF20,KLWZZ20}. This work further strengthens this interaction by emphasizing the appearance of a braided fusion 2-category in a concrete lattice model: the 3d toric code model. Throughout this work, we use $n$d to represent the spatial dimension and $n$D to represent the spacetime dimension.  

\medskip
It was long postulated that the fusion and braiding properties of quasi-particles in a 2d topological order can be precisely formulated in terms of a non-degenerate braided fusion 1-category \cite{FRS89,MS89,FG90} (see \cite[Appendix\ E]{Kit06} for a review). The demonstration of this idea through explicit physical or lattice models came later \cite{Wen91a,RS91,Kit03,LW05,KK12,HWW13,LW14,CCW17}. For 3d topological orders, it was also predicted that topological defects of codimension 2 and higher should form a non-degenerate braided fusion 2-category \cite{KW14,KWZ15,EN17,LKW18,KTZ20}. In particular, for a 3+1D twisted gauge theory with a finite gauge group $G$ (i.e. 3+1D Dijkgraaf-Witten theories), it was predicted that all topological defects of codimension 2 and higher should form the non-degenerate braided fusion 2-category $\Z(2\vect_G^\omega)$ that is the monoidal center of the category $2\vect_G^\omega$ of $G$-graded 2-vector spaces with a 4-cocycle twist $\omega \in H^4(G,U(1))$ \cite{EN17,LKW18,KTZ20}. On the one hand, the 2-category $\Z(2\vect_G^\omega)$ has been explicitly computed in \cite{KTZ20}. On the other hand, the lattice model realizations of the 3+1D Dijkgraaf-Witten theories were available. They were constructed by Wan, Wang and He in \cite{WWH15}. However, the problem of classifying all topological excitations in these models are highly non-trivial and remains an open problem until recently. By categorifying the ideas in \cite{Kit03,KK12,Kon13}, Bullivant and Delcamp successfully classified topological defects in these models as modules over certain higher tube algebras in \cite{BD19,BD20}. Interestingly, Bullivant and Delcamp's interpretation of $\Z(2\vect_G^\omega)$ as topological defects is different from that in \cite{EN17,LKW18,KTZ20} (see Remark~\ref{rem:alex}). The relation between these two interpretation remains a mystery to us. Moreover, the mathematical beauty and richness of Bullivant and Delcamp's results, however, make it harder for working physicists to see what is going on. Therefore, it is highly desirable to demonstrate the postulated relation between the topological defects and 2-categories as predicted in \cite{EN17,LKW18,KTZ20} in a simple lattice model.

In this work, in a very concrete and explicit way, we demonstrate the idea that topological defects form a non-degenerate braided fusion 2-category in the simplest case: the 3d toric code model \cite{HZW05}, which realizes the 3d $\Zb_2$ topological order. 
It was well known that there are $e$-particles and $m$-strings in the 3d toric code model. However, by the prediction that the topological defects in the 3d $\Zb_2$ topological order can be described by the braided fusion 2-category $\Z(2\vect_{\Zb_2})$ \cite{EN17,LKW18}, and by the recent mathematical computation of $\Z(2\vect_{\Zb_2})$ \cite[Example\ 3.8]{KTZ20}, there are additional string-like topological defects: $\ot$-strings and $\mt$-strings in the 3d $\Zb_2$ topological order. A $\ot$-string can be obtained from condensing a 1d $e$-particle gas \cite[Example\ 1,2]{KW14}, and was explicitly constructed in the 3d toric code model by Else and Nayak in \cite[Section\ III]{EN17} under the name of ``Cheshire charged loops''. Similarly, the $\mt$-strings can be obtained from condensing particles on the $m$-strings. Therefore, the $\ot$-strings and the $\mt$-strings are both condensation descendants \cite{KW14}. 

In this work, we explain why we work with strings instead of loops (see Section~\ref{sec:cat-defect}) and why we should include above condensation descendants in our categorical description of the 3d $\Zb_2$ topological order (see Section~\ref{sec:toric-cat}). Moreover, we show that all strings (including the trivial one) and 0d domain walls between strings form a braided fusion 2-category with non-degenerate double braidings. This braided fusion 2-category coincides precisely with $\Z(2\vect_{\Zb_2})$ as predicted in \cite{EN17,LKW18,KTZ20}\footnote{The physical interpretation of $\Z(2\vect_{\Zb_2})$ in \cite{KTZ20} is slightly different from that in \cite{EN17,LKW18} in their interpretations of the objects as either strings or loops (see Section~\ref{sec:cat-defect}).}. We also show how to compute loops and links from our categorical description of strings (see Section~\ref{sec:loop_link}).

\medskip
We briefly explain the layout of this paper. In Section~\ref{sec:cat-defect}, we provide a general categorical description of topological defects in a 3d topological order. In Section~\ref{sec:braiding_equivalent}, we show that the braiding of two strings, which is normally defined by a 1+0D defect in spacetime, can be deformed into a 0+1D defect that can be conveniently computed in the 3d toric code model. In Section~\ref{sec:3d_toric_code}, we briefly review the 3d toric code model. In Section~\ref{sec:topological_defects}, we explain that the topological defects in the 3d toric code model form a fusion 2-category $\toric$. In Section~\ref{sec:loop_link}, we compute the topological invariants associated to three loop-like and link-like defects. We also compute the double braidings in Section~\ref{sec:double-braidings}, but leave the study of the braiding structures to Section~\ref{sec:boundary+braiding}. The braiding structure on $\toric$ is gauge dependent, and a gauge can be fixed by selecting a gapped boundary of the 3d toric code model then determining the braidings via the half-braidings of the defects on the boundary. In Section~\ref{sec:smooth_rough_boundary}, we construct two gapped boundaries of the 3d toric code model: the smooth boundary and the rough boundary. In Section~\ref{sec:half_braiding}, we compute physically the half-braidings on the rough boundary and compare our result with the half-braidings computed mathematically in \cite{KTZ20}. In Appendix~\ref{sec:app}, we review the mathematical results of the fusion and braidings of $\Z(2\vect_{\Zb_2})$ in \cite{KTZ20}.


\medskip
\noindent {\bf Acknowledgement}: We would like to thank Chenjie Wang, Alex Bullivant and Xiao-Gang Wen for helpful discussion. We would also like to thank Stathis Vitouladitis for finding an unfinished sentence in the earlier version. LK is supported by Guangdong Provincial Key Laboratory (Grant No.~2019B121203002) and by NSFC under Grant No.~11971219. YT is partially supported by the NSFC grant No.~11971256. ZHZ is supported by Wu Wen-Tsun Key Laboratory of Mathematics at USTC of Chinese Academy of Sciences. 

\section{3+1D topological orders and braided fusion 2-categories}

\subsection{Categorical description of topological defects} \label{sec:cat-defect}

An $\nao$D topological order (defined on an open $\nao$-disk in spacetime) is characterized by its topological defects up to invertible topological orders. Topological defects are distinguished by their spacetime dimensions and their types. An adiabatic move and a deformation (in spacetime) of a topological defect do not change its spacetime dimension nor its type. 

\smallskip
In an anomaly-free 3+1D topological order, all types of topological defects of codimension 2 and higher form a structure called a braided monoidal 2-category $\CC$. A 2-category consists of objects, 1-morphisms between objects and 2-morphisms between 1-morphisms. Both 1-morphisms and 2-morphisms can be composed, and there are many coherent data and conditions. The braiding and monoidal structures are additional structures on a 2-category. We explain the physical meanings of all ingredients of $\CC$ in details below. 
\bit
\item 2-category structures: 
\bit

\item The objects are the types of 2D (spacetime dimension) topological defects.  A 2D defect in spacetime can be either the 1+1D world sheet of a string-like (1d spatial dimension) topological defects, or a 2+0D defect, which can be rotated in spacetime to become the 1+1D world sheet of a string-like topological defect. Therefore, we can also say that an object is a type of string-like topological defects, or a string for simplicity. 

\item The 1-morphisms from an object $X$ to an object $Y$ are 1D domain walls between two 2D defects $X$ and $Y$. Such a 1D domain wall can be either the world line of a particle-like (0d spatial dimension) topological defect or a 1+0D defect, which can be rotated to become the world line of a 0d domain wall between two strings. For example, consider a 0d domain wall $f$ between two strings $P,Q$ as illustrated in the following picture:  
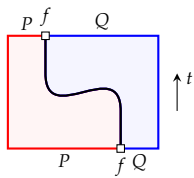
\[
\begin{tikzpicture}[scale=0.5]
\draw[red,thick,fill=red!5,fill opacity=0.7] (0,0)--(3,0)--(3,1) .. controls (3,2.5) and (1,0.5) .. (1,2)--(1,3)--(0,3)--cycle ;
\draw[blue,thick,fill=blue!5,fill opacity=0.7] (4,0)--(3,0)--(3,1) .. controls (3,2.5) and (1,0.5) .. (1,2)--(1,3)--(4,3)--cycle ;
\draw[thick] (3,0)--(3,1) .. controls (3,2.5) and (1,0.5) .. (1,2)--(1,3) ;
\draw[fill=white] (2.9,-0.1) rectangle (3.1,0.1) node[midway,below] {\scriptsize $f$} ;
\draw[fill=white] (0.9,2.9) rectangle (1.1,3.1) node[midway,above] {\scriptsize $f$} ;
\node[below] at (1.5,0.1) {\scriptsize $P$} ;
\node[below] at (3.5,0.1) {\scriptsize $Q$} ;
\node[above] at (0.5,2.9) {\scriptsize $P$} ;
\node[above] at (2.5,2.9) {\scriptsize $Q$} ;
\draw[-stealth,thin] (4.5,1)--(4.5,2) node[very near end,right] {\scriptsize $t$} ;
\end{tikzpicture}
\]
The 1D defect $f$ can be viewed as either a 0+1D defect or a 1+0D defect. The trivial 1D defect, i.e. no defect between 2D defect $X$ and itself is denoted by $1_X$.

\item The 2-morphisms are 0D (spacetime dimension) defects, which is also called instantons.

\item The instantons can be fused in spacetime. Thus for any two string-like topological defects $X,Y$, the 0d domain walls between them and associated instantons form a 1-category, denoted by $\Hom_\CC(X,Y)$.

\item The 0d domain walls can be fused along string-like topological defects. Thus for any string-like topological defects $X,Y$, both $\Hom_\CC(X,X)$ and $\Hom_\CC(Y,Y)$ are multi-fusion categories, and $\Hom_\CC(X,Y)$ is a $(\Hom_\CC(Y,Y),\Hom_\CC(X,X))$-bimodule. This description coincides with the intuition that a string-like topological defect can be viewed as a (potentially anomalous) 1d topological order.

\eit

\item Monoidal structures: Two parallel string-like topological defects can be fused in the 3d space. This equips the 2-category of topological defects with a monoidal structure. The fusion of two strings $X,Y$ is denoted by $X \otimes Y$. The tensor unit is the trivial string (i.e. no string), denoted by $\one$.

\item Braiding structures: The adiabatic move of one string around another parallel string in a spatial open 3-disk defines the braiding between two strings (see Figure~\ref{fig:braid_1}). 
\eit
As a consequence, the topological defects of codimension 2 and higher form a braided monoidal 2-category. For example, it was conjectured that the topological defects of codimension 2 and higher in a 3d twist gauge theory should form the braided monoidal 2-category $\Z(2\vect_G^\omega)$ for a finite group $G$ and $\omega \in H^4(G,U(1))$ \cite{EN17,LKW18,KTZ20}.

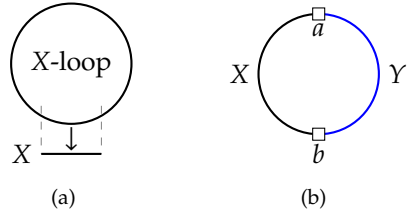
\begin{figure}
\centering
\subfigure[]{
\begin{tikzpicture}[scale=0.8]
\draw[thick] (0,0) circle (1) ;
\draw[help lines,dashed] (-0.5,-0.7)--(-0.5,-1.5) ;
\draw[help lines,dashed] (0.5,-0.7)--(0.5,-1.5) ;
\draw[thick] (-0.5,-1.5) node[left] {$X$} --(0.5,-1.5) ;
\node at (0,0) {$X$-loop} ;
\node at (0,-1.25) {$\downarrow$} ;
\end{tikzpicture}
}
\hspace{5ex}
\subfigure[]{
\begin{tikzpicture}[scale=0.8]
\draw[thick] (0,1) arc (90:270:1) ;
\draw[thick,blue] (0,1) arc (90:-90:1) ;
\draw[fill=white] (-0.1,0.9) rectangle (0.1,1.1) node[midway,below] {$a$} ;
\draw[fill=white] (-0.1,-0.9) rectangle (0.1,-1.1) node[midway,below] {$b$} ;
\node[left] at (-1,0) {$X$} ;
\node[right] at (1,0) {$Y$} ;
\end{tikzpicture}
}
\caption{(a): A loop-like topological defect looks like a string locally; (b): Two different string-like topological defects ($X,Y$) and 0d domain walls ($a,b$) between them constitute a loop-like topological defect.}
\label{fig:string_loop}
\end{figure}

\medskip
In physical literatures (see for example \cite{WL14,JMR14,EN17,LKW18}), physicists usually consider loop-like topological defects rather than string-like topological defects. There are some reasons why we use string-like topological defects to describe a 3d topological order.
\bnu
\item The string-like topological defects are more fundamental than the loop-like ones. For a loop-like, or even a knot-like topological defect, we can talk about its type. When we say two knot-like topological defects are of the same type, what we mean is that they are the same locally, i.e. on every open line segment. As depicted Figure~\ref{fig:string_loop} (a), the type $X$ of the string determines the type of the loop. This is similar to the fact that an $n$d topological order is a notion defined on an open $n$-disk in the infinite size limit \cite{AKZ17}. 
The physical observables defined on any closed manifold $M$, such as the ground state degeneracy on $M$, can be obtained by integrating observables on open $n$-disks over the entire $M$ \cite{AKZ17}. This process of integration is mathematically defined by factorization homology (see for example \cite{AF19} for a review).

\item Moreover, we can consider more general topological defects, an example of which is depicted in Figure~\ref{fig:string_loop} (b): two different string-like topological defects $X,Y$, together with two 0d domain walls $a,b$, constitute a loop-like topological defect. Note that Figure~\ref{fig:string_loop} (a) is just a special case of (b) when $X=Y=a=b$. It simply means that the string-like defects are more fundamental. 

\enu
It is possible to compute the topological invariant associated to a loop-like or knot-like topological defect. This computation amounts to a dimensional reduction process of shrinking the size of the loop-like or knot-like topological defect to nearly a point so that it becomes a 0d domain wall between two trivial strings $\one$. Therefore, a string-like topological defect in a 3+1D topological order $\CC$ determines a knot invariant valued in the symmetric fusion 1-category $\Hom_\CC(\one,\one)$. We illustrate this idea by some examples in Section~\ref{sec:loop_link}.

\begin{rem} \label{rem:theo}
We have not included topological defects of codimension 1 in our categorical description $\CC$. The reason is that all topological defects of codimension 1 in an anomaly-free topological order are all so-called condensation descendants that can be obtained from defects in $\CC$ via condensations \cite{KW14,KWZ15,KLWZZ20}. If we include those topological defects of codimension 1, we obtain a fusion 3-category \cite{JF20}, which is mathematically defined by the so-called delooping of $\CC$ (denoted by $\Sigma\CC$). Conversely, $\CC$ is the looping of $\Sigma\CC$. Importantly, it turns out that requiring $\CC$ to have non-degenerate braidings is equivalent to requiring $\Sigma\CC$ to have a trivial monoidal center. This fact was proved rigorously by Johnson-Freyd as the 3d case of a general result for arbitrary dimensions \cite{JF20}. It guarantees that $\CC$ is enough to characterize an anomaly-free stable 3d topological order. We will leave the study of topological defects of codimension 1 to the future. 
\end{rem}

\begin{rem} \label{rem:alex}
The physical interpretation of $\Z(2\vect_G^\omega)$ in \cite{BD19,BD20} is different from ours. For readers' convenience, we copied relevant sentences from \cite[Sec.\ 6.9]{BD20} here: ``objects of $\Z(2\vect_G^\omega)$ should be interpreted as particle excitations at the endpoints of a string-like excitation terminating at boundary components of the spatial manifold, the 1-morphisms as string-like topological excitations, and 2-morphisms as implementating the renormalisation of concatenated string-like excitations.'' Although we do have some guess of the possible relation between these two interpretations, it remains largely as a mystery at this stage. We will leave it to the future. 
\end{rem}

\subsection{Braiding structure} \label{sec:braiding_equivalent}

In this subsection, we provide an equivalent way of formulating the braiding between two strings that allows us to compute the braiding in lattice model efficiently.

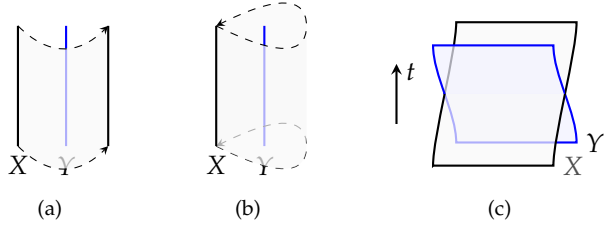
\begin{figure}[htbp]
\centering
\subfigure[]{
\begin{tikzpicture}[scale=0.8]
\draw[thick,blue] (0.8,0) node[below,black] {$Y$} --(0.8,2) ;
\fill[gray!5,opacity=0.7] (0,2)--(0,0) .. controls (0.5,-0.5) and (1,-0.5) .. (1.5,0)--(1.5,2) .. controls (1,1.5) and (0.5,1.5) .. (0,2) ;
\draw[-stealth,dashed] (0,0) .. controls (0.5,-0.5) and (1,-0.5) .. (1.5,0) ;
\draw[-stealth,dashed] (0,2) .. controls (0.5,1.5) and (1,1.5) .. (1.5,2) ;
\draw[thick] (0,0) node[below] {$X$} --(0,2) ;
\draw[thick] (1.5,0)--(1.5,2) ;
\end{tikzpicture}
}
\hspace{5ex}
\subfigure[]{
\begin{tikzpicture}[scale=0.8]
\fill[gray!5,opacity=0.7] (0,2)--(0,0) .. controls (1,0.5) and (1.5,0.5) .. (1.5,0)--(1.5,2) .. controls (1.5,2.5) and (1,2.5) .. (0,2) ;
\draw[thick,blue] (0.8,0) node[below,black] {$Y$} --(0.8,2) ;
\draw[-stealth,dashed] (0,0) .. controls (1,-0.5) and (1.5,-0.5) .. (1.5,0) .. controls (1.5,0.5) and (1,0.5) .. (0,0) ;
\fill[gray!5,opacity=0.7] (0,2)--(0,0) .. controls (1,-0.5) and (1.5,-0.5) .. (1.5,0)--(1.5,2) .. controls (1.5,1.5) and (1,1.5) .. (0,2) ;
\draw[-stealth,dashed] (0,2) .. controls (1,1.5) and (1.5,1.5) .. (1.5,2) .. controls (1.5,2.5) and (1,2.5) .. (0,2) ;
\draw[thick] (0,0) node[below] {$X$} --(0,2) ;
\end{tikzpicture}
}
\hspace{5ex}
\subfigure[]{
\begin{tikzpicture}[scale=0.8]
\begin{scope}
\clip (-0.1,-0.1,1) rectangle (2.4,1,0.5) ;
\draw[thick,blue,fill=blue!5,fill opacity=0.7] (0,0,0)--(2,0,0) .. controls (2,0.5,0) and (2,1.5,1) .. (2,2,1)--(0,2,1) .. controls (0,1.5,1) and (0,0.5,0) .. cycle ;
\end{scope}
\draw[thick,fill=gray!5,fill opacity=0.7] (0,0,1)--(2,0,1) node[right] {$X$} .. controls (2,0.5,1) and (2,1.5,0) .. (2,2,0)--(0,2,0) .. controls (0,1.5,0) and (0,0.5,1) .. cycle ;
\begin{scope}
\clip (-0.1,2.1,1) rectangle (2,1,0.5) ;
\draw[thick,blue,fill=blue!5,fill opacity=0.7] (0,0,0)--(2,0,0) .. controls (2,0.5,0) and (2,1.5,1) .. (2,2,1)--(0,2,1) .. controls (0,1.5,1) and (0,0.5,0) .. cycle ;
\end{scope}
\node[right] at (2,0,0) {$Y$} ;
\draw[-stealth,thick] (-0.8,0.5,0.5)--(-0.8,1.5,0.5) node[very near end,right] {$t$} ;
\end{tikzpicture}
}
\caption{(a): Moving a string $X$ around $Y$ in the 3d space is a braiding of $X$ and $Y$; (b): The trajectory of a double braiding is a cylinder; (c): The braiding is a 1+0D defect embedded in the 3+1D spacetime.}
\label{fig:braid_1}
\end{figure}

\medskip
Intuitively, the braiding of two strings $X$ and $Y$ is defined by an adiabatic move of the string $X$ around the string $Y$ in the 3d space, as depicted in Figure~\ref{fig:braid_1} (a). Therefore, a braiding is a 1+0D defect between the 1+1D world sheets of $X \otimes Y$ and $Y \otimes X$, as depicted in Figure~\ref{fig:braid_1} (c). Note that two world sheets of $X$ and $Y$ do not intersect in the 3+1D spacetime. Also Figure~\ref{fig:braid_1} (b) depicts a double braiding process, in which the trajectory of $X$ is a cylinder.

\begin{figure}[htbp]
\centering
\subfigure[]{
\begin{tikzpicture}[scale=0.9]
\draw[thick,blue] (0.5,0)--(0.5,2) ;
\fill[gray!5,opacity=0.7] (0,2)--(0,0) .. controls (0.3,-0.3) and (0.7,-0.3) .. (1,0)--(1,2) .. controls (0.7,1.7) and (0.3,1.7) .. (0,2) ;
\draw[-stealth,dashed] (0,0) .. controls (0.3,-0.3) and (0.7,-0.3) .. (1,0) ;
\draw[-stealth,dashed] (0,2) .. controls (0.3,1.7) and (0.7,1.7) .. (1,2) ;
\draw[thick] (0,0)--(0,2) node[above,scale=0.7] {$t=0$} ;
\draw[thick] (1,0)--(1,2) node[above,scale=0.7] {$t=1$} ;
\draw[help lines,dashed] (-0.5,0,-1)--(1.5,0,-1) ;
\draw[help lines,dashed] (-0.5,0,-1)--(-0.5,2,-1) ;
\draw[help lines,dashed] (-0.5,0,-1)--(-0.5,0,1) ;
\foreach \y/\z in {0/1,2/1,2/-1}
	\draw[help lines] (-0.5,\y,\z)--(1.5,\y,\z) ;
\foreach \x/\z in {-0.5/1,1.5/1,1.5/-1}
	\draw[help lines] (\x,0,\z)--(\x,2,\z) ;
\foreach \x/\y in {-0.5/2,1.5/2,1.5/0}
	\draw[help lines] (\x,\y,-1)--(\x,\y,1) ;
\draw[help lines,fill=green!5,fill opacity=0.3] (-0.5,0.4,-1)--(-0.5,0.4,1)--(1.5,0.4,1) node[above right,black,opacity=1] {$H_{z_0}$} --(1.5,0.4,-1)--cycle ;
\node at (-0.4,1) {$B_{1,t}$} ;
\draw[-stealth,thick] (-1.5,0.5,1)--(-0.7,0.5,1) node[very near end,above] {$x$} ;
\draw[-stealth,thick] (-1.5,0.5,1)--(-1.5,0.5,-0.2) node[very near end,above right] {$y$} ;
\draw[-stealth,thick] (-1.5,0.5,1)--(-1.5,1.3,1) node[very near end,right] {$z$} ;
\end{tikzpicture}
}
\hspace{5ex}
\subfigure[]{
\begin{tikzpicture}[scale=0.9]
\draw[thick,blue] (0.5,0)--(0.5,2) ;
\draw[white,line width=4pt] (0,0) .. controls (0,0.5) and (1,1.5) .. (1,2) ;
\draw[thick] (0,0) (0,0) .. controls (0,0.5) and (1,1.5) .. (1,2) ;
\draw[help lines,dashed] (-0.5,0,-1)--(1.5,0,-1) ;
\draw[help lines,dashed] (-0.5,0,-1)--(-0.5,2,-1) ;
\draw[help lines,dashed] (-0.5,0,-1)--(-0.5,0,1) ;
\foreach \y/\z in {0/1,2/1,2/-1}
	\draw[help lines] (-0.5,\y,\z)--(1.5,\y,\z) ;
\foreach \x/\z in {-0.5/1,1.5/1,1.5/-1}
	\draw[help lines] (\x,0,\z)--(\x,2,\z) ;
\foreach \x/\y in {-0.5/2,1.5/2,1.5/0}
	\draw[help lines] (\x,\y,-1)--(\x,\y,1) ;
\node at (-0.2,1.2) {$S_1 \cap H_{z_0}$} ;
\draw[-stealth,thick] (-1.5,0.5,1)--(-0.7,0.5,1) node[very near end,above] {$x$} ;
\draw[-stealth,thick] (-1.5,0.5,1)--(-1.5,0.5,-0.2) node[very near end,above right] {$y$} ;
\draw[-stealth,thick] (-1.5,0.5,1)--(-1.5,1.3,1) node[very near end,right] {$t$} ;
\end{tikzpicture}
}
\hspace{5ex}
\subfigure[]{
\begin{tikzpicture}[scale=0.9]
\draw[thick,blue] (1,0) .. controls (1,0.5) and (0,1.5) .. (0,2) ;
\draw[white,line width=4pt] (0,0) .. controls (0,0.5) and (1,1.5) .. (1,2) ;
\draw[thick] (0,0) .. controls (0,0.5) and (1,1.5) .. (1,2) ;
\draw[help lines,dashed] (-0.5,0,-1)--(1.5,0,-1) ;
\draw[help lines,dashed] (-0.5,0,-1)--(-0.5,2,-1) ;
\draw[help lines,dashed] (-0.5,0,-1)--(-0.5,0,1) ;
\foreach \y/\z in {0/1,2/1,2/-1}
	\draw[help lines] (-0.5,\y,\z)--(1.5,\y,\z) ;
\foreach \x/\z in {-0.5/1,1.5/1,1.5/-1}
	\draw[help lines] (\x,0,\z)--(\x,2,\z) ;
\foreach \x/\y in {-0.5/2,1.5/2,1.5/0}
	\draw[help lines] (\x,\y,-1)--(\x,\y,1) ;
\node at (0,1) {$B_2$} ;
\draw[-stealth,thick] (-1.5,0.5,1)--(-0.7,0.5,1) node[very near end,above] {$x$} ;
\draw[-stealth,thick] (-1.5,0.5,1)--(-1.5,0.5,-0.2) node[very near end,above right] {$y$} ;
\draw[-stealth,thick] (-1.5,0.5,1)--(-1.5,1.3,1) node[very near end,right] {$z$} ;
\end{tikzpicture}
}
\caption{(a): The time slice of the world sheet $S_1$ of braiding two strings at time $t$ is $B_{1,t}$; (b): The section of $S_1$ at $z = z_0$ is a braid $B_1' \coloneqq S_1 \cap H_{z_0}$; (c): After a rotation in $zt$-plane, the braid $B_1'$ is isotopic to $B_2$. }
\label{fig:braiding_equivalence}
\end{figure}
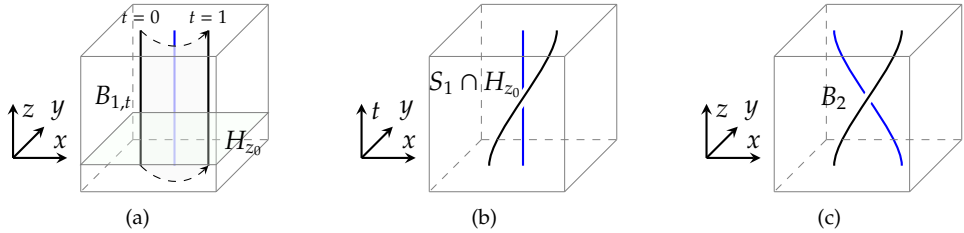

In order to compute the braiding structure explicitly in the 3d toric code model. We rotate this 1+0D defect of braiding to obtain a 0+1D defect. We denote $I \coloneqq [0,1]$ and $I^n \coloneqq [0,1]^n$. The world sheet of the braiding of two strings (see Figure~\ref{fig:braid_1} (c)) is denoted by $S_1 \subset I^4$. At each time $t$, the time slice of $S_1$ is a braid $B_{1,t}$ in the 3d space, as depicted in Figure~\ref{fig:braiding_equivalence} (a). Thus we have
\[
S_1 = \{(p,t) \in I^3 \times I \mid p \in B_{1,t}\} \subset I^3 \times I = I^4 ,
\]
where the first component $I^3$ represents the 3d space, and the second component $I$ represents the time axis.

Let us analyze the world sheet $S_1$ in $xyt$-spacetime by fixing the $z$-coordinate. Let $H_{z_0} \subset I^4$ be the hyperplane determined by $z = z_0$. Topologically $H_{z_0}$ is also a 3-disk. The section of $S_1$ at $z = z_0$ is $S_1 \cap H_{z_0} \subset H_{z_0} = I^3$, as depicted in Figure~\ref{fig:braiding_equivalence} (b). It is clear that for all values $z_0$ the sections $S_1 \cap H_{z_0}$ are the same. Denote $B_1' \coloneqq S_1 \cap H_{z_0} \subset I^3$, then we have
\[
S_1 = B_1' \times I \subset I^3 \times I = I^4 ,
\]
where the first component $I^3$ represents the $xyt$-spacetime (i.e. $H_{z_0}$), and the second component $I$ represents the $z$-axis.

After a rotation in $zt$-plane, the braid $B_1'$ is isotopic to $B_2$ depicted in Figure~\ref{fig:braiding_equivalence} (c). We denote
\[
S_2 \coloneqq B_2 \times I \subset I^3 \times I = I^4 ,
\]
where the first component $I^3$ represents the 3d space, and the second component $I$ represents the time axis. Then we know that $S_1$ and $S_2$ is isotopic in the 3+1D spacetime, up to a rotation.

\begin{figure}[htbp]
\centering
\subfigure[]{
\begin{tikzpicture}[scale=0.8]
\begin{scope}
\clip (1,0,0.5) rectangle (2.1,2.1,0) ;
\draw[thick,blue,fill=blue!5,fill opacity=0.7] (0,2,1)--(0,0,1) .. controls (0.5,0,1) and (1.5,0,0) .. (2,0,0)--(2,2,0) .. controls (1.5,2,0) and (0.5,2,1) .. cycle ;
\end{scope}
\draw[thick,fill=gray!5,fill opacity=0.7] (0,2,0)--(0,0,0) .. controls (0.5,0,0) and (1.5,0,1) .. (2,0,1) node[right] {$X$} --(2,2,1) .. controls (1.5,2,1) and (0.5,2,0) .. cycle ;
\begin{scope}
\clip (-0.1,0,1.1) rectangle (1,2,0.5) ;
\draw[thick,blue,fill=blue!5,fill opacity=0.7] (0,2,1)--(0,0,1) .. controls (0.5,0,1) and (1.5,0,0) .. (2,0,0)--(2,2,0) .. controls (1.5,2,0) and (0.5,2,1) .. cycle ;
\end{scope}
\node[right] at (2,0,0) {$Y$} ;
\draw[-stealth,thick] (-0.8,0.5,0.5)--(-0.8,1.5,0.5) node[very near end,right] {$t$} ;
\end{tikzpicture}
}
\hspace{5ex}
\subfigure[]{
\begin{tikzpicture}[scale=0.8]
\draw[thick,blue] (1,0) node[below,black] {$Y$} .. controls (1,0.5) and (0,1.5) .. (0,2) ;
\draw[white,line width=4pt] (0,0) .. controls (0,0.5) and (1,1.5) .. (1,2) ;
\draw[thick] (0,0) node[below] {$X$} .. controls (0,0.5) and (1,1.5) .. (1,2) ;
\draw[dashed,fill=gray!5,fill opacity=0.5] (-0.2,1.5) rectangle (1.2,0.5) ;
\node[right] at (1.2,1) {$R_{X,Y}$} ;
\end{tikzpicture}
}
\caption{(a) After a rotation, the braiding of two strings gives a 0+1D defect; (b): The braiding structure is a 0d domain wall between $X \otimes Y$ and $Y \otimes X$.}
\label{fig:braid_2}
\end{figure}
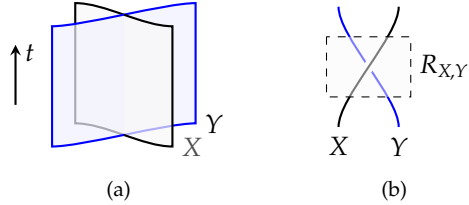

Since the time slices of $S_2$ are constant (the braid $B_2$), it can be viewed as a 0+1D defect. Figure~\ref{fig:braid_2} (a) depicts the world sheet $S_2$, and its time slice is depicted in Figure~\ref{fig:braid_2} (b). This braid in the 3d space is a 0d domain wall between two strings $X \otimes Y$ and $Y \otimes X$, denoted by $R_{X,Y}$.

In the categorical description of topological defects, the 0d domain wall $R_{X,Y}$ is a 1-morphism $R_{X,Y} : X \otimes Y \to Y \otimes X$. Since the braiding is a 0+1D defect, there are also additional 2-morphism data (0D defects) in the definition of a braiding. More precisely, let $f$ be a 0d domain wall between two strings $X$ and $X'$. Then moving $f$ across the braiding with $Y$ is an instanton (2-morphism), denoted by $R_{f,Y}$:
\be
\begin{array}{c}
\xymatrix@R=2em@C=3em{
X' \otimes Y \ar[r]^{R_{X',Y}} & Y \otimes X' \\
X \otimes Y \ar[r]_{R_{X,Y}} \ar[u]^{f \otimes 1_Y} \urtwocell<\omit>{\quad \, R_{f,Y}} & Y \otimes X \ar[u]_{1_Y \otimes f}
}
\end{array}
=
\begin{array}{c}
\begin{tikzpicture}[scale=0.7]
\draw[thick,blue] (1,0) node[below,black] {$Y$} .. controls (1,0.5) and (0,1.5) .. (0,2) node[above,black] {$Y$} ;
\draw[white,line width=4pt] (0,0) .. controls (0,0.5) and (1,1.5) .. (1,2) ;
\draw[thick] (0,0) node[below] {$X$} .. controls (0,0.5) and (1,1.5) .. (1,2) node[above] {$X'$} ;
\draw[fill=white] (0.15,0.7) rectangle (0.35,0.5) node[midway,left] {$f$} ;
\draw[-stealth,densely dashed] (0.4,0.6)--(0.8,1.2) ;
\end{tikzpicture}
\end{array}
\hspace{-1em} \xRightarrow{R_{f,Y}} 
\hspace{-1em}
\begin{array}{c}
\begin{tikzpicture}[scale=0.7]
\draw[thick,blue] (1,0) node[below,black] {$Y$} .. controls (1,0.5) and (0,1.5) .. (0,2) node[above,black] {$Y$} ;
\draw[white,line width=4pt] (0,0) .. controls (0,0.5) and (1,1.5) .. (1,2) ;
\draw[thick] (0,0) node[below] {$X$} .. controls (0,0.5) and (1,1.5) .. (1,2) node[above] {$X'$} ;
\draw[fill=white] (0.85,1.3) rectangle (0.65,1.5) node[midway,right] {$f$} ;
\end{tikzpicture}
\end{array} .
\ee
Clearly $R_{f,Y}$ is an isomorphism because the moving process is invertible up to a deformation in spacetime. Similarly we have another 2-morphism $R_{Y,f}$. The braiding structure, which consists of all 1-morphisms $R_{X,Y}$ and 2-morphisms $R_{f,Y}, R_{Y,f}$, is a natural transformation.

\begin{rem}
The braiding structure of a braided monoidal 2-category also involves some 2-morphisms $R_{(A \mid X,Y)} , R_{(X,Y \mid A)}$ (which form invertible modifications) for all objects $A,X,Y$ (see \cite{KTZ20} and references therein). In this work we do not discuss these 2-morphisms because they are all trivial in the 3d toric code model.
\end{rem}

\subsection{Dualities} \label{sec:dual}

In this subsection we show that each string-like defect in a 3+1D topological order admits a dual string.

\medskip
First we briefly review the left and right duals of objects in a monoidal 2-category $\CC$ \cite{DR18}. We say an object $X \in \CC$ is the left dual of $Y \in \CC$, or equivalently $Y$ is the right dual of $X$, if there exist 1-morphisms $\ev : X \otimes Y \to \one$ and $\coev: \one \to Y \otimes X$, called the evaluation and coevaluation 1-morphisms, such that $(\ev \otimes 1_X) \circ (1_X \otimes \coev) \simeq 1_X$ and $(1_Y \otimes \ev) \circ (\coev \otimes 1_Y) \simeq 1_Y$. We call these two equations the zig-zag equations.

\begin{figure}[htbp]
\begin{align*}
\begin{array}{c}
\begin{tikzpicture}
\draw[->-=0.65,thick] (-1,0.5)--(0,0.5) node[midway,above] {$X$} ;
\draw[thick] (0,0.5) arc (90:-90:0.25) ;
\draw[->-=0.65,thick] (0,0)--(-1,0) node[midway,below] {$X$} ;
\end{tikzpicture}
\end{array}
& =
\begin{array}{c}
\begin{tikzpicture}
\draw[->-=0.65,thick] (-1,0.5)--(0,0.5) node[midway,above] {$X$} ;
\draw[thick] (0,0.5) arc (90:-90:0.25) ;
\draw[->-=0.65,thick] (-1,0)--(0,0) node[midway,below] {$X^*$} ;
\draw[thick,dashed] (0.25,0.25)--(1.25,0.25) node[midway,above] {$\one$} ;
\end{tikzpicture}
\end{array}
= (\ev : X \otimes X^* \to \one) \\
\begin{array}{c}
\begin{tikzpicture}
\draw[->-=0.65,thick] (1,0.5)--(0,0.5) node[midway,above] {$X$} ;
\draw[thick] (0,0.5) arc (90:270:0.25) ;
\draw[->-=0.65,thick] (0,0)--(1,0) node[midway,below] {$X$} ;
\end{tikzpicture}
\end{array}
& =
\begin{array}{c}
\begin{tikzpicture}
\draw[->-=0.65,thick] (0,0.5)--(1,0.5) node[midway,above] {$X^*$} ;
\draw[thick] (0,0.5) arc (90:270:0.25) ;
\draw[->-=0.65,thick] (0,0)--(1,0) node[midway,below] {$X$} ;
\draw[thick,dashed] (-0.25,0.25)--(-1.25,0.25) node[midway,above] {$\one$} ;
\end{tikzpicture}
\end{array}
= (\coev : \one \to X^* \otimes X)
\end{align*}
\caption{The right dual $X^*$ of a string $X$ is given by reversing the orientation. Both the evaluation and coevaluation 1-morphisms are 0d domain walls.}
\label{fig:dual}
\end{figure}

\begin{figure}[htbp]
\centering
\begin{tikzpicture}
\fill[gray!5,opacity=0.7] (0,1) rectangle (3,2) ;
\fill[gray!5,opacity=0.7] (0,0)--(0,1)--(1.5,1) .. controls (1.8,0.8) and (2,0.5) .. (2,0.3) .. controls (2,0.5) and (1,0.5) .. (0,0) ;
\draw[thick] (0,0) .. controls (1,0.5) and (2,0.5) .. (2,0.3) ;
\draw[thick,gray!30] (1.5,1) .. controls (1.8,0.8) and (2,0.5) .. (2,0.3) ;
\fill[gray!5,opacity=0.7] (1.5,1) .. controls (1.8,0.8) and (2,0.5) .. (2,0.3) .. controls (2,0) and (1,0) .. (1,-0.3) .. controls (1,0) and (1.2,0.8) .. (1.5,1) ;
\draw[thick] (2,0.3) .. controls (2,0) and (1,0) .. (1,-0.3) ;
\fill[gray!5,opacity=0.7] (1.5,1) .. controls (1.2,0.8) and (1,0) .. (1,-0.3) .. controls (1,-0.5) and (2,-0.5) .. (3,0)--(3,1)--(1.5,1) ;
\draw[thick] (1,-0.3) .. controls (1,-0.5) and (2,-0.5) .. (3,0) node[midway,below] {$X$} ;
\draw[thick,gray!30] (1.5,1) .. controls (1.2,0.8) and (1,0) .. (1,-0.3) ;
\draw[thick] (0,0)--(0,2)--(3,2) node[midway,above] {$X$} --(3,0) ;
\draw[-stealth,thick] (-0.8,0.5)--(-0.8,1.5) node[very near end,right] {$t$} ;
\end{tikzpicture}
\caption{Straightening a string $X$ is an instanton.}
\label{fig:straighten}
\end{figure}
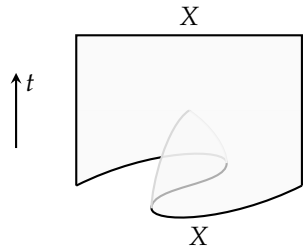

When $\CC$ describes topological defects of codimension 2 and higher in a 3+1D topological order, the physical intuition implies that each object in $\CC$ admits both left and right duals, and the left and right duals coincide with each other. We denote the string obtained by reversing the orientation of a string $X$ by $X^*$. As depicted in Figure~\ref{fig:dual}, if we bend the string $X$, the end of the double string can be viewed as the evaluation and coevaluation 1-morphisms associated to $X$. The zig-zag equations for these 1-morphisms are depicted as follows:
\be \label{eq:zig_zag}
\begin{array}{c}
\begin{tikzpicture}
\draw[->-=0.65,thick] (-1.5,0.5)--(0,0.5) node[midway,above] {$X$} ;
\draw[thick] (0,0.5) arc (90:-90:0.25) ;
\draw[->-=0.65,thick] (-0.5,0)--(0,0) node[midway,below] {$X^*$} ;
\draw[thick] (-0.5,0) arc (90:270:0.25) ;
\draw[->-=0.65,thick] (-0.5,-0.5)--(1,-0.5) node[midway,below] {$X$} ;
\end{tikzpicture}
\end{array}
\simeq
\begin{array}{c}
\begin{tikzpicture}
\draw[->-=0.65,thick] (-1.5,0)--(0,0) node[midway,above] {$X$} ;
\end{tikzpicture}
\end{array} ,
\qquad
\begin{array}{c}
\begin{tikzpicture}
\draw[->-=0.65,thick] (0,0.5)--(1.5,0.5) node[midway,above] {$X^*$} ;
\draw[thick] (0,0.5) arc (90:270:0.25) ;
\draw[->-=0.65,thick] (0,0)--(0.5,0) node[midway,below] {$X$} ;
\draw[thick] (0.5,0) arc (90:-90:0.25) ;
\draw[->-=0.65,thick] (-1,-0.5)--(0.5,-0.5) node[midway,below] {$X^*$} ;
\end{tikzpicture}
\end{array}
\simeq
\begin{array}{c}
\begin{tikzpicture}
\draw[->-=0.65,thick] (-1.5,0)--(0,0) node[midway,above] {$X^*$} ;
\end{tikzpicture}
\end{array} .
\ee
Physically these two isomorphisms are given by straightening the strings. Such an instanton is depicted in Figure~\ref{fig:straighten}, and the inverse of the instanton is given by its time-reversal. Therefore, $X^*$ is the right dual of $X$. Similarly, if we bend $X$ to another direction in Figure~\ref{fig:dual}, it is not hard to see that $X^*$ is also the left dual of $X$.

\begin{figure}[htbp]
\centering
\subfigure[]{
\begin{tikzpicture}[scale=0.9]
\fill[gray!5,opacity=0.7] (0,1,0)--(0,0,0) .. controls (0,-0.5,0) and (0,-1,0) .. (0,-1,0.5)--(2,-1,0.5) .. controls (2,-1,0) and (2,-0.5,0) .. (2,0,0)--(2,1,0) ;
\draw[thick,gray!30] (0,-1,0.5)--(2,-1,0.5) ;
\draw[thick] (2,-1,0.5) .. controls (2,-1,0) and (2,-0.5,0) .. (2,0,0)--(2,1,0)--(0,1,0) node[midway,below] {$X$} --(0,0,0) .. controls (0,-0.5,0) and (0,-1,0) .. (0,-1,0.5) ;
\fill[gray!5,opacity=0.7] (0,1,1)--(0,0,1) .. controls (0,-0.5,1) and (0,-1,1) .. (0,-1,0.5)--(2,-1,0.5) .. controls (2,-1,1) and (2,-0.5,1) .. (2,0,1)--(2,1,1) ;
\draw[thick] (2,-1,0.5) .. controls (2,-1,1) and (2,-0.5,1) .. (2,0,1)--(2,1,1)--(0,1,1) node[midway,below] {$X^*$} --(0,0,1) .. controls (0,-0.5,1) and (0,-1,1) .. (0,-1,0.5) ;
\draw[-stealth,thick] (-0.8,0,0.5)--(-0.8,1,0.5) node[very near end,right] {$t$} ;
\end{tikzpicture}
}
\hspace{5ex}
\subfigure[]{
\begin{tikzpicture}[scale=0.9]
\fill[gray!5,opacity=0.7] (0,1,0)--(0,2,0) .. controls (0,2.5,0) and (0,3,0) .. (0,3,0.5)--(2,3,0.5) .. controls (2,3,0) and (2,2.5,0) .. (2,2,0)--(2,1,0) ;
\draw[thick,gray!30] (0,3,0.5)--(2,3,0.5) ;
\draw[thick] (2,3,0.5) .. controls (2,3,0) and (2,2.5,0) .. (2,2,0)--(2,1,0)--(0,1,0) node[midway,above] {$X^*$} --(0,2,0) .. controls (0,2.5,0) and (0,3,0) .. (0,3,0.5) ;
\fill[gray!5,opacity=0.7] (0,1,1)--(0,2,1) .. controls (0,2.5,1) and (0,3,1) .. (0,3,0.5)--(2,3,0.5) .. controls (2,3,1) and (2,2.5,1) .. (2,2,1)--(2,1,1) ;
\draw[thick] (2,3,0.5) .. controls (2,3,1) and (2,2.5,1) .. (2,2,1)--(2,1,1)--(0,1,1) node[midway,above] {$X$} --(0,2,1) .. controls (0,2.5,1) and (0,3,1) .. (0,3,0.5) ;
\draw[-stealth,thick] (-0.8,1,0.5)--(-0.8,2,0.5) node[very near end,right] {$t$} ;
\end{tikzpicture}
}
\caption{(a): The creation of $X$ and $X^*$ is a 1+0D defect in spacetime; (b): The annihilation of $X$ and $X^*$ is a 1+0D defect in spacetime.}
\label{fig:dual_2}
\end{figure}
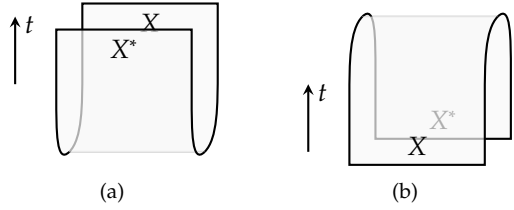

\begin{rem}
More intuitively, the evaluation and coevaluation 1-morphisms should be interpreted as 1+0D defects in spacetime. Figure~\ref{fig:dual_2} depicts the 1+0D defects of creation and annihilation of $X$ and its dual $X^*$. In a concrete lattice model, they can be realized as membrane operators. Similar to Section~\ref{sec:braiding_equivalent}, after a rotation these 1+0D defects become 0+1D (0d) defects as depicted in Figure~\ref{fig:dual}.
\end{rem}

\section{3d toric code model} \label{sec:3d_toric_code}

The 3d toric code model was introduced in \cite{HZW05} as an exactly solvable model realizing the 3+1D $\Zb_2$ gauge theory. In this section, we recall the construction of this model. 

\subsection{Hamiltonian}
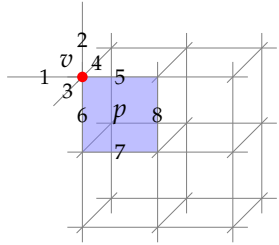
\begin{figure}[htbp]
\centering
\begin{tikzpicture}
\foreach \y in {0,...,2}
	\foreach \z in {0,...,1}
		\draw[help lines] (-0.2,\y,\z)--(2.2,\y,\z) ;
\foreach \x in {0,...,2}
	\foreach \z in {0,...,1}
		\draw[help lines] (\x,-0.2,\z)--(\x,2.2,\z) ;
\foreach \x in {0,...,2}
	\foreach \y in {0,...,2}
		\draw[help lines] (\x,\y,-0.2)--(\x,\y,1.2) ;
\draw[help lines] (-1,2,1)--(0,2,1) ;
\draw[help lines] (0,3,1)--(0,2,1) ;
\draw[help lines] (0,2,2)--(0,2,1) ;

\draw[help lines,fill=m_ext,opacity=0.5] (0,2,1)--(0,1,1)--(1,1,1)--(1,2,1)--cycle ;
\node at (0.5,1.5,1) {$p$} ;
\fill[e_ext] (0,2,1) circle (0.07) node[above left,black] {$v$} ;

\node[link_label] at (-0.5,2,1) {$1$} ;
\node[link_label] at (0,2.5,1) {$2$} ;
\node[link_label] at (0,2,1.5) {$3$} ;
\node[link_label] at (0,2,0.5) {$4$} ;
\node[link_label] at (0.5,2,1) {$5$} ;
\node[link_label] at (0,1.5,1) {$6$} ;
\node[link_label] at (0.5,1,1) {$7$} ;
\node[link_label] at (1,1.5,1) {$8$} ;
\end{tikzpicture}
\caption{the local operators in the 3d toric code model}
\label{fig:3d_toric_code}
\end{figure}

Similar to the 2d toric code model \cite{Kit03}, the 3d toric code model can be defined on any kind of lattice. But for simplicity, we consider only the cubic lattice as depicted in Figure~\ref{fig:3d_toric_code}.

There is a space of spin-$1/2$ on each edge (1-cell) of the lattice. In other words, the local degrees of freedom $\mathcal H_e$ on each edge $e$ form a two-dimensional Hilbert space $\Cb^2$, i.e. $\mathcal H_e=\Cb^2$. The total Hilbert space is $\mathcal H_{\text{tot}} \coloneqq \medotimes_e \mathcal H_e$. For any vertex $v$ and plaquette $p$ we define a vertex operator $A_v \coloneqq \prod_i \sigma_x^i$ and a plaquette operator $B_p \coloneqq \prod_j \sigma_z^j$ acting on adjacent edges. Here $\sigma_x^i$ means a $\sigma_x$ operator only acting on the $i$-th edge. For example, the operators in Figure~\ref{fig:3d_toric_code} are
\[
A_v = \sigma_x^1 \sigma_x^2 \sigma_x^3 \sigma_x^4 \sigma_x^5 \sigma_x^6 , \quad B_p = \sigma_z^5 \sigma_z^6 \sigma_z^7 \sigma_z^8 .
\]
The Hamiltonian of the 3d toric code model is defined by
\be \label{eq:Hamiltonian}
H \coloneqq \sum_v (1-A_v) + \sum_p (1-B_p) .
\ee
Since all $A_v$ and $B_p$ operators mutually commute, the total Hilbert space can be decomposed as the direct sum of common eigenspaces of all $A_v$ and $B_p$ operators, whose eigenvalues are $\pm 1$. The ground state subspace has energy $0$ and is the common eigenspace of all $A_v$ and $B_p$ operators with eigenvalues $+1$. The system is gapped because the first excited state has energy at least $2$ no matter how large the system size is.

\subsection{Ground state degeneracy}

Let us compute the ground state degeneracy (GSD) of the 3d toric code model defined on a 3-dimensional oriented closed manifold $M$. Suppose there are $V$ vertices (0-cells), $E$ edges (1-cells), $F$ plaquettes (2-cells) and $B$ cubes (3-cells) in the lattice. The Poincar\'{e} duality implies the Euler number of $M$ is $0 = V - E + F - B$, because $\dim(M) = 3$ is odd.

The total Hilbert space has dimension $2^E$. The constraints of the ground states are given by $A_v = +1$ and $B_p = +1$ for all vertices $v$ and plaquettes $p$. However, these constraints are not independent, and their relations are listed below. 
\bnu[(1)]
\item The product of all $A_v$ operators equals to $1$ because each $\sigma_x$ on the edge has been counted twice.
\item For any cube $c$, the product of all $B_p$ operators on the faces of $c$ equals to $1$ because each $\sigma_z$ on the edge of $c$ has been counted twice.
\item For any 2-dimensional closed surface $\Sigma \subset M$, the product of all $B_p$ operators on $\Sigma$ equals to $1$ because each $\sigma_z$ on the edge of $\Sigma$ has been counted twice.
\enu
There are $B$ relations in (2), but only $(B-1)$ of them are independent because
\[
\prod_c \biggl( \prod_{p \in \partial c} B_p \biggr) = 1 .
\]
Similarly, the relations in (3) are also not independent. If two surfaces $\Sigma$ and $\Sigma'$ are homologous, i.e. they encircle a 3-dimensional manifold $N$, then we have
\[
\biggl( \prod_{p \in \Sigma} B_p \biggr) \biggl( \prod_{p' \in \Sigma'} B_{p'} \biggr) = \prod_{c \in N} \biggl( \prod_{q \in \partial c} B_q \biggr) .
\]
Therefore, there is at most one independent relation associated to surfaces $\Sigma$ in each homology class in $H_2(M;\Zb_2)$. We denote
\[
d_k \coloneqq \dim_{\Zb_2} H_k(M;\Zb_2) ,
\]
then only $b_2$ relations in (3) are independent. So there are $(V+F-1-B+1-d_2)$ independent constraints, and each one halves the total Hilbert space. Therefore, we obtain
$$
\mbox{GSD on $M$} = 2^{E-(V+F-B-d_2)} = 2^{d_2},
$$ 
which is a topological invariant.

\begin{expl}
The GSD of the 3d toric code model on the 3-dimensional sphere $S^3$ is $2^0 = 1$, and the GSD on the 3-dimensional torus $T^3 = S^1 \times S^1 \times S^1$ is $2^3 = 8$.
\end{expl}



\section{Topological defects} \label{sec:topological_defects}

There are topological defects in the 3d toric code model. In this section, we find all topological defects of codimension 2 and higher and show that they form a fusion 2-category. 
A 1d (or 2-codimensional) defect is string-like and is called a string for simplicity; a 0d defect is in general a domain wall between two strings. 

\subsection{Four types of strings and 0d defects}
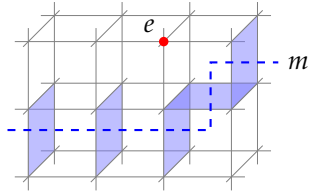
\begin{figure}[htbp]
\centering
\begin{tikzpicture}[scale=0.9]
\foreach \y in {0,...,2}
	\foreach \z in {0,...,1}
		\draw[help lines] (-1.2,\y,\z)--(2.2,\y,\z) ;
\foreach \x in {-1,...,2}
	\foreach \z in {0,...,1}
		\draw[help lines] (\x,-0.2,\z)--(\x,2.2,\z) ;
\foreach \x in {-1,...,2}
	\foreach \y in {0,...,2}
		\draw[help lines] (\x,\y,-0.2)--(\x,\y,1.2) ;


\foreach \x in {-1,...,1}
	\draw[help lines,fill=m_ext,opacity=0.5] (\x,0,0)--(\x,1,0)--(\x,1,1)--(\x,0,1)--cycle ;
\draw[help lines,fill=m_ext,opacity=0.5] (1,1,0)--(2,1,0)--(2,1,1)--(1,1,1)--cycle ;
\draw[help lines,fill=m_ext,opacity=0.5] (2,1,0)--(2,2,0)--(2,2,1)--(2,1,1)--cycle ;
\draw[m_dual_str] (-1.5,0.5,0.5)--(1.5,0.5,0.5)--(1.5,1.5,0.5)--(2.5,1.5,0.5) node[right,black] {$m$} ;

\fill[e_ext] (1,2,1) circle (0.07) node[above left,black] {$e$} ;
\end{tikzpicture}
\caption{an $e$-particle and an $m$-string in the 3d toric code model}
\label{fig:excitation_e_m}
\end{figure}

In the 3d toric code model, there are well-known topological excitations: the $e$-particle and the $m$-string as depicted in Figure~\ref{fig:excitation_e_m}. 

Similar to the 2d case, an $e$-particle at a vertex $v_0$ corresponds to the eigenspace that all $B_p,A_v$ act on as $+1$ except $A_{v_0}$ acting as $-1$. A pair of $e$-particles can be created by a string operator $\otimes_i \, \sigma_z^i$, where $i$'s are links along the string. An $e$-particle can be viewed as a 0d domain wall between two trivial strings, denoted by $\one$.

As in the 2d case, one may also expect that $B_p = -1$ at a plaquette $p$ defines an $m$-particle. In 3d, however, the constraint that
\[
\prod_{p \in c} B_p = 1
\]
for any cube $c$ implies there must be even plaquettes $p$ in a cube such that $B_p = -1$. Thus the correct topological excitation is the $m$-string (see Figure~\ref{fig:excitation_e_m}), which can not be broken by the above constraint. It is well-known that one can create a loop of $m$-strings by a membrane operator \cite{HZW05}.

\begin{figure}[htbp]
\centering
\begin{tikzpicture}[scale=0.9]
\foreach \y in {0,...,2}
	\foreach \z in {0,...,2}
		\draw[help lines] (-0.2,\y,\z)--(2.2,\y,\z) ;
\foreach \x in {0,...,2}
	\foreach \z in {0,...,2}
		\draw[help lines] (\x,-0.2,\z)--(\x,2.2,\z) ;
\foreach \x in {0,...,2}
	\foreach \y in {0,...,2}
		\draw[help lines] (\x,\y,-0.2)--(\x,\y,2.2) ;
\draw[white,line width=0.02cm] (2.2,1,0)--(1,1,0)--(1,1,2)--(1,-0.2,2) ;
\begin{scope}
\clip (0,0,2) rectangle (0.98,1,2) ;
\draw[help lines,fill=m_ext,fill opacity=0.5] (0,0,2) rectangle (1,1,2) node[midway,black,fill opacity=1] {$p$} ;
\end{scope}
\draw[densely dashed,thick] (2.2,1,0) node[right] {$\ot$} --(1,1,0)--(1,1,2)--(1,-0.2,2) ;
\node[link_label] at (0.5,1,2) {$1$} ;
\node[link_label] at (0,0.5,2) {$2$} ;
\node[link_label] at (0.5,0,2) {$3$} ;
\end{tikzpicture}
\caption{a $\ot$-string in the 3d toric code model}
\label{fig:excitation_T}
\end{figure}
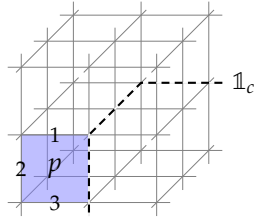

\medskip
Besides the $e$-particles and $m$-strings, there are other topological defects. We slightly modify the lattice model. 
In Figure~\ref{fig:excitation_T}, we depict a string-like topological defect, called a $\ot$-string, where the subscript ``${}_c$'' represents the ``condensation descendant'' (see Remark~\ref{rem:e-condense-on-T}). The local Hilbert spaces on the edges of the $\ot$-string (the dashed line) are one-dimensional $\mathcal H_e = \Cb$, i.e. there is no local degree of freedom along the $\ot$-string. Also, for vertices on the string there are no $A_v$ operators. The $B_p$ operator adjacent to the $\ot$-string, for example the one in Figure~\ref{fig:excitation_T}, is defined by $B_p = \sigma_z^1 \sigma_z^2 \sigma_z^3$.
The new Hamiltonian again consists of mutually commuting operators. An $m$-string can be fused with a $\ot$-string to give a new string-like defect denoted by $\mt$. An $\mt$-string can be microscopically realized by an $m$-string and an adjacent $\ot$-string. Hence we obtain four types of string-like topological defects: $\one,\ot,m,\mt$. 

\begin{rem} \label{rem:T-string}
It was known that condensing 1d $e$-particle gas leads to the $\ot$-string in the 3d $\Zb_2$ topological order (see \cite[Example\ 1,2]{KW14}). In the 3d toric code model, the $\ot$-string was explicitly constructed by Else and Nayak in \cite[Section\ III]{EN17}, where a $\ot$-string is defined by adding to the original Hamiltonian \eqref{eq:Hamiltonian} the $A_v$ terms and the projection operator onto the spin-$\uparrow$ subspace (i.e. $(1 + \sigma_z)/2$) on each 1-cell along the string.
\end{rem}

\begin{figure}[htbp]
\centering
\subfigure[]{
\begin{tikzpicture}[scale=0.9]
\foreach \y in {0,...,2}
	\foreach \z in {0,...,1}
		\draw[help lines] (-0.2,\y,\z)--(3.2,\y,\z) ;
\foreach \x in {0,...,3}
	\foreach \z in {0,...,1}
		\draw[help lines] (\x,-0.2,\z)--(\x,2.2,\z) ;
\foreach \x in {0,...,3}
	\foreach \y in {0,...,2}
		\draw[help lines] (\x,\y,-0.2)--(\x,\y,2.2) ;

\draw[white,line width=0.02cm] (-0.2,1,1)--(3.2,1,1) ;

\begin{scope}
\clip (1,1,0)--(2,1,0)--(2,1,0.98)--(1,1,0.98)--cycle ;
\draw[help lines,fill=m_ext,fill opacity=0.5] (1,1,0)--(2,1,0)--(2,1,1)--(1,1,1)--cycle ;
\end{scope}
\begin{scope}
\clip (1,2,1) rectangle (2,1.02,1) ;
\draw[help lines,fill=m_ext,fill opacity=0.5] (1,2,1) rectangle (2,1,1) ;
\end{scope}
\begin{scope}
\clip (1,0,1) rectangle (2,0.98,1) ;
\draw[help lines,fill=m_ext,fill opacity=0.5] (1,0,1) rectangle (2,1,1) ;
\end{scope}
\begin{scope}
\clip (1,1,2)--(2,1,2)--(2,1,1.02)--(1,1,1.02)--cycle ;
\draw[help lines,fill=m_ext,fill opacity=0.5] (1,1,2)--(2,1,2)--(2,1,1.02)--(1,1,1.02)--cycle ;
\end{scope}
		
\draw[densely dashed,thick] (-0.2,1,1)--(3.2,1,1) node[right] {$\ot$} ;

\foreach \y in {0,...,2}
	\draw[help lines] (-0.2,\y,2)--(3.2,\y,2) ;
\foreach \x in {0,...,3}
	\draw[help lines] (\x,-0.2,2)--(\x,2.2,2) ;

\node[below right] at (2,1,1) {$z$} ;
\end{tikzpicture}
}
\hspace{5ex}
\subfigure[]{
\begin{tikzpicture}[scale=0.9]
\draw[thick] (-2,0)--(0,0) ;
\draw[white,line width=4pt] (0,0) ellipse (0.5 and 1) ;
\draw[blue,thick] (0,0) ellipse (0.5 and 1) ;
\draw[white,line width=4pt] (0,0)--(2,0) ;
\draw[thick] (0,0)--(2,0) node[right] {$\ot$} ;
\node at (0.7,0.8) {$m$} ;
\node at (0,-1.2) {} ;
\end{tikzpicture}
}
\caption{(a): The microscopic realization of the nontrivial 0d defect between two $\ot$-strings; (b): This 0d defect can be obtained by shrinking an $m$-loop winding around a $\ot$-string.}
\label{fig:excitation_z}
\end{figure}
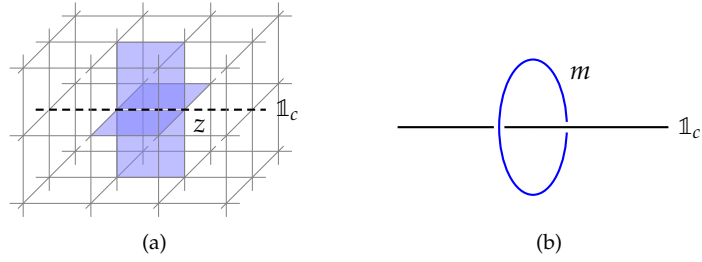

\begin{rem} \label{rem:e-condense-on-T}
Using string operators, we can move an $e$-particle onto a $\ot$-string. It is easy to see that the $e$-particles condense on the $\ot$-string. Moreover, one can also realize a $\ot$-string physically by introducing a 1d condensation on a $\one$-string defined by condensing the condensable algebra $1 \oplus e$ \cite{Kon14}. For this reason, $\ot$-strings can be viewed as condensation descendants of the $\one$-strings. Note that one can also introduce a 1d condensation on a $\ot$-string to obtain a $\one$-string by condensing the condensable algebra $1_{\ot} \oplus z$ on the $\ot$-strings, where $z$ is a 0d domain wall between two $\ot$-strings (see Figure~\ref{fig:excitation_z} (a)).
\end{rem}

 \begin{rem}
A physical explanation of condensation descendants in topological orders in all dimensions was given in \cite[Sec.\ V.B.7-8]{KW14}. A mathematical construction of all condensation descendants, called ``condensation completion'' (or ``Karoubi completion''), was given in the 2d case in \cite{DR18} and for all dimensions in \cite{GJF19}. An intuitive explanation of the condensation completion can be found in \cite[Sec.\ 3.3]{KLWZZ20}. 
\end{rem}

The 0d defects are 0d domain walls between two string-like defects. For example, the trivial particles (denoted by $1_\one$) and the $e$-particles can both be viewed as 0d domain walls between two trivial strings.

\begin{figure}[htbp]
\centering
\begin{tikzpicture}[scale=0.9]
\foreach \y in {0,...,2}
	\foreach \z in {0,...,1}
		\draw[help lines] (-3.2,\y,\z)--(4.2,\y,\z) ;
\foreach \x in {-3,...,4}
	\foreach \z in {0,...,1}
		\draw[help lines] (\x,-0.2,\z)--(\x,2.2,\z) ;
\foreach \x in {-3,...,4}
	\foreach \y in {0,...,2}
		\draw[help lines] (\x,\y,-0.2)--(\x,\y,2.2) ;
\draw[white,line width=0.02cm] (-1,1,1)--(2,1,1) ;
\draw[densely dashed,thick] (-1,1,1)--(2,1,1) node[midway,below] {$\ot$} ;
\node[right] at (4.2,1,1) {$\one$} ;
\node[left] at (-3.2,1,1) {$\one$} ;
\draw[very thick,fill=white] (2,1,1) circle (0.1) node[below right] {$x$} ;
\draw[very thick,fill=white] (-1,1,1) circle (0.1) node[below right] {$y$} ;
\foreach \y in {0,...,2}
	\draw[help lines] (-3.2,\y,2)--(4.2,\y,2) ;
\foreach \x in {-3,...,4}
	\draw[help lines] (\x,-0.2,2)--(\x,2.2,2) ;
\end{tikzpicture}
\caption{the 0d defects between a trivial string and a $\ot$-string}
\label{fig:excitation_xy}
\end{figure}
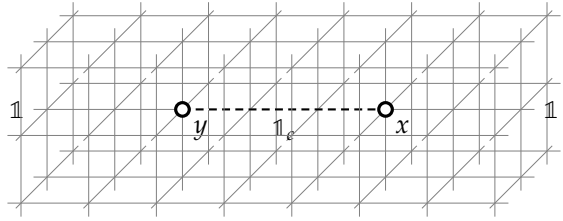

The ends of a $\ot$-string (see Figure~\ref{fig:excitation_xy}) are the 0d domain walls between a trivial string and a $\ot$-string. The 0d wall from $\one$ to $\ot$ is denoted by $x : \one \to \ot$, and by reversing its orientation, we get the 0d wall from $\ot$ to $\one$, denoted by $y : \ot \to \one$.

There are two kinds of 0d domain walls between two $\ot$-strings. One is the trivial 0d wall, denoted by $1_{\ot}$; the other one is depicted in Figure~\ref{fig:excitation_z}, denoted by $z : \ot \to \ot$. Microscopically, the 0d defect $z$ is realized by four plaquettes $B_p = -1$ around the $\ot$-string. It can be obtained by shrinking an $m$-loop winding around the $\ot$-string.
Note that if we shrink an $m$-loop around the trivial string, we simply get the trivial particle because the $m$-loop across the four plaquettes in Figure~\ref{fig:excitation_z} (a) (with $\ot$-string replaced by the trivial string) can be annihilated by a single $\sigma_x$ operator.

There is no 0d domain wall between a trivial string and an $m$-string due to the constraint that an $m$-string can not be broken. Similarly, there is no 0d domain wall between any one of $\{\one,\ot\}$ and any one of $\{m,\mt\}$. By fusing with an $m$-string, we can see that the 0d domain walls between the two strings of type either $m$ or $\mt$ can be obtained from those between the two strings of type either $\one$ or $\ot$.

\begin{rem}
Similar to the relation between $\one$ and $\ot$, the $\mt$-strings can also be viewed as condensation descendants of the $m$-strings. 
\end{rem}

\subsection{2-category \texorpdfstring{$\toric$}{Toric}} \label{sec:toric-cat}

In this subsection, we show that the  topological defects in the 3d toric code model form a finite semisimple 2-category denoted by $\toric$. 
\begin{itemize}
\item {\bf Objects = 1d defects (i.e. strings)}: There are four simple objects in the 2-category $\toric$ given by the four types of {\it simple strings} $\one,\ot,m,\mt$ in the model. General objects are direct sum of simple strings, and are called {\it composite strings}. When we discuss the fusion rule of $\ot \otimes \ot$ in the next subsection, we will see that the direct sums of simple strings are absolutely necessary.

\item {\bf 1-morphisms = 0d defects; 2-morphisms = instantons}: The 0d domain walls between any two strings $i$ and $j$ and instantons form a 1-category, denoted by $\Hom_\toric(i,j)$. We explain the ingredients of these 1-categories below for $i,j=\one,\ot,m,\mt$. 
\begin{itemize}
\item $\Hom_\toric(\one,\one)$: There are two types of simple particles: the trivial particle $1_\one$ and the $e$-particle in $\Hom_\toric(\one,\one)$. General particles are the direct sum of simple particles. For example, the 0d defect $1_\one \oplus e$ located at vertex $v_0$ can be obtained by throwing away the term $(1-A_{v_0})$ from the Hamiltonian.

As in the 2d case, two $e$-particles fuse into the trivial particle, i.e. $e \circ e = 1_\one$. Thus we have $\Hom_\toric(\one,\one) \simeq \rep(\Zb_2)$ as fusion 1-categories, where $\rep(\Zb_2)$ denotes the category of finite-dimensional representations of the $\Zb_2$-group. 

\item $\Hom_\toric(\ot,\ot)$: There are also two simple 0d defects $1_{\ot}$ and $z$ in $\Hom_\toric(\ot,\ot)$. Figure~\ref{fig:fusion_zz} shows that two $z$-particles fuse into $1_{\ot}$ because the operator $\sigma_x^1 \sigma_x^2 \sigma_x^3 \sigma_x^4$ annihilates two $m$-strings, i.e. $z \circ z = 1_{\ot}$. Thus we have $\Hom_\toric(\ot,\ot) \simeq \vect_{\Zb_2}$ as fusion 1-categories, where $\vect_{\Zb_2}$ denotes the category of finite-dimensional $\Zb_2$-graded vector spaces.

\begin{figure}[htbp]
\[
\begin{array}{c}
\begin{tikzpicture}[scale=0.9]
\foreach \y in {0,...,2}
	\foreach \z in {0,...,1}
		\draw[help lines] (-0.2,\y,\z)--(4.2,\y,\z) ;
\foreach \x in {0,...,4}
	\foreach \z in {0,...,1}
		\draw[help lines] (\x,-0.2,\z)--(\x,2.2,\z) ;
\foreach \x in {0,...,4}
	\foreach \y in {0,...,2}
		\draw[help lines] (\x,\y,-0.2)--(\x,\y,2.2) ;
\draw[white,line width=0.02cm] (-0.2,1,1)--(4.2,1,1) ;
\foreach \x in {1,2}{
	\begin{scope}
	\clip (\x,1,0)--(\x+1,1,0)--(\x+1,1,0.98)--(\x,1,0.98)--cycle ;
	\draw[help lines,fill=m_ext,fill opacity=0.5] (\x,1,0)--(\x+1,1,0)--(\x+1,1,1)--(\x,1,1)--cycle ;
	\end{scope}
}
\draw[m_str] (2,1,0)--(2,1,0.98) ;
\node[link_label] at (2,1,0.5) {$1$} ;
\foreach \x in {1,2}{
	\begin{scope}
	\clip (\x,2,1) rectangle (\x+1,1.02,1) ;
	\draw[help lines,fill=m_ext,fill opacity=0.5] (\x,2,1) rectangle (\x+1,1,1) ;
	\end{scope}
}
\draw[m_str] (2,2,1)--(2,1.02,1) ;
\node[link_label] at (2,1.5,1) {$2$} ; 
\foreach \x in {1,2}{
	\begin{scope}
	\clip (\x,0,1) rectangle (\x+1,0.98,1) ;
	\draw[help lines,fill=m_ext,fill opacity=0.5] (\x,0,1) rectangle (\x+1,1,1) ;
	\end{scope}
}
\draw[m_str] (2,0,1)--(2,0.98,1) ;
\node[link_label] at (2,0.5,1) {$3$} ; 
\foreach \x in {1,2}{
	\begin{scope}
	\clip (\x,1,2)--(\x+1,1,2)--(\x+1,1,1.02)--(\x,1,1.02)--cycle ;
	\draw[help lines,fill=m_ext,fill opacity=0.5] (\x,1,2)--(\x+1,1,2)--(\x+1,1,1)--(\x,1,1)--cycle ;
	\end{scope}
}
\draw[m_str] (2,1,2)--(2,1,1.02) ;
\node[link_label] at (2,1,1.5) {$4$} ;  
\draw[densely dashed,thick] (-0.2,1,1)--(4.2,1,1) node[right] {$\ot$} ;
\foreach \y in {0,...,2}
	\draw[help lines] (-0.2,\y,2)--(4.2,\y,2) ;
\foreach \x in {0,...,4}
	\draw[help lines] (\x,-0.2,2)--(\x,2.2,2) ;
\end{tikzpicture}
\end{array}
=
\begin{array}{c}
\begin{tikzpicture}[scale=0.9]
\foreach \y in {0,...,2}
	\foreach \z in {0,...,1}
		\draw[help lines] (-0.2,\y,\z)--(4.2,\y,\z) ;
\foreach \x in {0,...,4}
	\foreach \z in {0,...,1}
		\draw[help lines] (\x,-0.2,\z)--(\x,2.2,\z) ;
\foreach \x in {0,...,4}
	\foreach \y in {0,...,2}
		\draw[help lines] (\x,\y,-0.2)--(\x,\y,2.2) ;
		
\draw[white,line width=0.02cm] (-0.2,1,1)--(4.2,1,1) ;
\draw[densely dashed,thick] (-0.2,1,1)--(4.2,1,1) node[right] {$\ot$} ;

\foreach \y in {0,...,2}
	\draw[help lines] (-0.2,\y,2)--(4.2,\y,2) ;
\foreach \x in {0,...,4}
	\draw[help lines] (\x,-0.2,2)--(\x,2.2,2) ;
\end{tikzpicture}
\end{array}
\]
\caption{The fusion of two $z$-particles is $z \circ z = 1_{\ot}$.}
\label{fig:fusion_zz}
\end{figure}
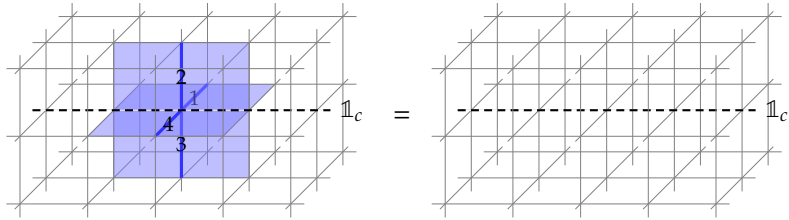

\item $\Hom_\toric(\one,\ot)$: There is only one simple 0d defect $x$ in $\Hom_\toric(\one,\ot)$. General 0d defects in $\Hom_\toric(\one,\ot)$ are the direct sums of $x$. Therefore, we have $\Hom_\toric(\one,\ot) \simeq \vect$ as 1-categories, where $\vect$ denotes the category of finite dimensional vector spaces. 

\item $\Hom_\toric(\ot,\one)$: There is only one simple 0d defect $y$ in $\Hom_\toric(\ot, \one)$. General 0d defects in $\Hom_\toric(\ot,\one)$ are the direct sums of $y$. Again we have $\Hom_\toric(\ot,\one) \simeq \vect$ as 1-categories. 

\item The fusion of 0d defects $y \circ x$ gives a particle in $\Hom_\toric(\one,\one)$. This particle $y\otimes x$ can be represented by a $\ot$-string of length zero, or equivalently, by deleting the $A_v$ operator at a vertex $v$ from the Hamiltonian. This particle is precisely $1_\one \oplus e$ because it corresponds to the eigenspaces with $A_v = \pm 1$ in the total Hilbert space. Thus $y \circ x = 1_\one \oplus e$. 

\item The fusion of $x \circ y$ gives a 0d defect in $\Hom_\toric(\ot,\ot)$. To compute it, we let $x$ and $y$ be close enough as depicted in Figure~\ref{fig:fusion_xy}. The spin on the edge between $x$ and $y$ can take the value $\uparrow$ or $\downarrow$. If the spin is $\uparrow$, we get a $\ot$-string (recall the convention in Remark~\ref{rem:T-string}); if the spin is $\downarrow$, we need to flip it (recall our convention in Remark~\ref{rem:T-string}) by applying a $\sigma_x$ operator, then four $B_p$ operators adjacent to this edge have to take the value $-1$, that is, we get a $z$-particle. Hence we have $x \circ y = 1_{\ot} \oplus z$.

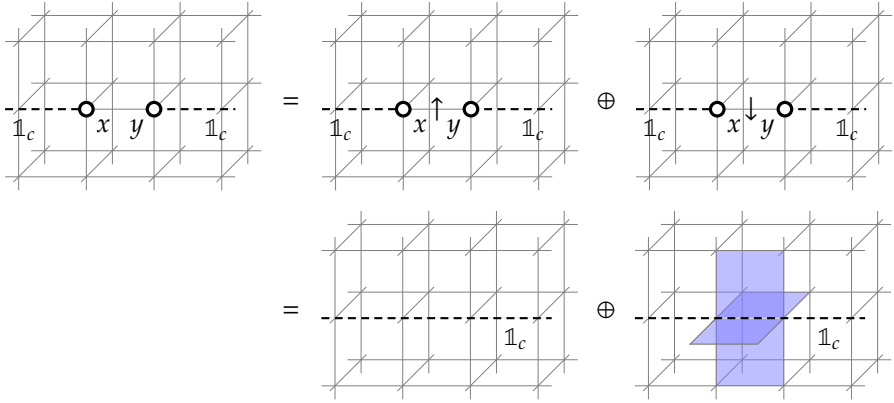
\begin{figure}[htbp]
\begin{align*}
\begin{array}{c}
\begin{tikzpicture}[scale=0.9]
\foreach \y in {0,...,2}
	\foreach \z in {0,...,1}
		\draw[help lines] (-0.2,\y,\z)--(3.2,\y,\z) ;
\foreach \x in {0,...,3}
	\foreach \z in {0,...,1}
		\draw[help lines] (\x,-0.2,\z)--(\x,2.2,\z) ;
\foreach \x in {0,...,3}
	\foreach \y in {0,...,2}
		\draw[help lines] (\x,\y,-0.2)--(\x,\y,1.2) ;
\draw[white,line width=0.02cm] (-0.2,1,1)--(1,1,1) ;
\draw[densely dashed,thick] (-0.2,1,1)--(1,1,1) node[midway,below left] {$\ot$} ;
\draw[very thick,fill=white] (1,1,1) circle (0.1) node[below right] {$x$} ;
\draw[white,line width=0.02cm] (2,1,1)--(3.2,1,1) ;
\draw[densely dashed,thick] (2,1,1)--(3.2,1,1) node[midway,below right] {$\ot$} ;
\draw[very thick,fill=white] (2,1,1) circle (0.1) node[below left] {$y$} ;
\end{tikzpicture}
\end{array}
& =
\begin{array}{c}
\begin{tikzpicture}[scale=0.9]
\foreach \y in {0,...,2}
	\foreach \z in {0,...,1}
		\draw[help lines] (-0.2,\y,\z)--(3.2,\y,\z) ;
\foreach \x in {0,...,3}
	\foreach \z in {0,...,1}
		\draw[help lines] (\x,-0.2,\z)--(\x,2.2,\z) ;
\foreach \x in {0,...,3}
	\foreach \y in {0,...,2}
		\draw[help lines] (\x,\y,-0.2)--(\x,\y,1.2) ;
\draw[white,line width=0.02cm] (-0.2,1,1)--(1,1,1) ;
\draw[densely dashed,thick] (-0.2,1,1)--(1,1,1) node[midway,below left] {$\ot$} ;
\draw[very thick,fill=white] (1,1,1) circle (0.1) node[below right] {$x$} ;
\draw[white,line width=0.02cm] (2,1,1)--(3.2,1,1) ;
\draw[densely dashed,thick] (2,1,1)--(3.2,1,1) node[midway,below right] {$\ot$} ;
\draw[very thick,fill=white] (2,1,1) circle (0.1) node[below left] {$y$} ;
\node at (1.5,1,1) {$\uparrow$} ;
\end{tikzpicture}
\end{array}
\oplus
\begin{array}{c}
\begin{tikzpicture}[scale=0.9]
\foreach \y in {0,...,2}
	\foreach \z in {0,...,1}
		\draw[help lines] (-0.2,\y,\z)--(3.2,\y,\z) ;
\foreach \x in {0,...,3}
	\foreach \z in {0,...,1}
		\draw[help lines] (\x,-0.2,\z)--(\x,2.2,\z) ;
\foreach \x in {0,...,3}
	\foreach \y in {0,...,2}
		\draw[help lines] (\x,\y,-0.2)--(\x,\y,1.2) ;
\draw[white,line width=0.02cm] (-0.2,1,1)--(1,1,1) ;
\draw[densely dashed,thick] (-0.2,1,1)--(1,1,1) node[midway,below left] {$\ot$} ;
\draw[very thick,fill=white] (1,1,1) circle (0.1) node[below right] {$x$} ;
\draw[white,line width=0.02cm] (2,1,1)--(3.2,1,1) ;
\draw[densely dashed,thick] (2,1,1)--(3.2,1,1) node[midway,below right] {$\ot$} ;
\draw[very thick,fill=white] (2,1,1) circle (0.1) node[below left] {$y$} ;
\node at (1.5,1,1) {$\downarrow$} ;
\end{tikzpicture}
\end{array} \\
& =
\begin{array}{c}
\begin{tikzpicture}[scale=0.9]
\foreach \y in {0,...,2}
	\foreach \z in {0,...,1}
		\draw[help lines] (-0.2,\y,\z)--(3.2,\y,\z) ;
\foreach \x in {0,...,3}
	\foreach \z in {0,...,1}
		\draw[help lines] (\x,-0.2,\z)--(\x,2.2,\z) ;
\foreach \x in {0,...,3}
	\foreach \y in {0,...,2}
		\draw[help lines] (\x,\y,-0.2)--(\x,\y,1.2) ;
\draw[white,line width=0.02cm] (-0.2,1,1)--(3.2,1,1) ;
\draw[densely dashed,thick] (-0.2,1,1)--(3.2,1,1) node[near end,below right] {$\ot$} ;
\end{tikzpicture}
\end{array}
\oplus
\begin{array}{c}
\begin{tikzpicture}[scale=0.9]
\foreach \y in {0,...,2}
	\foreach \z in {0,...,1}
		\draw[help lines] (-0.2,\y,\z)--(3.2,\y,\z) ;
\foreach \x in {0,...,3}
	\foreach \z in {0,...,1}
		\draw[help lines] (\x,-0.2,\z)--(\x,2.2,\z) ;
\foreach \x in {0,...,3}
	\foreach \y in {0,...,2}
		\draw[help lines] (\x,\y,-0.2)--(\x,\y,1.2) ;
\draw[white,line width=0.02cm] (-0.2,1,1)--(3.2,1,1) ;
\begin{scope}
\clip (1,1,0)--(2,1,0)--(2,1,0.98)--(1,1,0.98)--cycle ;
\draw[help lines,fill=m_ext,fill opacity=0.5] (1,1,0)--(2,1,0)--(2,1,1)--(1,1,1)--cycle ;
\end{scope}
\begin{scope}
\clip (1,2,1) rectangle (2,1.02,1) ;
\draw[help lines,fill=m_ext,fill opacity=0.5] (1,2,1) rectangle (2,1,1) ;
\end{scope}
\begin{scope}
\clip (1,0,1) rectangle (2,0.98,1) ;
\draw[help lines,fill=m_ext,fill opacity=0.5] (1,0,1) rectangle (2,1,1) ;
\end{scope}
\begin{scope}
\clip (1,1,2)--(2,1,2)--(2,1,1.02)--(1,1,1.02)--cycle ;
\draw[help lines,fill=m_ext,fill opacity=0.5] (1,1,2)--(2,1,2)--(2,1,1)--(1,1,1)--cycle ;
\end{scope}
\draw[help lines] (1,1,1.02)--(1,1,2)--(2,1,2)--(2,1,1.02) ;
\draw[densely dashed,thick] (-0.2,1,1)--(3.2,1,1) node[near end,below right] {$\ot$} ;
\end{tikzpicture}
\end{array}
\end{align*}
\caption{The fusion of an $x$-particles and a $y$-particle is $x \circ y = 1_{\ot} \oplus z$.}
\label{fig:fusion_xy}
\end{figure}

\item $\Hom_\toric(i,j)=0$ for $i=\one,\ot$ and $j=m,\mt$: This is because $m$-strings can not be broken. 

\item $\Hom_\toric(m,m) \simeq \Hom_\toric(\one,\one) \simeq \rep(\Zb_2)$: This is because when $e$-particles are moved close to an $m$-string, no condensation happens. We denote the trivial 0d domain wall on the $m$-string by $1_m$, and denote the $e$-particle on the $m$-string by $e_m \coloneqq e \otimes 1_m$. 

\item $\Hom_\toric(\mt,\mt) \simeq \Hom_\toric(\ot,\ot) \simeq \vect_{\Zb_2}$. We denote the trivial 0d defect on the $\mt$-string by $1_{\mt}$. Since $\mt = \ot \otimes m$, we denote the non-trivial 0d defect on the $\mt$-string by $z_m \coloneqq z \otimes 1_m$. 

\item $\Hom_\toric(m,\mt) \simeq \Hom_\toric(\one,\ot) \simeq \vect$;

\item $\Hom_\toric(\mt,m) \simeq \Hom_\toric(\ot,\one) \simeq \vect$. 
\end{itemize}
\item According to the mathematical definition of the semisimpleness and idempotent completeness of a 2-category \cite{DR18}, $\toric$ is a semisimple 2-category. 
\end{itemize}

\begin{rem}
Note that $\rep(\Zb_2) \simeq \vect_{\Zb_2}$ as fusion 1-categories, but we denote $\Hom_\toric(\one,\one) = \rep(\Zb_2)$ and $\Hom_\toric(\ot,\ot) = \vect_{\Zb_2}$. This is because in the 3+1D $G$-gauge theory where $G$ is a (possibly non-abelian) finite group, the particles on the trivial string $\one$ form a (symmetric) fusion 1-category $\Hom(\one,\one) = \rep(G)$; by condensing all particles on the trivial string we get a string $\ot$ with $\Hom(\ot,\ot) = \vect_G$. In general $\rep(G)$ and $\vect_G$ are not equivalent as fusion 1-categories. 
\end{rem}

\medskip
We can summarize all the ingredients of the 2-category $\toric$ heuristically by the following diagram:  
\be \label{eq:Z(2Rep(Z_2))}
\xymatrix{
\one \ar@(ul,ur)[]^{\rep(\Zb_2)}  \ar@/^/[rr]^{\vect} & & \ot \ar@(ul,ur)[]^{\vect_{\Zb_2}} \ar@/^/[ll]^{\vect}
& & m \ar@(ul,ur)[]^{\rep(\Zb_2)}  \ar@/^/[rr]^{\vect} & & \mt \ar@(ul,ur)[]^{\vect_{\Zb_2}} \ar@/^/[ll]^{\vect}
} .
\ee
Mathematically, the full 2-subcategory represented by the following subdiagram
\be \label{eq:2Rep(Z_2)}
\xymatrix{
\one \ar@(ul,ur)[]^{\rep(\Zb_2)} \ar@/^/[rr]^{\vect} & & \ot \ar@(ul,ur)[]^{\vect_{\Zb_2}} \ar@/^/[ll]^{\vect}
} 
\ee
is precisely the well-known 2-category $2\rep(\Zb_2)$, which is the 2-category of finite semisimple module categories over $\rep(\Zb_2)$ (or $\vect_{\Zb_2}$ equivalently). By \cite[Theorem\ 1.4.8]{DR18}, $2\rep(\Zb_2)$ is a finite semisimple 2-category. As a consequence, the complete 2-category $\toric$ of topological defects of the 3d toric code model is equivalent to the direct sum $2\rep(\Zb_2) \boxplus 2 \rep(\Zb_2)$, which is also finite semisimple. This is precisely the 2-category $\Z(2\vect_{\Zb_2})$ of topological excitations in the 3d $\Zb_2$ topological order predicted in \cite[Example\ 3.8]{KTZ20}, where $\Z(2\vect_{\Zb_2})$ denotes the Drinfeld center of the monoidal 2-category $2\vect_{\Zb_2}$ of $\Zb_2$-graded 2-vector spaces.

\medskip
We denote the full 2-subcategory consisting of $\one$ and $m$ by $\core$. According to \cite{DR18}, the 2-category $\toric$ is semisimple but $\core$ is not semisimple. According to \cite{KW14}, $\core$ is the core of $\toric$. By definition, the core is a subcategory consisting of elementary defects such that all the rest of defects can be obtained from the elementary ones by condensations. The non-elementary defects are called condensed excitations or condensation descendants (of the elementary ones). It is clear that the $\ot$-strings and the $\mt$-strings are both condensation descendants. The process of adding more defects to $\core$ to obtain $\toric$ is called ``condensation completion'' \cite{GJF19,KLWZZ20}, which is also called an idempotent completion \cite{DR18} or a Karoubi completion \cite{GJF19}.

\medskip
In the end of this subsection, we try to answer an obvious question: Since condensation descendants can be obtained from others via condensations, 1-codimensional defects can be ignored (recall Remark~\ref{rem:theo}). Why should we include the condensation descendants: the $\ot$-strings and the $\mt$-strings in our categorical description of the 3d $\Zb_2$ topological order? We provide a few reasons below.
\bnu

\item The $\ot$-strings and the $\mt$-strings are physically natural. We can not ignore them for physical processes involving condensations of particles. It is also unnatural to ignore them when we consider dimensional reduction processes. For example, by a dimensional reduction from 3d to 2d, both an $m$-string and an $\mt$-strings can be viewed as two gapped 1d boundaries of the same 2d topological order. 

\item The $\ot$-strings and the $\mt$-strings are mathematically natural. Without them, $\core$ is not idempotent complete (or semisimple) as shown in \cite{DR18}. 

\item We construct the rough boundary of the 3d toric code model in Section~\ref{sec:smooth_rough_boundary}. Although there is no condensation descendants on this boundary (with only the trivial strings and the $m$-strings), as shown in \cite{KTZ20}, the $\ot$-strings and the $\mt$-strings naturally pop up in the 3d bulk as they are demanded by the boundary-bulk relation \cite{KWZ17}. 

\item A gapped boundary of a topological order can be viewed as a consequence of a condensation of the topological excitations in the bulk. It turns out that the rough boundary of the 3d toric code model is a consequence of condensing the $\ot$-strings, which should be viewed as a condensable algebra in $\toric$ (see Remark~\ref{rem:condensing-T}). Using $\core$, there is no way to develop a condensation theory. 

\enu

\subsection{Monoidal structure on \texorpdfstring{$\toric$}{Toric}} \label{sec:monoidal}


The monoidal structure on $\toric$ is given by the fusion of string-like topological defects. The fusion of two strings $X,Y$ is denoted by $X \otimes Y$.

\medskip
By definition, a $\ot$-string and an $m$-string fuse into an $\mt$-string, i.e. $\mt = \ot \otimes m$.

It is not hard to see that two $m$-strings fuse into the trivial string, i.e. $m\otimes m =\one$. Figure~\ref{fig:fusion_zz} shows two $m$-loops can be annihilated by several $\sigma_x$ operators, and a similar argument applies to the fusion of two $m$-strings.

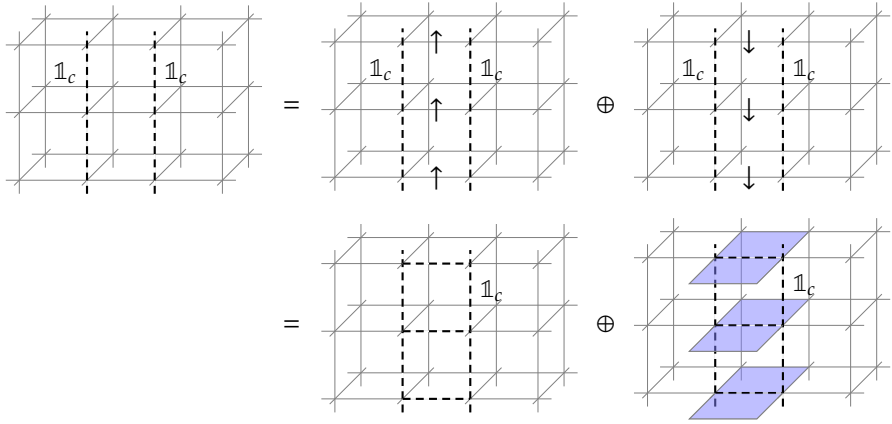
\begin{figure}[htbp]
\begin{align*}
\begin{array}{c}
\begin{tikzpicture}[scale=0.9]
\foreach \y in {0,...,2}
	\foreach \z in {0,...,1}
		\draw[help lines] (-0.2,\y,\z)--(3.2,\y,\z) ;
\foreach \x in {0,...,3}
	\foreach \z in {0,...,1}
		\draw[help lines] (\x,-0.2,\z)--(\x,2.2,\z) ;
\foreach \x in {0,...,3}
	\foreach \y in {0,...,2}
		\draw[help lines] (\x,\y,-0.2)--(\x,\y,1.2) ;
\draw[white,line width=0.02cm] (1,-0.2,1)--(1,2.2,1) ;
\draw[densely dashed,thick] (1,-0.2,1)--(1,2.2,1) node[very near end,below left] {$\ot$} ;
\draw[white,line width=0.02cm] (2,-0.2,1)--(2,2.2,1) ;
\draw[densely dashed,thick] (2,-0.2,1)--(2,2.2,1) node[very near end,below right] {$\ot$} ;
\end{tikzpicture}
\end{array}
& =
\begin{array}{c}
\begin{tikzpicture}[scale=0.9]
\foreach \y in {0,...,2}
	\foreach \z in {0,...,1}
		\draw[help lines] (-0.2,\y,\z)--(3.2,\y,\z) ;
\foreach \x in {0,...,3}
	\foreach \z in {0,...,1}
		\draw[help lines] (\x,-0.2,\z)--(\x,2.2,\z) ;
\foreach \x in {0,...,3}
	\foreach \y in {0,...,2}
		\draw[help lines] (\x,\y,-0.2)--(\x,\y,1.2) ;
\draw[white,line width=0.02cm] (1,-0.2,1)--(1,2.2,1) ;
\draw[densely dashed,thick] (1,-0.2,1)--(1,2.2,1) node[very near end,below left] {$\ot$} ;
\draw[white,line width=0.02cm] (2,-0.2,1)--(2,2.2,1) ;
\draw[densely dashed,thick] (2,-0.2,1)--(2,2.2,1) node[very near end,below right] {$\ot$} ;
\node at (1.5,0,1) {$\uparrow$} ;
\node at (1.5,1,1) {$\uparrow$} ;
\node at (1.5,2,1) {$\uparrow$} ;
\end{tikzpicture}
\end{array}
\oplus
\begin{array}{c}
\begin{tikzpicture}[scale=0.9]
\foreach \y in {0,...,2}
	\foreach \z in {0,...,1}
		\draw[help lines] (-0.2,\y,\z)--(3.2,\y,\z) ;
\foreach \x in {0,...,3}
	\foreach \z in {0,...,1}
		\draw[help lines] (\x,-0.2,\z)--(\x,2.2,\z) ;
\foreach \x in {0,...,3}
	\foreach \y in {0,...,2}
		\draw[help lines] (\x,\y,-0.2)--(\x,\y,1.2) ;
\draw[white,line width=0.02cm] (1,-0.2,1)--(1,2.2,1) ;
\draw[densely dashed,thick] (1,-0.2,1)--(1,2.2,1) node[very near end,below left] {$\ot$} ;
\draw[white,line width=0.02cm] (2,-0.2,1)--(2,2.2,1) ;
\draw[densely dashed,thick] (2,-0.2,1)--(2,2.2,1) node[very near end,below right] {$\ot$} ;
\node at (1.5,0,1) {$\downarrow$} ;
\node at (1.5,1,1) {$\downarrow$} ;
\node at (1.5,2,1) {$\downarrow$} ;
\end{tikzpicture}
\end{array} \\
& =
\begin{array}{c}
\begin{tikzpicture}[scale=0.9]
\foreach \y in {0,...,2}
	\foreach \z in {0,...,1}
		\draw[help lines] (-0.2,\y,\z)--(3.2,\y,\z) ;
\foreach \x in {0,...,3}
	\foreach \z in {0,...,1}
		\draw[help lines] (\x,-0.2,\z)--(\x,2.2,\z) ;
\foreach \x in {0,...,3}
	\foreach \y in {0,...,2}
		\draw[help lines] (\x,\y,-0.2)--(\x,\y,1.2) ;
\draw[white,line width=0.02cm] (1,-0.2,1)--(1,2.2,1) ;
\draw[densely dashed,thick] (1,-0.2,1)--(1,2.2,1) ;
\draw[white,line width=0.02cm] (2,-0.2,1)--(2,2.2,1) ;
\draw[densely dashed,thick] (2,-0.2,1)--(2,2.2,1) node[very near end,below right] {$\ot$} ;
\foreach \y in {0,...,2}{
	\draw[white,line width=0.02cm] (1,\y,1)--(2,\y,1) ;
	\draw[densely dashed,thick] (1,\y,1)--(2,\y,1) ;
}
\end{tikzpicture}
\end{array}
\oplus
\begin{array}{c}
\begin{tikzpicture}[scale=0.9]
\foreach \y in {0,...,2}
	\foreach \z in {0,...,1}
		\draw[help lines] (-0.2,\y,\z)--(3.2,\y,\z) ;
\foreach \x in {0,...,3}
	\foreach \z in {0,...,1}
		\draw[help lines] (\x,-0.2,\z)--(\x,2.2,\z) ;
\foreach \x in {0,...,3}
	\foreach \y in {0,...,2}
		\draw[help lines] (\x,\y,-0.2)--(\x,\y,1.2) ;
\draw[white,line width=0.02cm] (1,-0.2,1)--(1,2.2,1) ;
\draw[white,line width=0.02cm] (2,-0.2,1)--(2,2.2,1) ;
\foreach \y in {0,...,2}
	\draw[white,line width=0.02cm] (1,\y,1)--(2,\y,1) ;
\foreach \y in {0,...,2}{
	\begin{scope}
	\clip (1,\y,0)--(2,\y,0)--(2,\y,0.98)--(1,\y,0.98)--cycle ;
	\draw[help lines,fill=m_ext,fill opacity=0.5] (1,\y,0)--(2,\y,0)--(2,\y,1)--(1,\y,1)--cycle ;
	\end{scope}
}
\draw[densely dashed,thick] (1,-0.2,1)--(1,2.2,1) ;
\draw[densely dashed,thick] (2,-0.2,1)--(2,2.2,1) node[very near end,below right] {$\ot$} ;
\foreach \y in {0,...,2}
	\draw[densely dashed,thick] (1,\y,1)--(2,\y,1) ;
\foreach \y in {0,...,2}{
	\begin{scope}
	\clip (1,\y,2)--(2,\y,2)--(2,\y,1.02)--(1,\y,1.02)--cycle ;
	\draw[help lines,fill=m_ext,fill opacity=0.5] (1,\y,2)--(2,\y,2)--(2,\y,1)--(1,\y,1)--cycle ;
	\end{scope}
	\draw[help lines] (1,\y,1)--(1,\y,2)--(2,\y,2)--(2,\y,1) ;
}
\end{tikzpicture}
\end{array}
\end{align*}
\caption{The fusion of two $\ot$-strings is $\ot \otimes \ot = \ot \oplus \ot$.}
\label{fig:fusion_TT}
\end{figure}

Let us consider the fusion of two $\ot$-strings. As depicted in Figure~\ref{fig:fusion_TT}, if two $\ot$-strings are closed enough, all the spins on the edges between them have to be $\uparrow$ or $\downarrow$, due to the constraints that $B_p = +1$ for all plaquettes $p$ between two strings. If all of these spins take the value $\uparrow$, we get a single "thick" $\ot$-string. If all of these spins take the value $\downarrow$, we need to flip them (recall our convention in Remark~\ref{rem:T-string}) by applying $\sigma_x$ operators on these edges. The result is a "thick" $\ot$-string together with two $m$-strings, and clearly these strings fuse into a single $\ot$-string. Hence we obtain $\ot \otimes \ot = \ot \oplus \ot$, where two direct summands correspond to spin $\uparrow$ and spin $\downarrow$ respectively. It also implies that $\mt \otimes \mt = \ot \oplus \ot$ and $\ot \otimes \mt = \mt \oplus \mt$. 

We summarize all the non-trivial fusion rules as follows: 
\be \label{eq:fusion-rule}
m\otimes m = \one, \quad\quad \ot\otimes m = \mt, \quad\quad \ot\otimes \ot = \mt \otimes \mt = \ot\oplus \ot, \quad\quad \ot \otimes \mt = \mt \oplus \mt. 
\ee

\begin{rem}
Using the fact that $e$-particles condense on a $\ot$-string and $\ot \otimes \ot = \ot \oplus \ot$, it is easy to see that a $\ot$-string shrink to a particle-like excitation $1_\one \oplus e$ (see also Section~\ref{sec:loop_link}). 
\end{rem}

Comparing with the fusion rules of $\Z(2\vect_{\Zb_2})$ in \cite[Example\ 3.8]{KTZ20}, we see that $\toric \simeq \Z(2\vect_{\Zb_2})$ as monoidal semisimple 2-categories \cite{DR18}. 

\begin{rem}
Note that the full 2-subcategory $\core$ is a monoidal 2-subcategory of $\toric$. 
\end{rem}

\subsection{Loops and links} \label{sec:loop_link}

In this subsection we give some examples of loop-like topological defects in the 3d toric code model. By a dimensional reduction process, they shrink to particle-like topological defects, i.e. 0d domain walls between two trivial strings $\one$.

\begin{figure}[htbp]
\centering
\begin{tikzpicture}[scale=0.9]
\foreach \y in {0,...,3}
	\foreach \z in {0}
		\draw[help lines] (-0.2,\y,\z)--(3.2,\y,\z) ;
\foreach \x in {0,...,3}
	\foreach \z in {0}
		\draw[help lines] (\x,-0.2,\z)--(\x,3.2,\z) ;
\foreach \x in {0,...,3}
	\foreach \y in {0,...,3}
		\draw[help lines] (\x,\y,-0.2)--(\x,\y,1.2) ;
\draw[m_dual_str] (0.5,2.5,0.5)--(2.5,2.5,0.5)--(2.5,0.5,0.5) node[below,black] {$m$} --(0.5,0.5,0.5)--cycle ;
\foreach \x in {0,2}
	\foreach \y in {1,2}
		\draw[help lines,fill=m_ext,fill opacity=0.5] (\x,\y,0)--(\x+1,\y,0)--(\x+1,\y,1)--(\x,\y,1)--cycle ;
\foreach \x in {1,2}
	\foreach \y in {0,2}
		\draw[help lines,fill=m_ext,fill opacity=0.5] (\x,\y,0)--(\x,\y+1,0)--(\x,\y+1,1)--(\x,\y,1)--cycle ;
\draw[m_str] (1,2,0)--(1,2,1) node[midway,link_label] {$1$} ;
\draw[m_str] (1,1,0)--(1,1,1) node[midway,link_label] {$2$} ;
\draw[m_str] (2,1,0)--(2,1,1) node[midway,link_label] {$3$} ;
\draw[m_str] (2,2,0)--(2,2,1) node[midway,link_label] {$4$} ;
\foreach \y in {0,...,3}
	\foreach \z in {1}
		\draw[help lines] (-0.2,\y,\z)--(3.2,\y,\z) ;
\foreach \x in {0,...,3}
	\foreach \z in {1}
		\draw[help lines] (\x,-0.2,\z)--(\x,3.2,\z) ;
\end{tikzpicture}
\caption{An $m$-loop can be created/annihilated by a membrane operator.}
\label{fig:m_loop}
\end{figure}
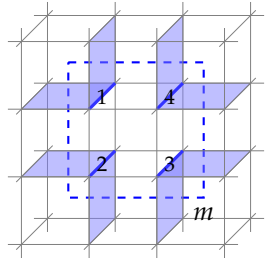

\medskip
The first example is an $m$-loop. As depicted in Figure~\ref{fig:m_loop}, an $m$-loop shirinks to a trivial particle $1_\one$ because it can be annihilated by a membrane operator $\sigma_x^1 \sigma_x^2 \sigma_x^3 \sigma_x^4$.

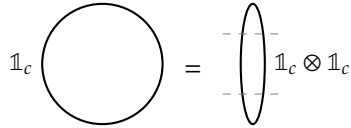
\begin{figure}[htbp]
\[
\begin{array}{c}
\begin{tikzpicture}[scale=0.8]
\draw[thick] (0,0) circle (1) ;
\node[left] at (-1,0) {$\ot$} ;
\end{tikzpicture}
\end{array}
=
\begin{array}{c}
\begin{tikzpicture}[scale=0.8]
\draw[thick] (0,0) ellipse (0.2 and 1) ;
\draw[help lines,dashed] (-0.5,0.5)--(0.5,0.5) ;
\draw[help lines,dashed] (-0.5,-0.5)--(0.5,-0.5) ;
\node[right] at (0.2,0) {$\ot \otimes \ot$} ;
\end{tikzpicture}
\end{array}
\]
\caption{A $\ot$-loop is equal to the composition of two 0d domain walls along a $\ot \otimes \ot$-string.}
\label{fig:T_loop}
\end{figure}

The second example is a $\ot$-loop. As depicted in Figure~\ref{fig:T_loop}, the dimensional reduction process can be done in two steps: first we horizontally squeeze the $\ot$-loop to get a $\ot \otimes \ot$-string with two ends; then we vertically squeeze the string to get a particle. In other words, the dimensional reduction of a $\ot$-loop is equal to the composition of two 0d domain walls along a $\ot \otimes \ot$-string.

\begin{figure}[htbp]
\[
\begin{array}{c}
\begin{tikzpicture}[scale=0.9]
\foreach \y in {0,...,3}
	\foreach \z in {0,...,1}
		\draw[help lines] (-0.2,\y,\z)--(3.2,\y,\z) ;
\foreach \x in {0,...,3}
	\foreach \z in {0,...,1}
		\draw[help lines] (\x,-0.2,\z)--(\x,3.2,\z) ;
\foreach \x in {0,...,3}
	\foreach \y in {0,...,3}
		\draw[help lines] (\x,\y,-0.2)--(\x,\y,1.2) ;
\draw[white,line width=0.02cm] (1,-0.2,1)--(1,2,1)--(2,2,1)--(2,-0.2,1) ;
\draw[densely dashed,thick] (1,-0.2,1)--(1,2,1)--(2,2,1)--(2,-0.2,1) node[midway,below right] {$\ot$} ;
\end{tikzpicture}
\end{array}
=
\begin{array}{c}
\begin{tikzpicture}[scale=0.9]
\foreach \y in {0,...,3}
	\foreach \z in {0,...,1}
		\draw[help lines] (-0.2,\y,\z)--(3.2,\y,\z) ;
\foreach \x in {0,...,3}
	\foreach \z in {0,...,1}
		\draw[help lines] (\x,-0.2,\z)--(\x,3.2,\z) ;
\foreach \x in {0,...,3}
	\foreach \y in {0,...,3}
		\draw[help lines] (\x,\y,-0.2)--(\x,\y,1.2) ;
\draw[white,line width=0.02cm] (1,-0.2,1)--(1,2,1)--(2,2,1)--(2,-0.2,1) ;
\draw[densely dashed,thick] (1,-0.2,1)--(1,2,1)--(2,2,1)--(2,-0.2,1) node[midway,below right] {$\ot$} ;
\node at (1.5,0,1) {$\uparrow$} ;
\node at (1.5,1,1) {$\uparrow$} ;
\end{tikzpicture}
\end{array}
=
\begin{array}{c}
\begin{tikzpicture}[scale=0.9]
\foreach \y in {0,...,3}
	\foreach \z in {0,...,1}
		\draw[help lines] (-0.2,\y,\z)--(3.2,\y,\z) ;
\foreach \x in {0,...,3}
	\foreach \z in {0,...,1}
		\draw[help lines] (\x,-0.2,\z)--(\x,3.2,\z) ;
\foreach \x in {0,...,3}
	\foreach \y in {0,...,3}
		\draw[help lines] (\x,\y,-0.2)--(\x,\y,1.2) ;
\draw[white,line width=0.02cm] (1,-0.2,1)--(1,2,1)--(2,2,1)--(2,-0.2,1) ;
\draw[densely dashed,thick] (1,-0.2,1)--(1,2,1)--(2,2,1)--(2,-0.2,1) node[midway,below right] {$\ot$} ;
\foreach \y in {0,1}{
	\draw[white,line width=0.02cm] (1,\y,1)--(2,\y,1) ;
	\draw[densely dashed,thick] (1,\y,1)--(2,\y,1) ;
}
\end{tikzpicture}
\end{array}
\]
\caption{The upper 0d domain wall in Figure~\ref{fig:T_loop} is equal to \eqref{eq:T_loop_upper}.}
\label{fig:T_loop_upper}
\end{figure}
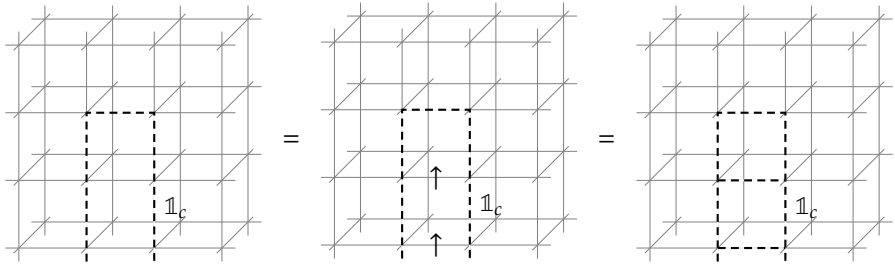

Let us compute the 0d domain walls at the two ends of the $\ot \otimes \ot$-string in Figure~\ref{fig:T_loop}. The upper one, which is viewed as a 1-morphism from $\one$ to $\ot \otimes \ot$, is depicted in Figure~\ref{fig:T_loop_upper}. All the spins on the edges between two $\ot$-strings should take the value $\uparrow$ or $\downarrow$. But at the end of the $\ot \otimes \ot$-string, the spins have to take the value $\uparrow$ because the $B_p$ operator at the end consists of a single $\sigma_z$ operator. Therefore, the result is a single $\ot$-string, which corresponds to the first direct summand of $\ot \otimes \ot = \ot \oplus \ot$. Recall that the end of this $\ot$-string, i.e. the 0d domain wall between the trivial string $\one$ and this $\ot$-string is nothing but $x : \one \to \ot$. Hence, the upper 0d domain wall in Figure~\ref{fig:T_loop} is equal to
\be \label{eq:T_loop_upper}
\one \xrightarrow{\begin{pmatrix} x \\ 0 \end{pmatrix}} \ot \oplus \ot = \ot \otimes \ot .
\ee

Similarly, the lower 0d domain wall in Figure~\ref{fig:T_loop}, viewed as a 1-morphism from $\ot \otimes \ot$ to $\one$, is equal to
\be \label{eq:T_loop_lower}
\ot \otimes \ot = \ot \oplus \ot \xrightarrow{\begin{pmatrix} y & 0 \end{pmatrix}} \one .
\ee
Then we have
\be
\biggl( \one \xrightarrow{\begin{pmatrix} x \\ 0 \end{pmatrix}} \ot \oplus \ot \xrightarrow{\begin{pmatrix} y & 0 \end{pmatrix}} \one \biggr) = \begin{pmatrix} y & 0 \end{pmatrix} \begin{pmatrix} x \\ 0 \end{pmatrix} = y \circ x = 1_\one \oplus e .
\ee
Hence, the dimensional reduction of a $\ot$-loop to a point gives a composite particle: $1_\one \oplus e$.

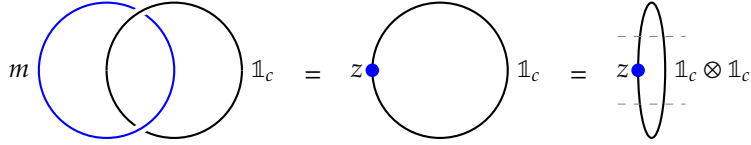
\begin{figure}[htbp]
\[
\begin{array}{c}
\begin{tikzpicture}[scale=0.9]
\draw[thick] (0,0) arc (180:360:1) ;
\draw[white,line width=4pt] (0,0) circle (1) ;
\draw[thick,blue] (0,0) circle (1) ;
\draw[white,line width=4pt] (0,0) arc (180:0:1) ;
\draw[thick] (0,0) arc (180:0:1) ;
\node[left] at (-1,0) {$m$} ;
\node[right] at (2,0) {$\ot$} ;
\end{tikzpicture}
\end{array}
=
\begin{array}{c}
\begin{tikzpicture}[scale=0.9]
\draw[thick] (0,0) circle (1) ;
\node[right] at (1,0) {$\ot$} ;
\fill[blue] (-1,0) circle (0.1) node[left,black] {$z$} ;
\end{tikzpicture}
\end{array}
=
\begin{array}{c}
\begin{tikzpicture}[scale=0.9]
\draw[thick] (0,0) ellipse (0.2 and 1) ;
\draw[help lines,dashed] (-0.5,0.5)--(0.5,0.5) ;
\draw[help lines,dashed] (-0.5,-0.5)--(0.5,-0.5) ;
\node[right] at (0.2,0) {$\ot \otimes \ot$} ;
\fill[blue] (-0.2,0) circle (0.1) node[left,black] {$z$} ;
\end{tikzpicture}
\end{array}
\]
\caption{An $m$-$\ot$-link is equal to the composition of three 0d domain walls.}
\label{fig:m_T_link}
\end{figure}

\medskip
The third example is a link consisting of an $m$-loop and a $\ot$-loop. As depicted in Figure~\ref{fig:m_T_link}, we first shrink the $m$-loop to get a $z$-particle on the $\ot$-loop (recall Figure~\ref{fig:excitation_z}), then horizontally squeeze the $\ot$-loop. Thus the link equals to the composition of three 0d domain walls, where the upper one and the lower one is known and the middle one is equal to $z \otimes 1_{\ot} : \ot \otimes \ot \to \ot \otimes \ot$.

\begin{figure}[htbp]
\begin{align*}
\begin{array}{c}
\begin{tikzpicture}[scale=0.8]
\foreach \y in {0,...,3}
	\foreach \z in {0,1}
		\draw[help lines] (-0.2,\y,\z)--(3.2,\y,\z) ;
\foreach \x in {0,...,3}
	\foreach \z in {0,1}
		\draw[help lines] (\x,-0.2,\z)--(\x,3.2,\z) ;
\foreach \x in {0,...,3}
	\foreach \y in {0,...,3}
		\draw[help lines] (\x,\y,-0.2)--(\x,\y,1.2) ;
\draw[white,line width=0.02cm] (1,-0.2,1)--(1,3.2,1) ;
\draw[densely dashed,thick] (1,-0.2,1)--(1,3.2,1) node[very near end,below left] {$\ot$} ;
\draw[white,line width=0.02cm] (2,-0.2,1)--(2,3.2,1) ;
\draw[densely dashed,thick] (2,-0.2,1)--(2,3.2,1) node[very near end,below right] {$\ot$} ;
\begin{scope}
\clip (1,1,0)--(1,2,0)--(1,2,0.98)--(1,1,0.98)--cycle ;
\draw[help lines,fill=m_ext,fill opacity=0.5] (1,1,0)--(1,2,0)--(1,2,1)--(1,1,1)--cycle ;
\end{scope}
\begin{scope}
\clip (0,1,1) rectangle (0.98,2,1) ;
\draw[help lines,fill=m_ext,fill opacity=0.5] (0,1,1) rectangle (1,2,1) ;
\end{scope}
\begin{scope}
\clip (1.02,1,1) rectangle (1.98,2,1) ;
\draw[help lines,fill=m_ext,fill opacity=0.5] (1,1,1) rectangle (2,2,1) ;
\end{scope}
\begin{scope}
\clip (1,1,2)--(1,2,2)--(1,2,1.02)--(1,1,1.02)--cycle ;
\draw[help lines,fill=m_ext,fill opacity=0.5] (1,1,2)--(1,2,2)--(1,2,1)--(1,1,1)--cycle ;
\end{scope}
\draw[help lines] (1,1,1.02)--(1,1,2)--(1,2,2)--(1,2,1.02) ;
\end{tikzpicture}
\end{array}
& =
\begin{array}{c}
\begin{tikzpicture}[scale=0.8]
\foreach \y in {0,...,3}
	\foreach \z in {0,...,1}
		\draw[help lines] (-0.2,\y,\z)--(3.2,\y,\z) ;
\foreach \x in {0,...,3}
	\foreach \z in {0,...,1}
		\draw[help lines] (\x,-0.2,\z)--(\x,3.2,\z) ;
\foreach \x in {0,...,3}
	\foreach \y in {0,...,3}
		\draw[help lines] (\x,\y,-0.2)--(\x,\y,1.2) ;
\draw[white,line width=0.02cm] (1,-0.2,1)--(1,3.2,1) ;
\draw[densely dashed,thick] (1,-0.2,1)--(1,3.2,1) node[very near end,below left] {$\ot$} ;
\draw[white,line width=0.02cm] (2,-0.2,1)--(2,3.2,1) ;
\draw[densely dashed,thick] (2,-0.2,1)--(2,3.2,1) node[very near end,below right] {$\ot$} ;
\begin{scope}
\clip (1,1,0)--(1,2,0)--(1,2,0.98)--(1,1,0.98)--cycle ;
\draw[help lines,fill=m_ext,fill opacity=0.5] (1,1,0)--(1,2,0)--(1,2,1)--(1,1,1)--cycle ;
\end{scope}
\begin{scope}
\clip (0,1,1) rectangle (0.98,2,1) ;
\draw[help lines,fill=m_ext,fill opacity=0.5] (0,1,1) rectangle (1,2,1) ;
\end{scope}
\begin{scope}
\clip (1.02,1,1) rectangle (1.98,2,1) ;
\draw[help lines,fill=m_ext,fill opacity=0.5] (1,1,1) rectangle (2,2,1) ;
\end{scope}
\begin{scope}
\clip (1,1,2)--(1,2,2)--(1,2,1.02)--(1,1,1.02)--cycle ;
\draw[help lines,fill=m_ext,fill opacity=0.5] (1,1,2)--(1,2,2)--(1,2,1)--(1,1,1)--cycle ;
\end{scope}
\draw[help lines] (1,1,1.02)--(1,1,2)--(1,2,2)--(1,2,1.02) ;
\node at (1.5,0,1) {$\downarrow$} ;
\node at (1.5,1,1) {$\downarrow$} ;
\node at (1.5,2,1) {$\uparrow$} ;
\node at (1.5,3,1) {$\uparrow$} ;
\end{tikzpicture}
\end{array}
\oplus
\begin{array}{c}
\begin{tikzpicture}[scale=0.8]
\foreach \y in {0,...,3}
	\foreach \z in {0,...,1}
		\draw[help lines] (-0.2,\y,\z)--(3.2,\y,\z) ;
\foreach \x in {0,...,3}
	\foreach \z in {0,...,1}
		\draw[help lines] (\x,-0.2,\z)--(\x,3.2,\z) ;
\foreach \x in {0,...,3}
	\foreach \y in {0,...,3}
		\draw[help lines] (\x,\y,-0.2)--(\x,\y,1.2) ;
\draw[white,line width=0.02cm] (1,-0.2,1)--(1,3.2,1) ;
\draw[densely dashed,thick] (1,-0.2,1)--(1,3.2,1) node[very near end,below left] {$\ot$} ;
\draw[white,line width=0.02cm] (2,-0.2,1)--(2,3.2,1) ;
\draw[densely dashed,thick] (2,-0.2,1)--(2,3.2,1) node[very near end,below right] {$\ot$} ;
\begin{scope}
\clip (1,1,0)--(1,2,0)--(1,2,0.98)--(1,1,0.98)--cycle ;
\draw[help lines,fill=m_ext,fill opacity=0.5] (1,1,0)--(1,2,0)--(1,2,1)--(1,1,1)--cycle ;
\end{scope}
\begin{scope}
\clip (0,1,1) rectangle (0.98,2,1) ;
\draw[help lines,fill=m_ext,fill opacity=0.5] (0,1,1) rectangle (1,2,1) ;
\end{scope}
\begin{scope}
\clip (1.02,1,1) rectangle (1.98,2,1) ;
\draw[help lines,fill=m_ext,fill opacity=0.5] (1,1,1) rectangle (2,2,1) ;
\end{scope}
\begin{scope}
\clip (1,1,2)--(1,2,2)--(1,2,1.02)--(1,1,1.02)--cycle ;
\draw[help lines,fill=m_ext,fill opacity=0.5] (1,1,2)--(1,2,2)--(1,2,1)--(1,1,1)--cycle ;
\end{scope}
\draw[help lines] (1,1,1.02)--(1,1,2)--(1,2,2)--(1,2,1.02) ;
\node at (1.5,0,1) {$\uparrow$} ;
\node at (1.5,1,1) {$\uparrow$} ;
\node at (1.5,2,1) {$\downarrow$} ;
\node at (1.5,3,1) {$\downarrow$} ;
\end{tikzpicture}
\end{array} \\
& =
\begin{array}{c}
\begin{tikzpicture}[scale=0.8]
\foreach \y in {0,...,3}
	\foreach \z in {0,...,1}
		\draw[help lines] (-0.2,\y,\z)--(3.2,\y,\z) ;
\foreach \x in {0,...,3}
	\foreach \z in {0,...,1}
		\draw[help lines] (\x,-0.2,\z)--(\x,3.2,\z) ;
\foreach \x in {0,...,3}
	\foreach \y in {0,...,3}
		\draw[help lines] (\x,\y,-0.2)--(\x,\y,1.2) ;
\draw[white,line width=0.02cm] (1,-0.2,1)--(1,3.2,1) ;
\draw[white,line width=0.02cm] (2,-0.2,1)--(2,3.2,1) ;
\foreach \y in {0,...,3}
	\draw[white,line width=0.02cm] (1,\y,1)--(2,\y,1) ;
\foreach \y in {0,1}{
	\begin{scope}
	\clip (1,\y,0)--(2,\y,0)--(2,\y,0.98)--(1,\y,0.98)--cycle ;
	\draw[help lines,fill=m_ext,fill opacity=0.5] (1,\y,0)--(2,\y,0)--(2,\y,1)--(1,\y,1)--cycle ;
	\end{scope}
}
\draw[densely dashed,thick] (1,-0.2,1)--(1,3.2,1) ;
\draw[densely dashed,thick] (2,-0.2,1)--(2,3.2,1) node[very near end,below right] {$\ot$} ;
\foreach \y in {0,...,3}
	\draw[densely dashed,thick] (1,\y,1)--(2,\y,1) ;
\foreach \y in {0,1}{
	\begin{scope}
	\clip (1,\y,2)--(2,\y,2)--(2,\y,1.02)--(1,\y,1.02)--cycle ;
	\draw[help lines,fill=m_ext,fill opacity=0.5] (1,\y,2)--(2,\y,2)--(2,\y,1)--(1,\y,1)--cycle ;
	\end{scope}
	\draw[help lines] (1,\y,1)--(1,\y,2)--(2,\y,2)--(2,\y,1) ;
}
\begin{scope}
\clip (1,1,0)--(1,2,0)--(1,2,0.98)--(1,1,0.98)--cycle ;
\draw[help lines,fill=m_ext,fill opacity=0.5] (1,1,0)--(1,2,0)--(1,2,1)--(1,1,1)--cycle ;
\end{scope}
\begin{scope}
\clip (0,1,1) rectangle (0.98,2,1) ;
\draw[help lines,fill=m_ext,fill opacity=0.5] (0,1,1) rectangle (1,2,1) ;
\end{scope}
\begin{scope}
\clip (1,1,2)--(1,2,2)--(1,2,1.02)--(1,1,1.02)--cycle ;
\draw[help lines,fill=m_ext,fill opacity=0.5] (1,1,2)--(1,2,2)--(1,2,1)--(1,1,1)--cycle ;
\end{scope}
\draw[help lines] (1,1,1.02)--(1,1,2)--(1,2,2)--(1,2,1.02) ;
\end{tikzpicture}
\end{array}
\oplus
\begin{array}{c}
\begin{tikzpicture}[scale=0.8]
\foreach \y in {0,...,3}
	\foreach \z in {0,...,1}
		\draw[help lines] (-0.2,\y,\z)--(3.2,\y,\z) ;
\foreach \x in {0,...,3}
	\foreach \z in {0,...,1}
		\draw[help lines] (\x,-0.2,\z)--(\x,3.2,\z) ;
\foreach \x in {0,...,3}
	\foreach \y in {0,...,3}
		\draw[help lines] (\x,\y,-0.2)--(\x,\y,1.2) ;
\draw[white,line width=0.02cm] (1,-0.2,1)--(1,3.2,1) ;
\draw[white,line width=0.02cm] (2,-0.2,1)--(2,3.2,1) ;
\foreach \y in {0,...,3}
	\draw[white,line width=0.02cm] (1,\y,1)--(2,\y,1) ;
\begin{scope}
\clip (1,1,0)--(1,2,0)--(1,2,0.98)--(1,1,0.98)--cycle ;
\draw[help lines,fill=m_ext,fill opacity=0.5] (1,1,0)--(1,2,0)--(1,2,1)--(1,1,1)--cycle ;
\end{scope}
\begin{scope}
\clip (0,1,1) rectangle (0.98,2,1) ;
\draw[help lines,fill=m_ext,fill opacity=0.5] (0,1,1) rectangle (1,2,1) ;
\end{scope}
\begin{scope}
\clip (1,1,2)--(1,2,2)--(1,2,1.02)--(1,1,1.02)--cycle ;
\draw[help lines,fill=m_ext,fill opacity=0.5] (1,1,2)--(1,2,2)--(1,2,1)--(1,1,1)--cycle ;
\end{scope}
\draw[help lines] (1,1,1.02)--(1,1,2)--(1,2,2)--(1,2,1.02) ;
\foreach \y in {2,3}{
	\begin{scope}
	\clip (1,\y,0)--(2,\y,0)--(2,\y,0.98)--(1,\y,0.98)--cycle ;
	\draw[help lines,fill=m_ext,fill opacity=0.5] (1,\y,0)--(2,\y,0)--(2,\y,1)--(1,\y,1)--cycle ;
	\end{scope}
}
\draw[densely dashed,thick] (1,-0.2,1)--(1,3.2,1) ;
\draw[densely dashed,thick] (2,-0.2,1)--(2,3.2,1) node[very near end,below right] {$\ot$} ;
\foreach \y in {0,...,3}
	\draw[densely dashed,thick] (1,\y,1)--(2,\y,1) ;
\foreach \y in {2,3}{
	\begin{scope}
	\clip (1,\y,2)--(2,\y,2)--(2,\y,1.02)--(1,\y,1.02)--cycle ;
	\draw[help lines,fill=m_ext,fill opacity=0.5] (1,\y,2)--(2,\y,2)--(2,\y,1)--(1,\y,1)--cycle ;
	\end{scope}
	\draw[help lines] (1,\y,1)--(1,\y,2)--(2,\y,2)--(2,\y,1) ;
}
\end{tikzpicture}
\end{array}
\end{align*}
\caption{The 0d domain wall $z \otimes 1_{\ot}$ is equal to \eqref{eq:m_T_link_middle}.}
\label{fig:m_T_link_middle}
\end{figure}
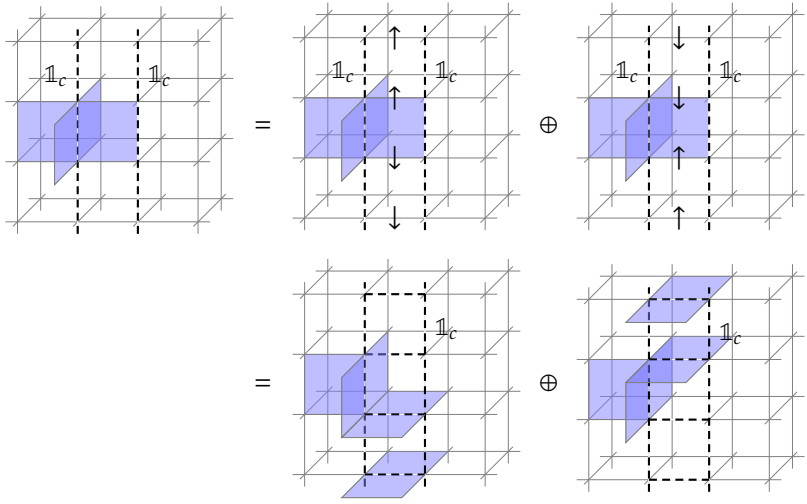

Let us compute the 0d domain wall $z \otimes 1_{\ot}$, as depicted in Figure~\ref{fig:m_T_link_middle}. Due to the middle blue plaquette with $B_p = -1$, there are two possibilities of the spins between two $\ot$-strings: the spins above the middle plaquette take the value $\uparrow$ and the spins below the middle plaquette take the value $\downarrow$, or the spins above the middle plaquette take the value $\downarrow$ and the spins below the middle plaquette take the value $\uparrow$. Therefore, the 0d domain wall $z \otimes 1_{\ot}$ exchanges two direct summands of $\ot \otimes \ot = \ot \oplus \ot$, and these two possibilities correspond to two direct summands of the 0d domain wall:
\be \label{eq:m_T_link_middle}
z \otimes 1_{\ot} = \biggl( \ot \otimes \ot = \ot \oplus \ot \xrightarrow{\begin{pmatrix} 0 & 1_{\ot} \\ 1_{\ot} & 0 \end{pmatrix}} \ot \oplus \ot = \ot \otimes \ot \biggr) .
\ee

Then we have
\be
\biggl( \one \xrightarrow{\begin{pmatrix} x \\ 0 \end{pmatrix}} \ot \oplus \ot \xrightarrow{\begin{pmatrix} 0 & 1_{\ot} \\ 1_{\ot} & 0 \end{pmatrix}} \ot \oplus \ot \xrightarrow{\begin{pmatrix} y & 0 \end{pmatrix}} \one \biggr) = \begin{pmatrix} y & 0 \end{pmatrix} \begin{pmatrix} 0 & 1_{\ot} \\ 1_{\ot} & 0 \end{pmatrix} \begin{pmatrix} x \\ 0 \end{pmatrix} = 0 .
\ee
Hence, the dimensional reduction of an $m$-$\ot$-link to a point gives $0$. This result means the configuration of an $m$-$\ot$-link is physically forbidden. One can see this fact explicitly in lattice model by reading Figure~\ref{fig:T_loop_upper} and Figure~\ref{fig:m_T_link_middle}. 

\begin{rem}
The fact that an $m$-$\ot$-link is physically forbidden was first pointed out by Else and Nayak in \cite[Section\ III]{EN17}. 
\end{rem}

\subsection{Dualities in \texorpdfstring{$\toric$}{Toric}}

In this subsection we prove that $\toric$ is a fusion 2-category. A fusion 2-category is a finite semisimple monoidal 2-category that has left and right duals for objects and a simple tensor unit $\one$ \cite{DR18}. Therefore, we only need to show that each string in $\toric$ admits left and right duals, which has been proved in Section~\ref{sec:dual} in an abstract way. In the following we explicitly list the evaluation and coevaluation 1-morphisms.

\medskip
In the 3d toric code model, it is clear that each object is self-dual, i.e. $X^* = X$ for $X = \one,\ot,m,\mt$. The evaluation and coevaluation 1-morphisms associated to $X = \one,m$ are given by $X \otimes X = \one \xrightarrow{1_\one} \one$ and $\one \xrightarrow{1_\one} \one = X \otimes X$. The evaluation and coevaluation 1-morphisms associated to $\ot$ are given by \eqref{eq:T_loop_lower} and \eqref{eq:T_loop_upper} respectively. Let us prove that the evaluation and coevaluation 1-morphisms associated to $\ot$ satisfy zig-zag equations \eqref{eq:zig_zag} by explicit computation. We only prove that
\be
\bigl( \ot = \ot \otimes \one \xrightarrow{1_{\ot} \otimes \coev} \ot \otimes \ot \otimes \ot \xrightarrow{\ev \otimes 1_{\ot}} \one \otimes \ot = \ot \bigr) = 1_{\ot} .
\ee
The proof of another equation is similar.

Note that $\ot \otimes \ot \otimes \ot = \ot \oplus \ot \oplus \ot \oplus \ot$. Here four direct summands correspond to four possibilities of spins between three $\ot$-strings (recall Figure~\ref{fig:fusion_TT}). Thus we denote
\[
\ot \otimes \ot \otimes \ot = \ot^{\uparrow \uparrow} \oplus \ot^{\uparrow \downarrow} \oplus \ot^{\downarrow \uparrow} \oplus \ot^{\downarrow \downarrow} ,
\]
where $\ot^{\alpha \beta}$ means the spins between the first and second $\ot$-strings take the value $\alpha$, and the spins between the second and third $\ot$-strings take the value $\beta$.

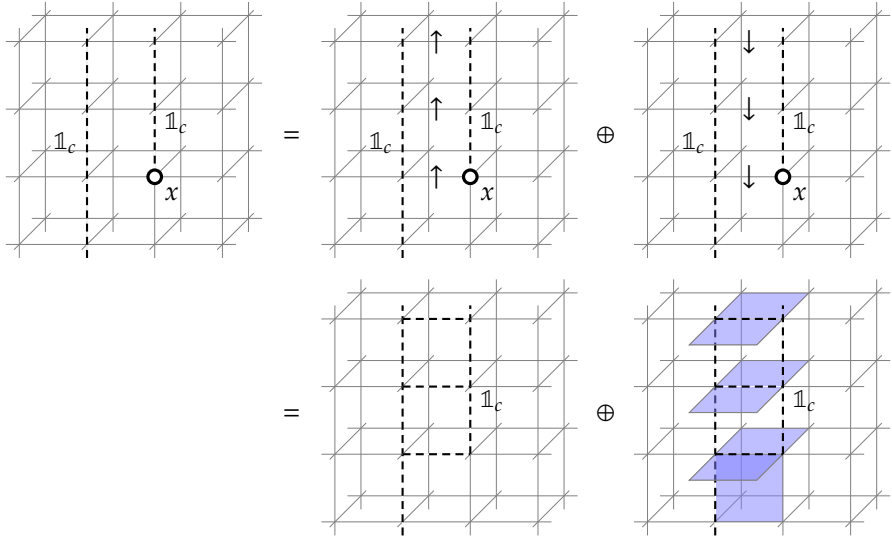
\begin{figure}[htbp]
\begin{align*}
\begin{array}{c}
\begin{tikzpicture}[scale=0.9]
\foreach \y in {0,...,3}
	\foreach \z in {0,...,1}
		\draw[help lines] (-0.2,\y,\z)--(3.2,\y,\z) ;
\foreach \x in {0,...,3}
	\foreach \z in {0,...,1}
		\draw[help lines] (\x,-0.2,\z)--(\x,3.2,\z) ;
\foreach \x in {0,...,3}
	\foreach \y in {0,...,3}
		\draw[help lines] (\x,\y,-0.2)--(\x,\y,1.2) ;
\draw[white,line width=0.02cm] (1,-0.2,1)--(1,3.2,1) ;
\draw[densely dashed,thick] (1,-0.2,1)--(1,3.2,1) node[midway,left] {$\ot$} ;
\draw[white,line width=0.02cm] (2,1,1)--(2,3.2,1) ;
\draw[densely dashed,thick] (2,1,1)--(2,3.2,1) node[midway,below right] {$\ot$} ;
\draw[very thick,fill=white] (2,1,1) circle (0.1) node[below right] {$x$} ;
\end{tikzpicture}
\end{array}
& =
\begin{array}{c}
\begin{tikzpicture}[scale=0.9]
\foreach \y in {0,...,3}
	\foreach \z in {0,...,1}
		\draw[help lines] (-0.2,\y,\z)--(3.2,\y,\z) ;
\foreach \x in {0,...,3}
	\foreach \z in {0,...,1}
		\draw[help lines] (\x,-0.2,\z)--(\x,3.2,\z) ;
\foreach \x in {0,...,3}
	\foreach \y in {0,...,3}
		\draw[help lines] (\x,\y,-0.2)--(\x,\y,1.2) ;
\draw[white,line width=0.02cm] (1,-0.2,1)--(1,3.2,1) ;
\draw[densely dashed,thick] (1,-0.2,1)--(1,3.2,1) node[midway,left] {$\ot$} ;
\draw[white,line width=0.02cm] (2,1,1)--(2,3.2,1) ;
\draw[densely dashed,thick] (2,1,1)--(2,3.2,1) node[midway,below right] {$\ot$} ;
\draw[very thick,fill=white] (2,1,1) circle (0.1) node[below right] {$x$} ;
\node at (1.5,1,1) {$\uparrow$} ;
\node at (1.5,2,1) {$\uparrow$} ;
\node at (1.5,3,1) {$\uparrow$} ;
\end{tikzpicture}
\end{array}
\oplus
\begin{array}{c}
\begin{tikzpicture}[scale=0.9]
\foreach \y in {0,...,3}
	\foreach \z in {0,...,1}
		\draw[help lines] (-0.2,\y,\z)--(3.2,\y,\z) ;
\foreach \x in {0,...,3}
	\foreach \z in {0,...,1}
		\draw[help lines] (\x,-0.2,\z)--(\x,3.2,\z) ;
\foreach \x in {0,...,3}
	\foreach \y in {0,...,3}
		\draw[help lines] (\x,\y,-0.2)--(\x,\y,1.2) ;
\draw[white,line width=0.02cm] (1,-0.2,1)--(1,3.2,1) ;
\draw[densely dashed,thick] (1,-0.2,1)--(1,3.2,1) node[midway,left] {$\ot$} ;
\draw[white,line width=0.02cm] (2,1,1)--(2,3.2,1) ;
\draw[densely dashed,thick] (2,1,1)--(2,3.2,1) node[midway,below right] {$\ot$} ;
\draw[very thick,fill=white] (2,1,1) circle (0.1) node[below right] {$x$} ;
\node at (1.5,1,1) {$\downarrow$} ;
\node at (1.5,2,1) {$\downarrow$} ;
\node at (1.5,3,1) {$\downarrow$} ;
\end{tikzpicture}
\end{array} \\
& =
\begin{array}{c}
\begin{tikzpicture}[scale=0.9]
\foreach \y in {0,...,3}
	\foreach \z in {0,...,1}
		\draw[help lines] (-0.2,\y,\z)--(3.2,\y,\z) ;
\foreach \x in {0,...,3}
	\foreach \z in {0,...,1}
		\draw[help lines] (\x,-0.2,\z)--(\x,3.2,\z) ;
\foreach \x in {0,...,3}
	\foreach \y in {0,...,3}
		\draw[help lines] (\x,\y,-0.2)--(\x,\y,1.2) ;
\draw[white,line width=0.02cm] (1,-0.2,1)--(1,3.2,1) ;
\draw[densely dashed,thick] (1,-0.2,1)--(1,3.2,1) ;
\draw[white,line width=0.02cm] (2,1,1)--(2,3.2,1) ;
\draw[densely dashed,thick] (2,1,1)--(2,3.2,1) node[midway,below right] {$\ot$} ;
\foreach \y in {1,2,3}{
	\draw[white,line width=0.02cm] (1,\y,1)--(2,\y,1) ;
	\draw[densely dashed,thick] (1,\y,1)--(2,\y,1) ;
}
\end{tikzpicture}
\end{array}
\oplus
\begin{array}{c}
\begin{tikzpicture}[scale=0.9]
\foreach \y in {0,...,3}
	\foreach \z in {0,...,1}
		\draw[help lines] (-0.2,\y,\z)--(3.2,\y,\z) ;
\foreach \x in {0,...,3}
	\foreach \z in {0,...,1}
		\draw[help lines] (\x,-0.2,\z)--(\x,3.2,\z) ;
\foreach \x in {0,...,3}
	\foreach \y in {0,...,3}
		\draw[help lines] (\x,\y,-0.2)--(\x,\y,1.2) ;
\draw[white,line width=0.02cm] (1,-0.2,1)--(1,3.2,1) ;
\draw[white,line width=0.02cm] (2,1,1)--(2,3.2,1) ;
\foreach \y in {1,2,3}{
	\draw[white,line width=0.02cm] (1,\y,1)--(2,\y,1) ;
}
\begin{scope}
\clip (1.02,0,1) rectangle (2,1,1) ;
\draw[help lines,fill=m_ext,fill opacity=0.5] (1,0,1) rectangle (2,1,1) ;
\end{scope}
\foreach \y in {1,...,3}{
	\begin{scope}
	\clip (1,\y,0)--(2,\y,0)--(2,\y,0.98)--(1,\y,0.98)--cycle ;
	\draw[help lines,fill=m_ext,fill opacity=0.5] (1,\y,0)--(2,\y,0)--(2,\y,1)--(1,\y,1)--cycle ;
	\end{scope}
}
\draw[densely dashed,thick] (1,-0.2,1)--(1,3.2,1) ;
\draw[densely dashed,thick] (2,1,1)--(2,3.2,1) node[midway,below right] {$\ot$} ;
\foreach \y in {1,...,3}{
	\begin{scope}
	\clip (1,\y,2)--(2,\y,2)--(2,\y,1.02)--(1,\y,1.02)--cycle ;
	\draw[help lines,fill=m_ext,fill opacity=0.5] (1,\y,2)--(2,\y,2)--(2,\y,1)--(1,\y,1)--cycle ;
	\end{scope}
	\draw[help lines] (1,\y,1)--(1,\y,2)--(2,\y,2)--(2,\y,1) ;
}
\foreach \y in {1,2,3}{
	\draw[densely dashed,thick] (1,\y,1)--(2,\y,1) ;
}
\end{tikzpicture}
\end{array}
\end{align*}
\caption{The 1-morphism $1_{\ot} \otimes x$ is equal to \eqref{eq:fusion_Tx}.}
\label{fig:fusion_Tx}
\end{figure}

The 1-morphism $\ot \otimes \one \xrightarrow{1_{\ot} \otimes \coev} \ot \otimes \ot \otimes \ot$ is equal to
\[
\ot = \ot \otimes \one \xrightarrow{1_{\ot} \otimes \begin{pmatrix} x \\ 0 \end{pmatrix}} \ot \otimes (\ot \oplus \ot) = \ot \otimes \ot \otimes \ot .
\]
Therefore, if the spins between the second and the third $\ot$-strings take the value $\downarrow$, this 1-morphism restricts to $0$; if the spins between the second and the third $\ot$-strings take the value $\uparrow$, this 1-morphism restricts to $\ot \xrightarrow{1_{\ot} \otimes x} \ot \otimes \ot = \ot^{\uparrow \uparrow} \oplus \ot^{\downarrow \uparrow}$. Figure~\ref{fig:fusion_Tx} shows that
\be \label{eq:fusion_Tx}
1_{\ot} \otimes x = \biggl( \ot \xrightarrow{\begin{pmatrix}1_{\ot} \\ 1_{\ot}\end{pmatrix}} \ot \oplus \ot \biggr) .
\ee
Hence we have
\be
\bigl( \ot \otimes \one \xrightarrow{1_{\ot} \otimes \coev} \ot \otimes \ot \otimes \ot \bigr) = \biggl( \ot \xrightarrow{\begin{pmatrix}1_{\ot} \\ 0 \\ 1_{\ot} \\ 0\end{pmatrix}} \ot^{\uparrow \uparrow} \oplus \ot^{\uparrow \downarrow} \oplus \ot^{\downarrow \uparrow} \oplus \ot^{\downarrow \downarrow} \biggr) .
\ee

Similarly, the 1-morphism $\ev \otimes 1_{\ot}$ is equal to
\be
\bigl( \ot \otimes \ot \otimes \ot \xrightarrow{\ev \otimes 1_{\ot}} \one \otimes \ot = \ot \bigr) = \biggl( \ot^{\uparrow \uparrow} \oplus \ot^{\uparrow \downarrow} \oplus \ot^{\downarrow \uparrow} \oplus \ot^{\downarrow \downarrow} \xrightarrow{\begin{pmatrix}1_{\ot} & 1_{\ot} & 0 & 0\end{pmatrix}} \ot \biggr) .
\ee
Then we have
\be
\bigl( \ot = \ot \otimes \one \xrightarrow{1_{\ot} \otimes \coev} \ot \otimes \ot \otimes \ot \xrightarrow{\ev \otimes 1_{\ot}} \one \otimes \ot = \ot \bigr) = \begin{pmatrix}1_{\ot} & 1_{\ot} & 0 & 0\end{pmatrix} \begin{pmatrix}1_{\ot} \\ 0 \\ 1_{\ot} \\ 0\end{pmatrix} = 1_{\ot} .
\ee
This completes the proof. Note that only the $\ot^{\uparrow \uparrow}$ direct summand of $\ot \otimes \ot \otimes \ot$ contributes a nonzero 1-morphism.

Similarly, it is not hard to see that the evaluation and coevaluation 1-morphisms associated to $\mt$ are given by
\be
\mt \otimes \mt = \ot \oplus \ot \xrightarrow{\begin{pmatrix} y & 0 \end{pmatrix}} \one ,
\ee
and
\be
\one \xrightarrow{\begin{pmatrix} x \\ 0 \end{pmatrix}} \ot \oplus \ot = \mt \otimes \mt .
\ee

\subsection{Double braidings in \texorpdfstring{$\toric$}{Toric}} \label{sec:double-braidings}

It is obvious that the fusion product is commutative, 
i.e. $i\otimes j = j\otimes i$ for $i,j=\one,\ot,m,\mt$. In the 3d toric model, one can compute the double braidings as higher codimensional defects (see Section~\ref{sec:braiding_equivalent}).
\bnu
\item The double braiding between two $\one$-strings (including the particles on them) is obviously trivial. This is compatible with the physical fact that $e$ is a boson, and with the mathematical fact that $\rep(\Zb_2)$ is a symmetric fusion 1-category. 

\item The double braiding between a $\one$-string and a $\ot$-string (including all 0d defects on them) is obviously trivial. Note that this is compatible with the mathematical fact that $2\rep(\Zb_2)$ is a symmetric fusion 2-category. 

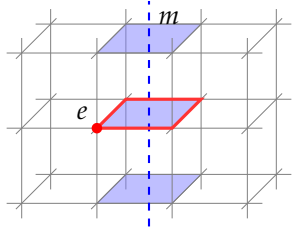
\begin{figure}[htbp]
\centering
\begin{tikzpicture}
\foreach \y in {0,...,2}
	\foreach \z in {0}
		\draw[help lines] (-0.2,\y,\z)--(3.2,\y,\z) ;
\foreach \x in {0,...,3}
	\foreach \z in {0}
		\draw[help lines] (\x,-0.2,\z)--(\x,2.2,\z) ;
\foreach \x in {0,...,3}
	\foreach \y in {0,...,2}
		\draw[help lines] (\x,\y,-0.2)--(\x,\y,1.2) ;
\foreach \y in {0,...,2}
	\draw[help lines,fill=m_ext,fill opacity=0.5] (1,\y,0)--(2,\y,0)--(2,\y,1)--(1,\y,1)--cycle ;
\foreach \y in {0,...,2}
	\foreach \z in {1}
		\draw[help lines] (-0.2,\y,\z)--(3.2,\y,\z) ;
\foreach \x in {0,...,3}
	\foreach \z in {1}
		\draw[help lines] (\x,-0.2,\z)--(\x,2.2,\z) ;
\draw[m_dual_str] (1.5,-0.5,0.5)--(1.5,2.5,0.5) node[at end,below right,black] {$m$} ;
\draw[e_str] (1,1,1)--(1,1,0)--(2,1,0)--(2,1,1)--cycle ;
\fill[e_ext] (1,1,1) circle (0.07) node[above left,black] {$e$} ;
\end{tikzpicture}
\caption{The double braiding of an $e$-particle and an $m$-string is $-1$.}
\label{fig:double_braiding_e_m}
\end{figure}

\item The double braiding between a $\one$-string (including particles $1_\one$ or $e$) and an $m$-string is non-trivial because the double braiding between an $e$-particle and an $m$-string is $-1$ as illustrated in Figure~\ref{fig:double_braiding_e_m}. 

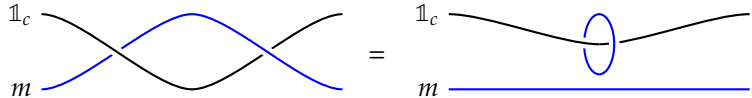
\begin{figure}[htbp]
\[
\begin{array}{c}
\begin{tikzpicture}
\draw[thick,blue] (0,0) node[left,black] {$m$} .. controls (0.5,0) and (1.5,1) .. (2,1) ;
\draw[white,line width=4pt] (0,1) .. controls (0.5,1) and (1.5,0) .. (2,0) ;
\draw[thick] (0,1) node[left] {$\ot$} .. controls (0.5,1) and (1.5,0) .. (2,0) .. controls (2.5,0) and (3.5,1) .. (4,1) ;
\draw[white,line width=4pt] (2,1) .. controls (2.5,1) and (3.5,0) .. (4,0) ;
\draw[thick,blue] (2,1) .. controls (2.5,1) and (3.5,0) .. (4,0) ;
\end{tikzpicture}
\end{array}
=
\begin{array}{c}
\begin{tikzpicture}
\draw[thick] (2,0.6) .. controls (2.5,0.6) and (3.5,1) .. (4,1) ;
\draw[white,line width=4pt] (2,0.6) ellipse (0.2 and 0.4) ;
\draw[thick,blue] (2,0.6) ellipse (0.2 and 0.4) ;
\draw[white,line width=4pt] (0,1) .. controls (0.5,1) and (1.5,0.6) .. (2,0.6) ;
\draw[thick] (0,1) node[left] {$\ot$} .. controls (0.5,1) and (1.5,0.6) .. (2,0.6) ;
\draw[thick,blue] (0,0) node[left,black] {$m$} --(4,0) ;
\end{tikzpicture}
\end{array}
\]
\caption{The double braiding of the $\ot$-string and the $m$-string is $z \otimes 1_m$.}
\label{fig:double_braiding_T_m}
\end{figure}

\item The double braiding between a $\ot$-string and an $\mt$-string is non-trivial and is given by $z_m=z \otimes 1_m$ as illustrated in Figure~\ref{fig:double_braiding_T_m}.

\item The double braiding between two $m$-strings is trivial unless we attach an $e$-particle on one of the $m$-strings, in which case the double braiding gives $-1$. 

\item All the rest double braidings can be obtained from above double braidings using the fusion properties. 

\enu
These double braidings coincide with those of $\Z(2\vect_{\Zb_2})$, which was implicitly given in \cite{KTZ20}. See more detail in Appendix~\ref{sec:app}. 

\medskip
The double braidings are physical and independent of any gauge choices. But the braiding structure is gauge dependent. Namely, different choices of the braidings can be braided equivalent. One way to fix the gauge is to select a gapped boundary of the 3d toric code model, then determine the braidings via the half-braidings of the defects on the boundary, as we do in Section~\ref{sec:half_braiding}.

\section{Gapped boundaries and braidings} \label{sec:boundary+braiding}

In this section, we first construct two gapped boundaries of 3d toric code model, then use the boundary-bulk relation to realize the braiding structure in $\toric$. 

\subsection{Smooth and rough gapped boundaries} \label{sec:smooth_rough_boundary}

Similar to the 2d case, there are two obvious gapped boundaries of the 3d toric code model: the smooth and rough boundaries. 

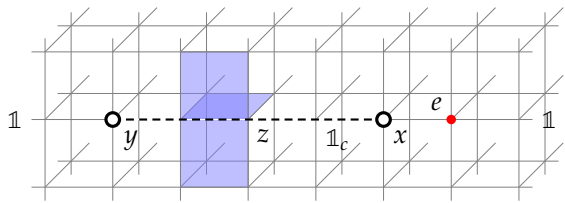
\begin{figure}[htbp]
\centering
\begin{tikzpicture}[scale=0.9]
\foreach \y in {0,...,2}
	\foreach \z in {0,...,1}
		\draw[help lines] (-3.2,\y,\z)--(4.2,\y,\z) ;
\foreach \x in {-3,...,4}
	\foreach \z in {0,...,1}
		\draw[help lines] (\x,-0.2,\z)--(\x,2.2,\z) ;
\foreach \x in {-3,...,4}
	\foreach \y in {0,...,2}
		\draw[help lines] (\x,\y,-0.7)--(\x,\y,1) ;
\draw[white,line width=0.02cm] (-2,1,1)--(2,1,1) ;
\begin{scope}
\clip (-1,1,0)--(0,1,0)--(0,1,0.98)--(-1,1,0.98)--cycle ;
\draw[help lines,fill=m_ext,fill opacity=0.5] (-1,1,0)--(0,1,0)--(0,1,1)--(-1,1,1)--cycle ;
\end{scope}
\begin{scope}
\clip (-1,2,1) rectangle (0,1.02,1) ;
\draw[help lines,fill=m_ext,fill opacity=0.5] (-1,2,1) rectangle (0,1,1) ;
\end{scope}
\begin{scope}
\clip (-1,0,1) rectangle (0,0.98,1) ;
\draw[help lines,fill=m_ext,fill opacity=0.5] (-1,0,1) rectangle (0,1,1) ;
\end{scope}
\node[below right] at (0,1,1) {$z$} ;
\draw[densely dashed,thick] (-2,1,1)--(2,1,1) node[near end,below right] {$\ot$} ;
\node[right] at (4.2,1,1) {$\one$} ;
\node[left] at (-3.2,1,1) {$\one$} ;
\draw[very thick,fill=white] (2,1,1) circle (0.1) node[below right] {$x$} ;
\draw[very thick,fill=white] (-2,1,1) circle (0.1) node[below right] {$y$} ;
\fill[e_ext] (3,1,1) circle (0.07) node[above left,black] {$e$} ;
\end{tikzpicture}
\caption{the topological defects on the smooth boundary}
\label{fig:excitation_smooth}
\end{figure}

\medskip
The smooth boundary is depicted in Figure~\ref{fig:excitation_smooth}. The $A_v$ operators on the smooth boundary is the product of five $\sigma_x$ operators, and the other operators are the same as the original 3d toric code model. Thus the Hamiltonian is still the sum of mutually commuting operators.

\begin{figure}[htbp]
\centering
\begin{tikzpicture}[scale=0.9]
\foreach \y in {0,...,2}
	\foreach \z in {0,...,1}
		\draw[help lines] (-1.2,\y,\z)--(2.2,\y,\z) ;
\foreach \x in {-1,...,2}
	\foreach \z in {0,...,1}
		\draw[help lines] (\x,-0.2,\z)--(\x,2.2,\z) ;
\foreach \x in {-1,...,2}
	\foreach \y in {0,...,2}
		\draw[help lines] (\x,\y,-0.7)--(\x,\y,1) ;
\foreach \x in {-1,...,1}
	\draw[help lines,fill=m_ext,fill opacity=0.5] (\x,0,0)--(\x,1,0)--(\x,1,1)--(\x,0,1)--cycle ;
\foreach \y in {1,2}
	\draw[help lines,fill=m_ext,fill opacity=0.5] (1,\y,0)--(2,\y,0)--(2,\y,1)--(1,\y,1)--cycle ;
\draw[m_dual_str] (-1.5,0.5,0.5) node[left,black] {$m$} --(1.5,0.5,0.5)--(1.5,2.5,0.5) ;
\draw[m_str] (1,2,1)--(2,2,1) node[midway,link_label] {$1$} ;
\draw[m_str] (1,1,1)--(2,1,1) node[midway,link_label] {$2$} ;
\draw[m_str] (1,1,1)--(1,0,1) node[midway,link_label] {$3$} ;
\draw[m_str] (0,1,1)--(0,0,1) node[midway,link_label] {$4$} ;
\draw[m_str] (-1,1,1)--(-1,0,1) node[midway,link_label] {$5$} ;
\end{tikzpicture}
\caption{An $m$-string can be created/annihilated by local operators on the smooth boundary.}
\label{fig:condensation_m}
\end{figure}
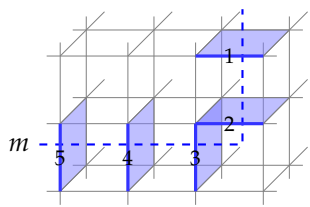

The 1d and 0d topological defects on the smooth boundary are also depicted in Figure~\ref{fig:excitation_smooth}. Figure~\ref{fig:condensation_m} shows that the $m$-strings are condensed on the smooth boundary, because they can be created/annihilated by local operators on the boundary (for example the operator $\sigma_x^1 \sigma_x^2 \sigma_x^3 \sigma_x^4 \sigma_x^5$ in the figure). The $e$-particle and $\ot$-string survive on the smooth boundary, and the $\mt$-string condenses to $\ot$ on the smooth boundary. Hence the topological defects on the smooth boundary form a fusion 2-category $2\rep(\Zb_2)$.

\begin{rem}
The smooth boundary can also be understood as condensing $A=\one\oplus m$, which should be viewed as a condensable algebra in $\toric$. From this point of view, the fusion 2-category $2\rep(\Zb_2)$ of boundary defects can also be viewed as the 2-category $\mathrm{RMod}_A(\toric)$ of right $A$-modules in $\toric$. The 2-category $\mathrm{RMod}_A(\toric)$ has two simple objects $\one \oplus m$ and $\ot \oplus \mt$. It is easy to check that $\ot \oplus \mt$ is not a local $A$-module. Therefore, we obtain that the 2-category of local $A$-modules in $\toric$ is $2\mathrm{Vec}$ and describes the trivial 3d topological order. This story is completely parallel to the condensation theory in 2d topological order \cite{Kon14}. 
\end{rem}

\begin{figure}[htbp]
\centering
\begin{tikzpicture}[scale=0.9]
\foreach \y in {0,...,2}
	\foreach \z in {-1,0}
		\draw[help lines] (-1.2,\y,\z)--(2.2,\y,\z) ;
\foreach \x in {-1,...,2}
	\foreach \z in {-1,0}
		\draw[help lines] (\x,-0.2,\z)--(\x,2.2,\z) ;
\foreach \x in {-1,...,2}
	\foreach \y in {0,...,2}
		\draw[help lines] (\x,\y,-1.2)--(\x,\y,0.99) ;
\foreach \x in {-1,...,2}{
	\begin{scope}
	\clip (\x,0,0)--(\x,1,0)--(\x,1,0.98)--(\x,0,0.98)--cycle ;
	\draw[help lines,fill=m_ext,fill opacity=0.5] (\x,0,0)--(\x,1,0)--(\x,1,1)--(\x,0,1)--cycle ;
	\end{scope}
}
\draw[m_dual_str] (-1.5,0.5,0.5) node[left,black] {$m$} --(2.5,0.5,0.5) ;
\draw[e_str] (1,2,0)--(1,2,1) node[midway,link_label] {$1$} ;
\fill[e_ext] (1,2,0) circle (0.07) node[above left,black] {$e$} ;
\foreach \y in {0,...,2}
	\foreach \z in {1}
		\draw[help lines,dashed] (-1.2,\y,\z)--(2.2,\y,\z) ;
\foreach \x in {-1,...,2}
	\foreach \z in {1}
		\draw[help lines,dashed] (\x,-0.2,\z)--(\x,2.2,\z) ;
\end{tikzpicture}
\caption{The only nontrivial topological defects on the rough boundary are the $m$-strings. The $e$-particles are condensed.}
\label{fig:excitation_rough}
\end{figure}
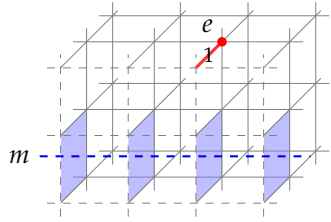

The rough boundary is depicted in Figure~\ref{fig:excitation_rough}. There is no spin on the dashed edges on the boundary, and the $B_p$ operators near the boundary is the product of three $\sigma_z$ operators. On the rough boundary, the $e$-particles are condensed. For example, the $e$-particle Figure~\ref{fig:excitation_rough} can be created/annihilated by $\sigma_z^1$. Also, Figure~\ref{fig:T_condense} shows that $\ot$-strings are condensed on the rough boundary. Thus there are only two simple 1d topological defects on the rough boundary: the trivial string $\one$ and the $m$-string, and there is no nontrivial 0d topological defects. Hence the topological defects on the rough boundary form a fusion 2-category $2\vect_{\Zb_2}$, which is equivalent to $2\vect \boxplus 2\vect$ as 2-categories.

\begin{rem} \label{rem:condensing-T}
The rough boundary is an example of the so-called string-only boundary of a 3d topological order \cite{LKW18}. It can be obtained by condensing all particles in the bulk. Note that $\ot$-strings also condense because they are condensation descendants of the particles. Mathematically, the rough boundary can be obtained by condensing $\ot$, which should be viewed as a condensable algebra in $\toric$ with the multiplication defined by the projection onto the first direct summand of $\ot \otimes \ot = \ot \oplus \ot$. From this point of view, the fusion 2-category $2\vect_{\Zb_2}$ of boundary defects can also be realized by the 2-category $\mathrm{RMod}_{\ot}(\toric)$ of right $\ot$-modules. It has two simple objects: the trivial $\ot$-module $\ot$ and a non-local $\ot$-module $\mt$. Therefore, the 2-category of local $\ot$-modules is $2\vect$ and describes the trivial 3d topological order. This story is again parallel to the condensation theory in 2d topological order \cite{Kon14}. We will develop the condensation theory for 3d topological orders elsewhere. 
\end{rem}

\begin{rem}
More general construction of gapped boundaries of 3d twisted gauge theories can be found in \cite{WLHW18}.
\end{rem}

\subsection{Braiding structure on \texorpdfstring{$\toric$}{Toric}} \label{sec:half_braiding}

According to the boundary-bulk relation \cite{KWZ15,KWZ17}, the topological defects in the bulk should be described by the center of those on a gapped/gapless boundary. In the case of 3d toric code model, it means that we should have the following two braided monoidal equivalences $\Z(2\rep(\Zb_2)) \simeq \toric \simeq \Z(2\vect_{\Zb_2})$. In particular, the boundary-bulk relation of the rough boundary, i.e. $\toric \simeq \Z(2\vect_{\Zb_2})$, tells us how to determine the braiding structures of $\toric$ via the half-braidings in the category $2\vect_{\Zb_2}$ as it was done in \cite{KTZ20}. To realize the braiding physically with respect to the rough boundary, we only need to check if the physical half-braidings of topological defects on the rough boundary coincide with the mathematical half-braidings in $2\vect_{\Zb_2}$ computed in \cite{KTZ20} (see more detail for Appendix~\ref{sec:app}). 

\medskip
A string-like topological defect $X$ in the bulk can be moved to the boundary and becomes a boundary topological defect, denoted by $F(X)$. Since $F(X)$ comes from the bulk, it has half-braidings 1-morphism $R_{F(X),Y} : F(X) \otimes Y \to Y \otimes F(X)$ with boundary topological defects $Y$. Similar to the fact that a braiding can be realized by a 0+1D defect in spacetime as discussed in Section~\ref{sec:braiding_equivalent}, this half-braiding is a 1-morphism (0d domain wall) depicted as follows:
\be
\bigl( F(X) \otimes Y \xrightarrow{R_{F(X),Y}} Y \otimes F(X) \bigr) =
\begin{array}{c}
\begin{tikzpicture}[scale=0.7]
\draw[thick,blue] (1,0) node[below,black] {$Y$} .. controls (1,0.5) and (0,1.5) .. (0,2) node[above,black] {$Y$} ;
\draw[white,line width=4pt] (0,0) .. controls (0,0.5) and (1,1.5) .. (1,2) ;
\draw[thick] (0,0) node[below] {$F(X)$} .. controls (0,0.5) and (1,1.5) .. (1,2) node[above] {$F(X)$} ;
\draw[dashed,fill=gray!5,fill opacity=0.5] (-0.2,1.5) rectangle (1.2,0.5) ;
\node[right] at (1.2,1) {$R_{F(X),Y}$} ;
\end{tikzpicture}
\end{array}
\ee

Similarly, a 0d domain wall $f$ in the bulk between $X$ and $X'$ can be moved to the boundary and becomes a boundary topological defect $F(f)$ between $F(X)$ and $F(X')$. Then $F$ should be a functor from the category of bulk topological defects to the category of boundary topological defects, called the bulk-to-boundary map. The 0d domain wall $F(f)$ also has half-braidings $R_{F(f),Y}$ with all boundary topological defects $Y$. This half-braiding is a 2-morphism (an instanton) depicted as follows:
\be
\begin{array}{c}
\xymatrix@R=2em@C=3em{
F(X') \otimes Y \ar[r]^{R_{F(X'),Y}} & Y \otimes F(X') \\
F(X) \otimes Y \ar[r]^{R_{F(X),Y}} \ar[u]^{F(f) \otimes 1_Y} \urtwocell<\omit>{\quad \, R_{F(f),Y}} & Y \otimes F(X) \ar[u]_{1_Y \otimes F(f)}
}
\end{array}
=
\begin{array}{c}
\begin{tikzpicture}[scale=0.7]
\draw[thick,blue] (1,0) node[below,black] {$Y$} .. controls (1,0.5) and (0,1.5) .. (0,2) node[above,black] {$Y$} ;
\draw[white,line width=4pt] (0,0) .. controls (0,0.5) and (1,1.5) .. (1,2) ;
\draw[thick] (0,0) node[below] {$F(X)$} .. controls (0,0.5) and (1,1.5) .. (1,2) node[above] {$F(X')$} ;
\draw[fill=white] (0.15,0.7) rectangle (0.35,0.5) node[midway,left] {$F(f)$} ;
\draw[-stealth,densely dashed] (0.4,0.6)--(0.8,1.2) ;
\end{tikzpicture}
\end{array}
\hspace{-1em} \xRightarrow{R_{F(f),Y}} \hspace{-1em}
\begin{array}{c}
\begin{tikzpicture}[scale=0.7]
\draw[thick,blue] (1,0) node[below,black] {$Y$} .. controls (1,0.5) and (0,1.5) .. (0,2) node[above,black] {$Y$} ;
\draw[white,line width=4pt] (0,0) .. controls (0,0.5) and (1,1.5) .. (1,2) ;
\draw[thick] (0,0) node[below] {$F(X)$} .. controls (0,0.5) and (1,1.5) .. (1,2) node[above] {$F(X')$} ;
\draw[fill=white] (0.85,1.3) rectangle (0.65,1.5) node[midway,right] {$F(f)$} ;
\end{tikzpicture}
\end{array}
\ee
These half-braidings should satisfy some compatibility conditions \cite[Section~2]{KTZ20}.

\medskip
Let us compute the half-braidings of bulk strings with boundary topological defects. There are only two simple strings $\one,m$ on the boundary, and clearly the half-braiding of a bulk topological defect with the trivial string $\one$ on the boundary is trivial. So we only need to compute the half-braidings of bulk topological defects with the $m$-string on the boundary.

It is clear that the $\one$-string and $m$-string in the bulk have trivial half-braiding with the $m$-string on the boundary. More precisely, we have $F(\one) = \one$ and $F(m) = m$, and the half-braidings are
\be
R_{F(\one),m} = \bigl( F(\one) \otimes m = \one \otimes m = m \xrightarrow{1_m} m = m \otimes \one = m \otimes F(\one) \bigr) ,
\ee
and
\be
R_{F(m),m} = \bigl( F(m) \otimes m = m \otimes m = \one \xrightarrow{1_\one} \one = m \otimes m = m \otimes F(m) \bigr) .
\ee

To compute the half-braidings of the $\ot$-string and $\mt$-string with the $m$-string on the rough boundary, we need to compute $F(\ot)$ and $F(\mt)$. Figure~\ref{fig:T_condense} shows that $F(\ot) = \one \oplus \one$ because all the spins on the edges between the $\ot$-string and the rough boundary have to be $\uparrow$ or $\downarrow$.

\begin{rem}
The computation of $F(\ot)$ is similar to the fusion rule of two $\ot$-strings in the bulk. Indeed, $\ot$ is a condensable algebra in $\toric$, and the rough boundary can be obtained by condensing the $\ot$-strings. 
\end{rem}

\begin{figure}
\begin{align*}
\begin{array}{c}
\begin{tikzpicture}[scale=0.9]
\foreach \y in {0,...,2}
	\foreach \z in {-1,0}
		\draw[help lines] (-1.2,\y,\z)--(2.2,\y,\z) ;
\foreach \x in {-1,...,2}
	\foreach \z in {-1,0}
		\draw[help lines] (\x,-0.2,\z)--(\x,2.2,\z) ;
\foreach \x in {-1,...,2}
	\foreach \y in {0,...,2}
		\draw[help lines] (\x,\y,-1.2)--(\x,\y,0.99) ;
\draw[white,line width=0.02cm] (-1.2,1,0)--(2.2,1,0) ;
\draw[densely dashed,thick] (-1.2,1,0)--(2.2,1,0) node[at end,right] {$\ot$} ;
\foreach \y in {0,...,2}
	\foreach \z in {1}
		\draw[help lines,dashed] (-1.2,\y,\z)--(2.2,\y,\z) ;
\foreach \x in {-1,...,2}
	\foreach \z in {1}
		\draw[help lines,dashed] (\x,-0.2,\z)--(\x,2.2,\z) ;
\end{tikzpicture}
\end{array}
& =
\begin{array}{c}
\begin{tikzpicture}[scale=0.9]
\foreach \y in {0,...,2}
	\foreach \z in {-1,0}
		\draw[help lines] (-1.2,\y,\z)--(2.2,\y,\z) ;
\foreach \x in {-1,...,2}
	\foreach \z in {-1,0}
		\draw[help lines] (\x,-0.2,\z)--(\x,2.2,\z) ;
\foreach \x in {-1,...,2}
	\foreach \y in {0,...,2}
		\draw[help lines] (\x,\y,-1.2)--(\x,\y,0.99) ;
\draw[white,line width=0.02cm] (-1.2,1,0)--(2.2,1,0) ;
\draw[densely dashed,thick] (-1.2,1,0)--(2.2,1,0) node[at end,above] {$\ot$} ;
\foreach \x in {-1,...,2}
	\node at (\x,1,0.5) {$\uparrow$} ;
\foreach \y in {0,...,2}
	\foreach \z in {1}
		\draw[help lines,dashed] (-1.2,\y,\z)--(2.2,\y,\z) ;
\foreach \x in {-1,...,2}
	\foreach \z in {1}
		\draw[help lines,dashed] (\x,-0.2,\z)--(\x,2.2,\z) ;
\end{tikzpicture}
\end{array}
\oplus
\begin{array}{c}
\begin{tikzpicture}[scale=0.9]
\foreach \y in {0,...,2}
	\foreach \z in {-1,0}
		\draw[help lines] (-1.2,\y,\z)--(2.2,\y,\z) ;
\foreach \x in {-1,...,2}
	\foreach \z in {-1,0}
		\draw[help lines] (\x,-0.2,\z)--(\x,2.2,\z) ;
\foreach \x in {-1,...,2}
	\foreach \y in {0,...,2}
		\draw[help lines] (\x,\y,-1.2)--(\x,\y,0.99) ;
\draw[white,line width=0.02cm] (-1.2,1,0)--(2.2,1,0) ;
\draw[densely dashed,thick] (-1.2,1,0)--(2.2,1,0) node[at end,above] {$\ot$} ;
\foreach \x in {-1,...,2}
	\node at (\x,1,0.5) {$\downarrow$} ;
\foreach \y in {0,...,2}
	\foreach \z in {1}
		\draw[help lines,dashed] (-1.2,\y,\z)--(2.2,\y,\z) ;
\foreach \x in {-1,...,2}
	\foreach \z in {1}
		\draw[help lines,dashed] (\x,-0.2,\z)--(\x,2.2,\z) ;
\end{tikzpicture}
\end{array} \\
& =
\begin{array}{c}
\begin{tikzpicture}[scale=0.9]
\foreach \y in {0,...,2}
	\foreach \z in {-1,0}
		\draw[help lines] (-1.2,\y,\z)--(2.2,\y,\z) ;
\foreach \x in {-1,...,2}
	\foreach \z in {-1,0}
		\draw[help lines] (\x,-0.2,\z)--(\x,2.2,\z) ;
\foreach \x in {-1,...,2}
	\foreach \y in {0,...,2}
		\draw[help lines] (\x,\y,-1.2)--(\x,\y,0.99) ;
\draw[white,line width=0.02cm] (-1.2,1,0)--(2.2,1,0) ;
\draw[densely dashed,thick] (-1.2,1,0)--(2.2,1,0) node[at end,below] {$\one$} ;
\foreach \x in {-1,...,2}{
	\draw[white,line width=0.02cm] (\x,1,0)--(\x,1,1) ;
	\draw[densely dashed,thick] (\x,1,0)--(\x,1,1) ;
}
\foreach \y in {0,...,2}
	\foreach \z in {1}
		\draw[help lines,dashed] (-1.2,\y,\z)--(2.2,\y,\z) ;
\foreach \x in {-1,...,2}
	\foreach \z in {1}
		\draw[help lines,dashed] (\x,-0.2,\z)--(\x,2.2,\z) ;
\end{tikzpicture}
\end{array}
\oplus
\begin{array}{c}
\begin{tikzpicture}[scale=0.9]
\foreach \y in {0,...,2}
	\foreach \z in {-1,0}
		\draw[help lines] (-1.2,\y,\z)--(2.2,\y,\z) ;
\foreach \x in {-1,...,2}
	\foreach \z in {-1,0}
		\draw[help lines] (\x,-0.2,\z)--(\x,2.2,\z) ;
\foreach \x in {-1,...,2}
	\foreach \y in {0,...,2}
		\draw[help lines] (\x,\y,-1.2)--(\x,\y,0.99) ;
\draw[white,line width=0.02cm] (-1.2,1,0)--(2.2,1,0) ;
\foreach \x in {-1,...,2}{
	\draw[white,line width=0.02cm] (\x,1,0)--(\x,1,1) ;
}
\draw[densely dashed,thick] (-1.2,1,0)--(2.2,1,0) node[at end,below] {$\one$} ;
\foreach \x in {-1,...,2}{
	\begin{scope}
	\clip (\x,1.02,0)--(\x,2,0)--(\x,2,0.98)--(\x,1.02,0.98)--cycle ;
	\draw[help lines,fill=m_ext,fill opacity=0.5] (\x,1,0)--(\x,2,0)--(\x,2,1)--(\x,1,1)--cycle ;
	\end{scope}
}
\foreach \x in {-1,...,2}{
	\begin{scope}
	\clip (\x,0.98,0)--(\x,0,0)--(\x,0,0.98)--(\x,0.98,0.98)--cycle ;
	\draw[help lines,fill=m_ext,fill opacity=0.5] (\x,1,0)--(\x,0,0)--(\x,0,1)--(\x,1,1)--cycle ;
	\end{scope}
}
\foreach \x in {-1,...,2}{
	\draw[densely dashed,thick] (\x,1,0)--(\x,1,1) ;
}
\foreach \y in {0,...,2}
	\foreach \z in {1}
		\draw[help lines,dashed] (-1.2,\y,\z)--(2.2,\y,\z) ;
\foreach \x in {-1,...,2}
	\foreach \z in {1}
		\draw[help lines,dashed] (\x,-0.2,\z)--(\x,2.2,\z) ;
\end{tikzpicture}
\end{array} .
\end{align*}
\caption{The $\ot$-strings are condensed on the rough boundary. We have $F(\ot) = \one \oplus \one$.}
\label{fig:T_condense}
\end{figure}

Similarly we have $F(\mt) = m \oplus m$:
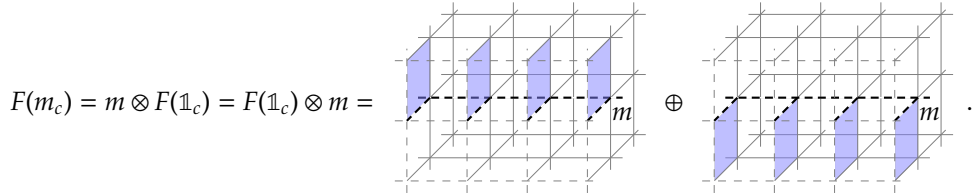
\[
F(\mt) = m \otimes F(\ot) = F(\ot) \otimes m =
\begin{array}{c}
\begin{tikzpicture}[scale=0.8]
\foreach \y in {0,...,2}
	\foreach \z in {-1,0}
		\draw[help lines] (-1.2,\y,\z)--(2.2,\y,\z) ;
\foreach \x in {-1,...,2}
	\foreach \z in {-1,0}
		\draw[help lines] (\x,-0.2,\z)--(\x,2.2,\z) ;
\foreach \x in {-1,...,2}
	\foreach \y in {0,...,2}
		\draw[help lines] (\x,\y,-1.2)--(\x,\y,0.99) ;
\draw[white,line width=0.02cm] (-1.2,1,0)--(2.2,1,0) ;
\draw[densely dashed,thick] (-1.2,1,0)--(2.2,1,0) node[at end,below] {$m$} ;
\foreach \x in {-1,...,2}{
	\begin{scope}
	\clip (\x,1.02,0)--(\x,2,0)--(\x,2,0.98)--(\x,1.02,0.98)--cycle ;
	\draw[help lines,fill=m_ext,fill opacity=0.5] (\x,1,0)--(\x,2,0)--(\x,2,1)--(\x,1,1)--cycle ;
	\end{scope}
}
\foreach \x in {-1,...,2}{
	\draw[white,line width=0.02cm] (\x,1,0)--(\x,1,1) ;
	\draw[densely dashed,thick] (\x,1,0)--(\x,1,1) ;
}
\foreach \y in {0,...,2}
	\foreach \z in {1}
		\draw[help lines,dashed] (-1.2,\y,\z)--(2.2,\y,\z) ;
\foreach \x in {-1,...,2}
	\foreach \z in {1}
		\draw[help lines,dashed] (\x,-0.2,\z)--(\x,2.2,\z) ;
\end{tikzpicture}
\end{array}
\oplus
\begin{array}{c}
\begin{tikzpicture}[scale=0.8]
\foreach \y in {0,...,2}
	\foreach \z in {-1,0}
		\draw[help lines] (-1.2,\y,\z)--(2.2,\y,\z) ;
\foreach \x in {-1,...,2}
	\foreach \z in {-1,0}
		\draw[help lines] (\x,-0.2,\z)--(\x,2.2,\z) ;
\foreach \x in {-1,...,2}
	\foreach \y in {0,...,2}
		\draw[help lines] (\x,\y,-1.2)--(\x,\y,0.99) ;
\draw[white,line width=0.02cm] (-1.2,1,0)--(2.2,1,0) ;
\foreach \x in {-1,...,2}{
	\draw[white,line width=0.02cm] (\x,1,0)--(\x,1,1) ;
}
\draw[densely dashed,thick] (-1.2,1,0)--(2.2,1,0) node[at end,below] {$m$} ;
\foreach \x in {-1,...,2}{
	\begin{scope}
	\clip (\x,0.98,0)--(\x,0,0)--(\x,0,0.98)--(\x,0.98,0.98)--cycle ;
	\draw[help lines,fill=m_ext,fill opacity=0.5] (\x,1,0)--(\x,0,0)--(\x,0,1)--(\x,1,1)--cycle ;
	\end{scope}
}
\foreach \x in {-1,...,2}{
	\draw[densely dashed,thick] (\x,1,0)--(\x,1,1) ;
}
\foreach \y in {0,...,2}
	\foreach \z in {1}
		\draw[help lines,dashed] (-1.2,\y,\z)--(2.2,\y,\z) ;
\foreach \x in {-1,...,2}
	\foreach \z in {1}
		\draw[help lines,dashed] (\x,-0.2,\z)--(\x,2.2,\z) ;
\end{tikzpicture}
\end{array} .
\]

\begin{figure}[htbp]
\begin{align*}
\begin{array}{c}
\begin{tikzpicture}[scale=0.9]
\foreach \y in {0,...,2}
	\foreach \z in {-1,0}
		\draw[help lines] (-1.2,\y,\z)--(2.2,\y,\z) ;
\foreach \x in {-1,...,2}
	\foreach \z in {-1,0}
		\draw[help lines] (\x,-0.2,\z)--(\x,2.2,\z) ;
\foreach \x in {-1,...,2}
	\foreach \y in {0,...,2}
		\draw[help lines] (\x,\y,-1.2)--(\x,\y,0.99) ;
\draw[white,line width=0.02cm] (-1.2,1,0)--(2.2,1,0) ;
\draw[densely dashed,thick] (-1.2,1,0)--(2.2,1,0) node[at end,right] {$\ot$} ;
\begin{scope}
\clip (0,1,-0.02)--(0,1,-1)--(1,1,-1)--(1,1,-0.02)--cycle ;
\draw[help lines,fill=m_ext,fill opacity=0.5] (0,1,0)--(0,1,-1)--(1,1,-1)--(1,1,0)--cycle ;
\end{scope}
\begin{scope}
\clip (0,2,0) rectangle (1,1.02,0) ;
\draw[help lines,fill=m_ext,fill opacity=0.5] (0,2,0) rectangle (1,1,0) ;
\end{scope}
\begin{scope}
\clip (0,0,0) rectangle (1,0.98,0) ;
\draw[help lines,fill=m_ext,fill opacity=0.5] (0,0,0) rectangle (1,1,0) ;
\end{scope}
\foreach \x in {-1,0}{
	\begin{scope}
	\clip (\x,1,0)--(\x,2,0)--(\x,2,0.98)--(\x,1,0.98)--cycle ;
	\draw[help lines,fill=m_ext,fill opacity=0.5] (\x,1,0)--(\x,2,0)--(\x,2,1)--(\x,1,1)--cycle ;
	\end{scope}
}
\foreach \x in {1,2}{
	\begin{scope}
	\clip (\x,1,0)--(\x,0,0)--(\x,0,0.98)--(\x,1,0.98)--cycle ;
	\draw[help lines,fill=m_ext,fill opacity=0.5] (\x,1,0)--(\x,0,0)--(\x,0,1)--(\x,1,1)--cycle ;
	\end{scope}
}
\draw[m_dual_str] (-1.5,1.5,0.5)--(0.5,1.5,0.5)--(0.5,1.5,-0.5)--(0.5,0.5,-0.5)--(0.5,0.5,0.5)--(2.5,0.5,0.5) node[above,black] {$m$} ;
\foreach \y in {0,...,2}
	\foreach \z in {1}
		\draw[help lines,dashed] (-1.2,\y,\z)--(2.2,\y,\z) ;
\foreach \x in {-1,...,2}
	\foreach \z in {1}
		\draw[help lines,dashed] (\x,-0.2,\z)--(\x,2.2,\z) ;
\end{tikzpicture}
\end{array}
& =
\begin{array}{c}
\begin{tikzpicture}[scale=0.9]
\foreach \y in {0,...,2}
	\foreach \z in {-1,0}
		\draw[help lines] (-1.2,\y,\z)--(2.2,\y,\z) ;
\foreach \x in {-1,...,2}
	\foreach \z in {-1,0}
		\draw[help lines] (\x,-0.2,\z)--(\x,2.2,\z) ;
\foreach \x in {-1,...,2}
	\foreach \y in {0,...,2}
		\draw[help lines] (\x,\y,-1.2)--(\x,\y,0.99) ;
\draw[white,line width=0.02cm] (-1.2,1,0)--(2.2,1,0) ;
\draw[densely dashed,thick] (-1.2,1,0)--(2.2,1,0) node[at end,above] {$\ot$} ;
\begin{scope}
\clip (0,1,-0.02)--(0,1,-1)--(1,1,-1)--(1,1,-0.02)--cycle ;
\draw[help lines,fill=m_ext,fill opacity=0.5] (0,1,0)--(0,1,-1)--(1,1,-1)--(1,1,0)--cycle ;
\end{scope}
\begin{scope}
\clip (0,2,0) rectangle (1,1.02,0) ;
\draw[help lines,fill=m_ext,fill opacity=0.5] (0,2,0) rectangle (1,1,0) ;
\end{scope}
\begin{scope}
\clip (0,0,0) rectangle (1,0.98,0) ;
\draw[help lines,fill=m_ext,fill opacity=0.5] (0,0,0) rectangle (1,1,0) ;
\end{scope}
\foreach \x in {-1,0}{
	\begin{scope}
	\clip (\x,1,0)--(\x,2,0)--(\x,2,0.98)--(\x,1,0.98)--cycle ;
	\draw[help lines,fill=m_ext,fill opacity=0.5] (\x,1,0)--(\x,2,0)--(\x,2,1)--(\x,1,1)--cycle ;
	\end{scope}
}
\foreach \x in {1,2}{
	\begin{scope}
	\clip (\x,1,0)--(\x,0,0)--(\x,0,0.98)--(\x,1,0.98)--cycle ;
	\draw[help lines,fill=m_ext,fill opacity=0.5] (\x,1,0)--(\x,0,0)--(\x,0,1)--(\x,1,1)--cycle ;
	\end{scope}
}
\foreach \x in {-1,...,2}
	\node at (\x,1,0.5) {$\uparrow$} ;
\foreach \y in {0,...,2}
	\foreach \z in {1}
		\draw[help lines,dashed] (-1.2,\y,\z)--(2.2,\y,\z) ;
\foreach \x in {-1,...,2}
	\foreach \z in {1}
		\draw[help lines,dashed] (\x,-0.2,\z)--(\x,2.2,\z) ;
\end{tikzpicture}
\end{array}
\oplus
\begin{array}{c}
\begin{tikzpicture}[scale=0.9]
\foreach \y in {0,...,2}
	\foreach \z in {-1,0}
		\draw[help lines] (-1.2,\y,\z)--(2.2,\y,\z) ;
\foreach \x in {-1,...,2}
	\foreach \z in {-1,0}
		\draw[help lines] (\x,-0.2,\z)--(\x,2.2,\z) ;
\foreach \x in {-1,...,2}
	\foreach \y in {0,...,2}
		\draw[help lines] (\x,\y,-1.2)--(\x,\y,0.99) ;
\draw[white,line width=0.02cm] (-1.2,1,0)--(2.2,1,0) ;
\draw[densely dashed,thick] (-1.2,1,0)--(2.2,1,0) node[at end,above] {$\ot$} ;
\begin{scope}
\clip (0,1,-0.02)--(0,1,-1)--(1,1,-1)--(1,1,-0.02)--cycle ;
\draw[help lines,fill=m_ext,fill opacity=0.5] (0,1,0)--(0,1,-1)--(1,1,-1)--(1,1,0)--cycle ;
\end{scope}
\begin{scope}
\clip (0,2,0) rectangle (1,1.02,0) ;
\draw[help lines,fill=m_ext,fill opacity=0.5] (0,2,0) rectangle (1,1,0) ;
\end{scope}
\begin{scope}
\clip (0,0,0) rectangle (1,0.98,0) ;
\draw[help lines,fill=m_ext,fill opacity=0.5] (0,0,0) rectangle (1,1,0) ;
\end{scope}
\foreach \x in {-1,0}{
	\begin{scope}
	\clip (\x,1,0)--(\x,2,0)--(\x,2,0.98)--(\x,1,0.98)--cycle ;
	\draw[help lines,fill=m_ext,fill opacity=0.5] (\x,1,0)--(\x,2,0)--(\x,2,1)--(\x,1,1)--cycle ;
	\end{scope}
}
\foreach \x in {1,2}{
	\begin{scope}
	\clip (\x,1,0)--(\x,0,0)--(\x,0,0.98)--(\x,1,0.98)--cycle ;
	\draw[help lines,fill=m_ext,fill opacity=0.5] (\x,1,0)--(\x,0,0)--(\x,0,1)--(\x,1,1)--cycle ;
	\end{scope}
}
\foreach \x in {-1,...,2}
	\node at (\x,1,0.5) {$\downarrow$} ;
\foreach \y in {0,...,2}
	\foreach \z in {1}
		\draw[help lines,dashed] (-1.2,\y,\z)--(2.2,\y,\z) ;
\foreach \x in {-1,...,2}
	\foreach \z in {1}
		\draw[help lines,dashed] (\x,-0.2,\z)--(\x,2.2,\z) ;
\end{tikzpicture}
\end{array} \\
& =
\begin{array}{c}
\begin{tikzpicture}[scale=0.9]
\foreach \y in {0,...,2}
	\foreach \z in {-1,0}
		\draw[help lines] (-1.2,\y,\z)--(2.2,\y,\z) ;
\foreach \x in {-1,...,2}
	\foreach \z in {-1,0}
		\draw[help lines] (\x,-0.2,\z)--(\x,2.2,\z) ;
\foreach \x in {-1,...,2}
	\foreach \y in {0,...,2}
		\draw[help lines] (\x,\y,-1.2)--(\x,\y,0.99) ;
\draw[white,line width=0.02cm] (-1.2,1,0)--(2.2,1,0) ;
\draw[densely dashed,thick] (-1.2,1,0)--(2.2,1,0) node[at end,below] {$m$} ;
\begin{scope}
\clip (0,1,-0.02)--(0,1,-1)--(1,1,-1)--(1,1,-0.02)--cycle ;
\draw[help lines,fill=m_ext,fill opacity=0.5] (0,1,0)--(0,1,-1)--(1,1,-1)--(1,1,0)--cycle ;
\end{scope}
\begin{scope}
\clip (0,2,0) rectangle (1,1.02,0) ;
\draw[help lines,fill=m_ext,fill opacity=0.5] (0,2,0) rectangle (1,1,0) ;
\end{scope}
\begin{scope}
\clip (0,0,0) rectangle (1,0.98,0) ;
\draw[help lines,fill=m_ext,fill opacity=0.5] (0,0,0) rectangle (1,1,0) ;
\end{scope}
\foreach \x in {-1,0}{
	\begin{scope}
	\clip (\x,1.02,0)--(\x,2,0)--(\x,2,0.98)--(\x,1.02,0.98)--cycle ;
	\draw[help lines,fill=m_ext,fill opacity=0.5] (\x,1,0)--(\x,2,0)--(\x,2,1)--(\x,1,1)--cycle ;
	\end{scope}
}
\foreach \x in {1,2}{
	\begin{scope}
	\clip (\x,0.98,0)--(\x,0,0)--(\x,0,0.98)--(\x,0.98,0.98)--cycle ;
	\draw[help lines,fill=m_ext,fill opacity=0.5] (\x,1,0)--(\x,0,0)--(\x,0,1)--(\x,1,1)--cycle ;
	\end{scope}
}
\foreach \x in {-1,...,2}{
	\draw[white,line width=0.02cm] (\x,1,0)--(\x,1,1) ;
	\draw[densely dashed,thick] (\x,1,0)--(\x,1,1) ;
}
\foreach \y in {0,...,2}
	\foreach \z in {1}
		\draw[help lines,dashed] (-1.2,\y,\z)--(2.2,\y,\z) ;
\foreach \x in {-1,...,2}
	\foreach \z in {1}
		\draw[help lines,dashed] (\x,-0.2,\z)--(\x,2.2,\z) ;
\end{tikzpicture}
\end{array}
\oplus
\begin{array}{c}
\begin{tikzpicture}[scale=0.9]
\foreach \y in {0,...,2}
	\foreach \z in {-1,0}
		\draw[help lines] (-1.2,\y,\z)--(2.2,\y,\z) ;
\foreach \x in {-1,...,2}
	\foreach \z in {-1,0}
		\draw[help lines] (\x,-0.2,\z)--(\x,2.2,\z) ;
\foreach \x in {-1,...,2}
	\foreach \y in {0,...,2}
		\draw[help lines] (\x,\y,-1.2)--(\x,\y,0.99) ;
\draw[white,line width=0.02cm] (-1.2,1,0)--(2.2,1,0) ;
\foreach \x in {-1,...,2}{
	\draw[white,line width=0.02cm] (\x,1,0)--(\x,1,1) ;
}
\draw[densely dashed,thick] (-1.2,1,0)--(2.2,1,0) node[at end,below] {$m$} ;
\begin{scope}
\clip (0,1,-0.02)--(0,1,-1)--(1,1,-1)--(1,1,-0.02)--cycle ;
\draw[help lines,fill=m_ext,fill opacity=0.5] (0,1,0)--(0,1,-1)--(1,1,-1)--(1,1,0)--cycle ;
\end{scope}
\begin{scope}
\clip (0,2,0) rectangle (1,1.02,0) ;
\draw[help lines,fill=m_ext,fill opacity=0.5] (0,2,0) rectangle (1,1,0) ;
\end{scope}
\begin{scope}
\clip (0,0,0) rectangle (1,0.98,0) ;
\draw[help lines,fill=m_ext,fill opacity=0.5] (0,0,0) rectangle (1,1,0) ;
\end{scope}
\foreach \x in {-1,0}{
	\begin{scope}
	\clip (\x,0.98,0)--(\x,0,0)--(\x,0,0.98)--(\x,0.98,0.98)--cycle ;
	\draw[help lines,fill=m_ext,fill opacity=0.5] (\x,1,0)--(\x,0,0)--(\x,0,1)--(\x,1,1)--cycle ;
	\end{scope}
}
\foreach \x in {1,2}{
	\begin{scope}
	\clip (\x,1.02,0)--(\x,2,0)--(\x,2,0.98)--(\x,1.02,0.98)--cycle ;
	\draw[help lines,fill=m_ext,fill opacity=0.5] (\x,1,0)--(\x,2,0)--(\x,2,1)--(\x,1,1)--cycle ;
	\end{scope}
}
\foreach \x in {-1,...,2}{
	\draw[densely dashed,thick] (\x,1,0)--(\x,1,1) ;
}
\foreach \y in {0,...,2}
	\foreach \z in {1}
		\draw[help lines,dashed] (-1.2,\y,\z)--(2.2,\y,\z) ;
\foreach \x in {-1,...,2}
	\foreach \z in {1}
		\draw[help lines,dashed] (\x,-0.2,\z)--(\x,2.2,\z) ;
\end{tikzpicture}
\end{array} .
\end{align*}
\caption{the half-braiding of the $\ot$-string with the $m$-string on the rough boundary}
\label{fig:braiding_T_m}
\end{figure}
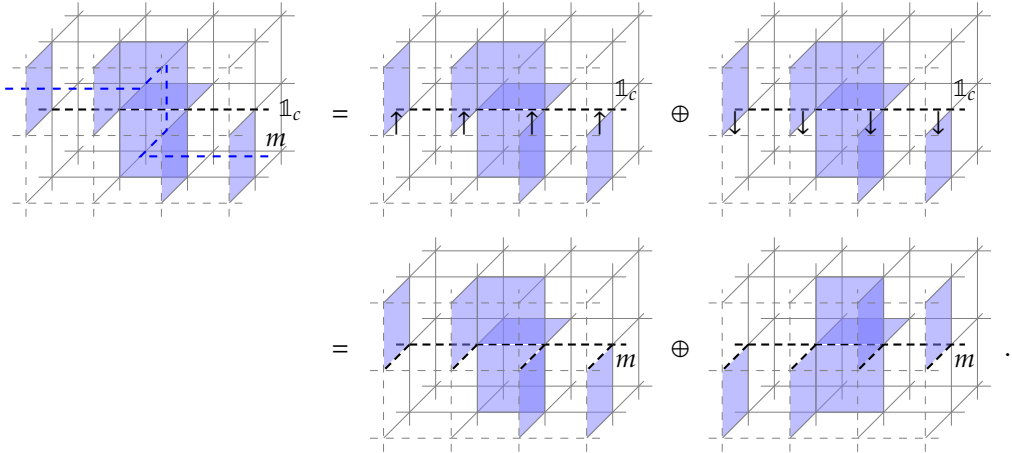

Then we can compute the half-braiding of the $\ot$-string with the $m$-string on the boundary, as depicted in Figure~\ref{fig:braiding_T_m}. Similarly, all the spins on the edges between the $\ot$-string and the rough boundary have to be $\uparrow$ or $\downarrow$. Thus the 0d domain wall in Figure~\ref{fig:braiding_T_m} exchanges two direct summands of $F(\mt) = m \oplus m$:
\be
R_{F(\ot),m} = \bigl( F(\ot) \otimes m = m \oplus m \xrightarrow{\begin{pmatrix} 0 & 1_m \\ 1_m & 0 \end{pmatrix}} m \oplus m = m \otimes F(\ot) \bigr) .
\ee

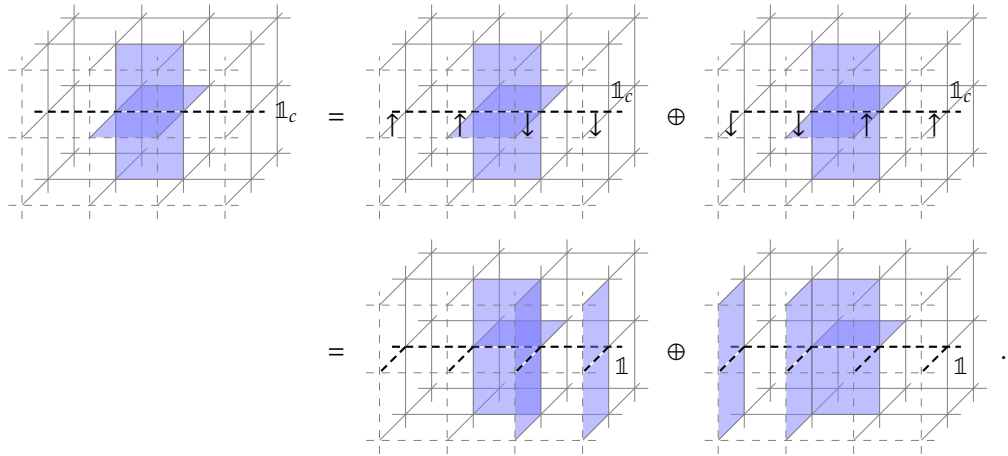
\begin{figure}[htbp]
\begin{align*}
\begin{array}{c}
\begin{tikzpicture}[scale=0.9]
\foreach \y in {0,...,2}
	\foreach \z in {-1,0}
		\draw[help lines] (-1.2,\y,\z)--(2.2,\y,\z) ;
\foreach \x in {-1,...,2}
	\foreach \z in {-1,0}
		\draw[help lines] (\x,-0.2,\z)--(\x,2.2,\z) ;
\foreach \x in {-1,...,2}
	\foreach \y in {0,...,2}
		\draw[help lines] (\x,\y,-1.2)--(\x,\y,0.99) ;
\draw[white,line width=0.02cm] (-1.2,1,0)--(2.2,1,0) ;
\draw[densely dashed,thick] (-1.2,1,0)--(2.2,1,0) node[at end,right] {$\ot$} ;
\begin{scope}
\clip (0,1,-0.02)--(0,1,-1)--(1,1,-1)--(1,1,-0.02)--cycle ;
\draw[help lines,fill=m_ext,fill opacity=0.5] (0,1,0)--(0,1,-1)--(1,1,-1)--(1,1,0)--cycle ;
\end{scope}
\begin{scope}
\clip (0,2,0) rectangle (1,1.02,0) ;
\draw[help lines,fill=m_ext,fill opacity=0.5] (0,2,0) rectangle (1,1,0) ;
\end{scope}
\begin{scope}
\clip (0,0,0) rectangle (1,0.98,0) ;
\draw[help lines,fill=m_ext,fill opacity=0.5] (0,0,0) rectangle (1,1,0) ;
\end{scope}
\begin{scope}
\clip (0,1,0.02)--(0,1,0.98)--(1,1,0.98)--(1,1,0.02)--cycle ;
\draw[help lines,fill=m_ext,fill opacity=0.5] (0,1,0)--(0,1,1)--(1,1,1)--(1,1,0)--cycle ;
\end{scope}
\foreach \y in {0,...,2}
	\foreach \z in {1}
		\draw[help lines,dashed] (-1.2,\y,\z)--(2.2,\y,\z) ;
\foreach \x in {-1,...,2}
	\foreach \z in {1}
		\draw[help lines,dashed] (\x,-0.2,\z)--(\x,2.2,\z) ;
\end{tikzpicture}
\end{array}
& =
\begin{array}{c}
\begin{tikzpicture}[scale=0.9]
\foreach \y in {0,...,2}
	\foreach \z in {-1,0}
		\draw[help lines] (-1.2,\y,\z)--(2.2,\y,\z) ;
\foreach \x in {-1,...,2}
	\foreach \z in {-1,0}
		\draw[help lines] (\x,-0.2,\z)--(\x,2.2,\z) ;
\foreach \x in {-1,...,2}
	\foreach \y in {0,...,2}
		\draw[help lines] (\x,\y,-1.2)--(\x,\y,0.99) ;
\draw[white,line width=0.02cm] (-1.2,1,0)--(2.2,1,0) ;
\draw[densely dashed,thick] (-1.2,1,0)--(2.2,1,0) node[at end,above] {$\ot$} ;
\begin{scope}
\clip (0,1,-0.02)--(0,1,-1)--(1,1,-1)--(1,1,-0.02)--cycle ;
\draw[help lines,fill=m_ext,fill opacity=0.5] (0,1,0)--(0,1,-1)--(1,1,-1)--(1,1,0)--cycle ;
\end{scope}
\begin{scope}
\clip (0,2,0) rectangle (1,1.02,0) ;
\draw[help lines,fill=m_ext,fill opacity=0.5] (0,2,0) rectangle (1,1,0) ;
\end{scope}
\begin{scope}
\clip (0,0,0) rectangle (1,0.98,0) ;
\draw[help lines,fill=m_ext,fill opacity=0.5] (0,0,0) rectangle (1,1,0) ;
\end{scope}
\begin{scope}
\clip (0,1,0.02)--(0,1,0.98)--(1,1,0.98)--(1,1,0.02)--cycle ;
\draw[help lines,fill=m_ext,fill opacity=0.5] (0,1,0)--(0,1,1)--(1,1,1)--(1,1,0)--cycle ;
\end{scope}
\foreach \x in {-1,0}
	\node at (\x,1,0.5) {$\uparrow$} ;
\foreach \x in {1,2}
	\node at (\x,1,0.5) {$\downarrow$} ;
\foreach \y in {0,...,2}
	\foreach \z in {1}
		\draw[help lines,dashed] (-1.2,\y,\z)--(2.2,\y,\z) ;
\foreach \x in {-1,...,2}
	\foreach \z in {1}
		\draw[help lines,dashed] (\x,-0.2,\z)--(\x,2.2,\z) ;
\end{tikzpicture}
\end{array}
\oplus
\begin{array}{c}
\begin{tikzpicture}[scale=0.9]
\foreach \y in {0,...,2}
	\foreach \z in {-1,0}
		\draw[help lines] (-1.2,\y,\z)--(2.2,\y,\z) ;
\foreach \x in {-1,...,2}
	\foreach \z in {-1,0}
		\draw[help lines] (\x,-0.2,\z)--(\x,2.2,\z) ;
\foreach \x in {-1,...,2}
	\foreach \y in {0,...,2}
		\draw[help lines] (\x,\y,-1.2)--(\x,\y,0.99) ;
\draw[white,line width=0.02cm] (-1.2,1,0)--(2.2,1,0) ;
\draw[densely dashed,thick] (-1.2,1,0)--(2.2,1,0) node[at end,above] {$\ot$} ;
\begin{scope}
\clip (0,1,-0.02)--(0,1,-1)--(1,1,-1)--(1,1,-0.02)--cycle ;
\draw[help lines,fill=m_ext,fill opacity=0.5] (0,1,0)--(0,1,-1)--(1,1,-1)--(1,1,0)--cycle ;
\end{scope}
\begin{scope}
\clip (0,2,0) rectangle (1,1.02,0) ;
\draw[help lines,fill=m_ext,fill opacity=0.5] (0,2,0) rectangle (1,1,0) ;
\end{scope}
\begin{scope}
\clip (0,0,0) rectangle (1,0.98,0) ;
\draw[help lines,fill=m_ext,fill opacity=0.5] (0,0,0) rectangle (1,1,0) ;
\end{scope}
\begin{scope}
\clip (0,1,0.02)--(0,1,0.98)--(1,1,0.98)--(1,1,0.02)--cycle ;
\draw[help lines,fill=m_ext,fill opacity=0.5] (0,1,0)--(0,1,1)--(1,1,1)--(1,1,0)--cycle ;
\end{scope}
\foreach \x in {-1,0}
	\node at (\x,1,0.5) {$\downarrow$} ;
\foreach \x in {1,2}
	\node at (\x,1,0.5) {$\uparrow$} ;
\foreach \y in {0,...,2}
	\foreach \z in {1}
		\draw[help lines,dashed] (-1.2,\y,\z)--(2.2,\y,\z) ;
\foreach \x in {-1,...,2}
	\foreach \z in {1}
		\draw[help lines,dashed] (\x,-0.2,\z)--(\x,2.2,\z) ;
\end{tikzpicture}
\end{array} \\
& =
\begin{array}{c}
\begin{tikzpicture}[scale=0.9]
\foreach \y in {0,...,2}
	\foreach \z in {-1,0}
		\draw[help lines] (-1.2,\y,\z)--(2.2,\y,\z) ;
\foreach \x in {-1,...,2}
	\foreach \z in {-1,0}
		\draw[help lines] (\x,-0.2,\z)--(\x,2.2,\z) ;
\foreach \x in {-1,...,2}
	\foreach \y in {0,...,2}
		\draw[help lines] (\x,\y,-1.2)--(\x,\y,0.99) ;
\draw[white,line width=0.02cm] (-1.2,1,0)--(2.2,1,0) ;
\draw[densely dashed,thick] (-1.2,1,0)--(2.2,1,0) node[at end,below] {$\one$} ;
\begin{scope}
\clip (0,1,-0.02)--(0,1,-1)--(1,1,-1)--(1,1,-0.02)--cycle ;
\draw[help lines,fill=m_ext,fill opacity=0.5] (0,1,0)--(0,1,-1)--(1,1,-1)--(1,1,0)--cycle ;
\end{scope}
\begin{scope}
\clip (0,2,0) rectangle (1,1.02,0) ;
\draw[help lines,fill=m_ext,fill opacity=0.5] (0,2,0) rectangle (1,1,0) ;
\end{scope}
\begin{scope}
\clip (0,0,0) rectangle (1,0.98,0) ;
\draw[help lines,fill=m_ext,fill opacity=0.5] (0,0,0) rectangle (1,1,0) ;
\end{scope}
\foreach \x in {1,2}{
	\begin{scope}
	\clip (\x,1.02,0)--(\x,2,0)--(\x,2,0.98)--(\x,1.02,0.98)--cycle ;
	\draw[help lines,fill=m_ext,fill opacity=0.5] (\x,1,0)--(\x,2,0)--(\x,2,1)--(\x,1,1)--cycle ;
	\end{scope}
}
\foreach \x in {1,2}{
	\begin{scope}
	\clip (\x,0.98,0)--(\x,0,0)--(\x,0,0.98)--(\x,0.98,0.98)--cycle ;
	\draw[help lines,fill=m_ext,fill opacity=0.5] (\x,1,0)--(\x,0,0)--(\x,0,1)--(\x,1,1)--cycle ;
	\end{scope}
}
\foreach \x in {-1,...,2}{
	\draw[white,line width=0.02cm] (\x,1,0)--(\x,1,1) ;
	\draw[densely dashed,thick] (\x,1,0)--(\x,1,1) ;
}
\foreach \y in {0,...,2}
	\foreach \z in {1}
		\draw[help lines,dashed] (-1.2,\y,\z)--(2.2,\y,\z) ;
\foreach \x in {-1,...,2}
	\foreach \z in {1}
		\draw[help lines,dashed] (\x,-0.2,\z)--(\x,2.2,\z) ;
\end{tikzpicture}
\end{array}
\oplus
\begin{array}{c}
\begin{tikzpicture}[scale=0.9]
\foreach \y in {0,...,2}
	\foreach \z in {-1,0}
		\draw[help lines] (-1.2,\y,\z)--(2.2,\y,\z) ;
\foreach \x in {-1,...,2}
	\foreach \z in {-1,0}
		\draw[help lines] (\x,-0.2,\z)--(\x,2.2,\z) ;
\foreach \x in {-1,...,2}
	\foreach \y in {0,...,2}
		\draw[help lines] (\x,\y,-1.2)--(\x,\y,0.99) ;
\draw[white,line width=0.02cm] (-1.2,1,0)--(2.2,1,0) ;
\foreach \x in {-1,...,2}{
	\draw[white,line width=0.02cm] (\x,1,0)--(\x,1,1) ;
}
\draw[densely dashed,thick] (-1.2,1,0)--(2.2,1,0) node[at end,below] {$\one$} ;
\begin{scope}
\clip (0,1,-0.02)--(0,1,-1)--(1,1,-1)--(1,1,-0.02)--cycle ;
\draw[help lines,fill=m_ext,fill opacity=0.5] (0,1,0)--(0,1,-1)--(1,1,-1)--(1,1,0)--cycle ;
\end{scope}
\begin{scope}
\clip (0,2,0) rectangle (1,1.02,0) ;
\draw[help lines,fill=m_ext,fill opacity=0.5] (0,2,0) rectangle (1,1,0) ;
\end{scope}
\begin{scope}
\clip (0,0,0) rectangle (1,0.98,0) ;
\draw[help lines,fill=m_ext,fill opacity=0.5] (0,0,0) rectangle (1,1,0) ;
\end{scope}
\foreach \x in {-1,0}{
	\begin{scope}
	\clip (\x,1.02,0)--(\x,2,0)--(\x,2,0.98)--(\x,1.02,0.98)--cycle ;
	\draw[help lines,fill=m_ext,fill opacity=0.5] (\x,1,0)--(\x,2,0)--(\x,2,1)--(\x,1,1)--cycle ;
	\end{scope}
}
\foreach \x in {-1,0}{
	\begin{scope}
	\clip (\x,0.98,0)--(\x,0,0)--(\x,0,0.98)--(\x,0.98,0.98)--cycle ;
	\draw[help lines,fill=m_ext,fill opacity=0.5] (\x,1,0)--(\x,0,0)--(\x,0,1)--(\x,1,1)--cycle ;
	\end{scope}
}
\foreach \x in {-1,...,2}{
	\draw[densely dashed,thick] (\x,1,0)--(\x,1,1) ;
}
\foreach \y in {0,...,2}
	\foreach \z in {1}
		\draw[help lines,dashed] (-1.2,\y,\z)--(2.2,\y,\z) ;
\foreach \x in {-1,...,2}
	\foreach \z in {1}
		\draw[help lines,dashed] (\x,-0.2,\z)--(\x,2.2,\z) ;
\end{tikzpicture}
\end{array} .
\end{align*}
\caption{The 0d domain wall $F(z)$ is equal to the half-braiding \eqref{eq:half_braiding_Tm_m}.}
\label{fig:z_condense}
\end{figure}

Similarly, the half-braiding of the $\mt$-string and the $m$-string is
\be \label{eq:half_braiding_Tm_m}
R_{F(\mt),m} = \bigl( F(\mt) \otimes m = \one \oplus \one \xrightarrow{\begin{pmatrix} 0 & 1_\one \\ 1_\one & 0 \end{pmatrix}} \one \oplus \one = m \otimes F(\mt) \bigr) .
\ee
This half-braiding $R_{F(\mt),m}$ is also equal to the 0d defect $F(z)$, which is depicted in Figure~\ref{fig:z_condense}.

\medskip
Finally we compute the half-braidings of 0d domain walls in the bulk with boundary topological defects. The only nontrivial one is the half-braiding of the $e$-particle in the bulk with the $m$-string, as depicted in the following figure: 
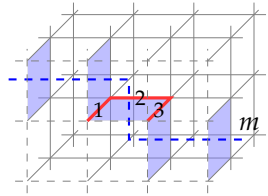
\[
\begin{tikzpicture}[scale=0.8]
\foreach \y in {0,...,2}
	\foreach \z in {-1,0}
		\draw[help lines] (-1.2,\y,\z)--(2.2,\y,\z) ;
\foreach \x in {-1,...,2}
	\foreach \z in {-1,0}
		\draw[help lines] (\x,-0.2,\z)--(\x,2.2,\z) ;
\foreach \x in {-1,...,2}
	\foreach \y in {0,...,2}
		\draw[help lines] (\x,\y,-1.2)--(\x,\y,0.99) ;
\foreach \x in {-1,0}{
	\begin{scope}
	\clip (\x,1,0)--(\x,2,0)--(\x,2,0.98)--(\x,1,0.98)--cycle ;
	\draw[help lines,fill=m_ext,opacity=0.5] (\x,1,0)--(\x,2,0)--(\x,2,1)--(\x,1,1)--cycle ;
	\end{scope}
}
\begin{scope}
\clip (0,1,0.02)--(0,1,0.98)--(1,1,0.98)--(1,1,0.02)--cycle ;
\draw[help lines,fill=m_ext,opacity=0.5] (0,1,0)--(0,1,1)--(1,1,1)--(1,1,0)--cycle ;
\end{scope}
\foreach \x in {1,2}{
	\begin{scope}
	\clip (\x,1,0)--(\x,0,0)--(\x,0,0.98)--(\x,1,0.98)--cycle ;
	\draw[help lines,fill=m_ext,opacity=0.5] (\x,1,0)--(\x,0,0)--(\x,0,1)--(\x,1,1)--cycle ;
	\end{scope}
}
\draw[m_dual_str] (-1.5,1.5,0.5)--(0.5,1.5,0.5)--(0.5,0.5,0.5)--(2.5,0.5,0.5) node[above,black] {$m$} ;
\draw[e_str] (0,1,1)--(0,1,0) node[midway,link_label] {$1$} -- (1,1,0) node[midway,link_label] {$2$} -- (1,1,1) node[midway,link_label] {$3$} ;
\foreach \y in {0,...,2}
	\foreach \z in {1}
		\draw[help lines,dashed] (-1.2,\y,\z)--(2.2,\y,\z) ;
\foreach \x in {-1,...,2}
	\foreach \z in {1}
		\draw[help lines,dashed] (\x,-0.2,\z)--(\x,2.2,\z) ;
\end{tikzpicture}
\]
This half-braiding can be realized by the operator $\sigma_z^1 \sigma_z^2 \sigma_z^3$: first an $e$-particle is created from the rough boundary, then moved across the $m$-string and annihilated to the rough boundary. This operator is nothing but the $B_p$ operator on the middle plaquette, thus it equals to $-1$ due to the existence of the $m$-string. In other words, we have
\[
R_{F(e),m} = \bigl( R_{F(\one),m} \circ (F(e) \otimes 1_m) = 1_m \xrightarrow{-1} 1_m = (1_m \otimes F(e)) \circ R_{F(\one),m} \bigr) .
\]

\subsection{Nondegeneracy}
Now we check the nondegeneracy of the braidings of the braided fusion 2-category $\toric$. Mathematically, $\toric$ is nondegenerate means its sylleptic center is trivial, i.e. equivalent to $2\vect$. The non-degeneracy of $\toric \simeq \Z(2\vect_{\Zb_2})$ has been proved in \cite{KTZ20}. Physically, the non-degeneracy of the braidings means:
\bnu
\item If a simple string $X$ has trivial double braiding with all strings, i.e. $R_{Y,X} \circ R_{X,Y} = 1_{X \otimes Y}$ for all $Y$, then $X = \one$ is the trivial string.
\item If a simple particle $f : \one \to \one$ on the trivial string has trivial double braiding with all strings, i.e. $\bigl( R_{\one,Y} \circ (f \otimes 1_Y) \xrightarrow{R_{f,Y}} (1_Y \otimes f) \circ R_{\one,Y} \xrightarrow{R_{Y,f}} R_{\one,Y} \circ (f \otimes 1_Y) \bigr) = \id_{R_{\one,Y} \circ (f \otimes 1_Y)}$ for all $Y$, then $f = 1_\one$ is the trivial particle.
\enu
We have shown that the double braiding between an $m$-string and a $\ot$-string is nontrivial, and the double braiding between an $e$-particle and an $m$-string is nontrivial. Therefore, we conclude that the braidings in $\toric$ are non-degenerate.

\begin{rem}
The three-loop braidings are important data to distinguish 4-cocyles in 3+1D Dijkgraaf-Witten theories \cite{WL14}, and have a deep relation with 3d modular data \cite{JMR14,WW15a}, and are perhaps also related to the non-degeneracy of braidings. It is an interesting project to compute the three-loop braidings explicitly. We leave it to the future. 
\end{rem}

\appendix
\appendixpage

\section{Braided monoidal 2-category \texorpdfstring{$\Z(2\vect_{\Zb_2})$}{Z(2VecZ2)}} \label{sec:app}

We recall the braiding monoidal structures of the 2-category $\Z(2\vect_{\Zb_2})$ from \cite{KTZ20}. By construction, $\one,m = \vect$ and $\ot,\mt = \vect_{\Zb_2}$ as $\vect_{\Zb_2}$-modules. The 2-category structure of $\Z(2\vect_{\Zb_2})$ can be conveniently represented by the following quiver.  
\be \label{eq:Z(2Rep(Z_2))-2}
\xymatrix{
\one \ar@(ul,ur)[]^{\rep(\Zb_2)}  \ar@/^/[rr]^{\vect} & & \ot \ar@(ul,ur)[]^{\vect_{\Zb_2}} \ar@/^/[ll]^{\vect}
& & m \ar@(ul,ur)[]^{\rep(\Zb_2)}  \ar@/^/[rr]^{\vect} & & \mt \ar@(ul,ur)[]^{\vect_{\Zb_2}} \ar@/^/[ll]^{\vect}
}
\ee
The non-trivial fusion rules are given by
\be
m\otimes m \simeq \one, \quad\quad \ot\otimes m \simeq \mt, \quad\quad \ot\otimes \ot \simeq \mt \otimes \mt \simeq \ot\oplus \ot; \quad\quad \ot\otimes \mt \simeq \mt \oplus \mt. 
\ee

We focus on the braiding structure below. For $X,Y=\one,\ot,m,\mt$ and $\Zb_2=\{1,s\}$, we have
\begin{align*}
R_{X,Y}: X\otimes Y &\to Y\otimes X \\
(x,y) &\mapsto (y, \rho_g(x)),
\end{align*}
where $g=1$ for $Y=\one,\ot$ and $g=s$ for $Y=m,\mt$ and $\rho_g: X \to X$ is the action of $\delta_g \in \vect_{\Zb_2}$ on $X$, where $\delta_g$ is a simple object of $\vect_{\Zb_2}$ with the $g$-grading. More explicitly, 
\be
R_{X,Y} = 
\begin{cases}
(x,y) \mapsto (y,\rho_s(x)) & \mbox{if $X=\ot,\mt$ and $Y=m,\mt$};  \\
(x,y) \mapsto (y,x) & \text{if otherwise}. 
\end{cases}
\ee
It follows that there are seven non-trivial double-braiding 1-isomorphisms:
\be
X \otimes Y \xrightarrow{R_{Y,X} \circ R_{X,Y}} X \otimes Y =
\begin{cases}
(x,y) \mapsto (\rho_s(x),y) & \mbox{for $X \otimes Y = \ot \otimes m, \ot \otimes \mt, \mt \otimes m$} ; \\
(x,y) \mapsto (x,\rho_s(y)) & \mbox{for $X\otimes Y = m\otimes\ot, m\otimes \mt, \mt\otimes\ot$} ; \\
(x,y) \mapsto (\rho_s(x),\rho_s(y)) & \mbox{for $X\otimes Y = \mt\otimes\mt$} .
\end{cases}
\ee
Using the compatibility between the fusions and braidings, all of the seven can be reduced to only two cases: $X\otimes Y = \ot\otimes m, m\otimes\ot$. Namely, it is 
enough to only check the double braiding between $\ot$ and $m$. 

\smallskip
When the braiding involves non-trivial 1-morphisms $f: X\to X'$, we also have the following 2-isomorphisms: 
$$
\xymatrix{
X\otimes Y \ar[d]_{f \otimes 1} \ar[r]^{R_{X,Y}} & Y\otimes X \ar[d]^{1\otimes f}\\
X'\otimes Y  \ar[r]_{R_{X',Y}}  \ar@{}[ur]|{\Rightarrow R_{f,Y}} &   Y\otimes X' 
}
\quad\quad\quad\quad\quad\quad
\xymatrix{
X\otimes Y \ar[d]_{1 \otimes g} \ar[r]^{R_{X,Y}} & Y\otimes X \ar[d]^{g\otimes 1}\\
X\otimes Y'  \ar[r]_{R_{X',Y}}  \ar@{}[ur]|{\Rightarrow (R_{X,g}=1)} &   Y'\otimes X 
}
$$
When $X=X'$, we have 
\be
R_{f,Y} = 
\begin{cases}
1 & \mbox{if $X=\one,m$ and $f=e$ and $Y=\one, \ot$}; \\
-1 & \mbox{if $X=\one,m$ and $f=e$ and $Y=m, \mt$}; \\
1 & \mbox{if $X=\ot,\mt$ and $f=z$ and $Y=\one, \ot$}; \\
1 & \mbox{if $X=\ot,\mt$ and $f=z$ and $Y=m, \mt$}. 
\end{cases}
\ee
It follows that, there are only four non-trivial double-braiding 2-isomorphisms: 
\be
R_{Y,f} \circ R_{f,Y} = -1 \quad \mbox{for $X \otimes Y = \one \otimes m, \one \otimes \mt, m \otimes m, m \otimes \mt$ and $f=e$} .
\ee
Using the compatibility between the fusions and braidings, all of the four can be reduced to only one case $X\otimes Y=\one\otimes m$ and $f=e$. When $X\nsimeq X'$, the 2-isomorphism $R_{f,Y}$ is always the identity, so is the associated double braidings.

\bibliography{Top}

\end{document}